\documentclass[twocolumn,showpacs,floatfix,aps,pra,showkeys]{revtex4}

\usepackage[dvips]{graphicx,color}
\usepackage{dcolumn}
\usepackage{bm}
\usepackage{amsmath, amssymb, mathrsfs}
\newcommand{\rr}{\mathbf{r}}
\newcommand{\nn}{\mathbf{n}}
\newcommand{\kk}{\mathbf{k}}
\newcommand{\be}{\begin{equation}}
\newcommand{\ee}{\end{equation}}
\newcommand{\bea}{\begin{eqnarray}}
\newcommand{\eea}{\end{eqnarray}}

\begin{document}
\title{Quantitative study of two- and three-dimensional strong localization of matter waves by atomic scatterers}

\author{Mauro Antezza}
\affiliation{Laboratoire Kastler Brossel, \'{E}cole Normale Sup\'{e}rieure,
CNRS and UPMC, 24 rue Lhomond, 75231 Paris, France}
\author{Yvan Castin}
\affiliation{Laboratoire Kastler Brossel, \'{E}cole Normale Sup\'{e}rieure,
CNRS and UPMC, 24 rue Lhomond, 75231 Paris, France}

\author{David A. W. Hutchinson}
\affiliation{The Jack Dodd Centre for Quantum Technology, Department of Physics,
University of Otago, Dunedin 9016, New Zealand}

\date{\today}

\begin{abstract}
We study the strong localization of atomic matter waves in a disordered potential created by atoms pinned at the nodes of a lattice, for both three-dimensional (3D) and two-dimensional (2D) systems. The localization length of the matter wave, the density of localized states, and the occurrence of energy mobility edges (for the 3D system), are numerically investigated as a function of the effective scattering length between the atomic matter wave and the pinned atoms. Both positive and negative matter wave energies are explored. Interesting features of the density of states are discovered at negative energies, where maxima in the density of bound states for the system can be interpreted in terms of bound states of a matter wave atom with a few pinned atomic scatterers. In 3D we found evidence of up to three mobility edges, one at positive energies, and two at negative energies, the latter corresponding to transitions between extended and localized bound states. In 2D, no mobility edge is found, and a rapid exponential-like increase of the localization length is observed at high energy.
\end{abstract}



\maketitle

\section{Introduction \label{sec:Intro}}
The propagation of waves in disordered systems is a
rich physical phenomenon, the object of enduring research interest. Its complexity is due to the fact that the scattering of a wave by a random potential depends on several features:
energy and type of the wave, internal and external degrees of freedom of the scattering potential,
dimensionality and symmetry of the physical system, and possible presence of interaction among the propagating waves. Such variety is the reason for a wide experimental and theoretical studies of the diffusion of several kind of classical and quantum waves by many kinds of disordered potentials. This field of research was started by P.W. Anderson \cite{Anderson58}, who predicted in 3D the localization of a quantum particle experiencing short range
hoping between discrete sites, when the on-site energies are sufficiently random.
Several theories have been developed, leading to a general consensus on the fact that a wave is localized in 1D and 2D infinite disordered systems, independently of its energy and of the strength of disorder. In 3D systems a metal-insulator phase transition induced by sufficiently strong disorder may take place: there exists one (or more) critical energy $E_c$ (called the mobility edge) which separates two energy regions. Waves with an energy on one side of $E_c$ are spatially localized (transport is absent), while waves with an energy on the other side are extended over the entire space (transport is diffusive). The localized waves are characterized by an amplitude which decreases exponentially in space at large distances from a central region, defining a typical length, called the \emph{localization length} $\xi$. The richness of the phase transition appears in the critical region around the critical energy $E_c$: here the physics is supposed to be universal, depending only on the symmetries of the system, and not explicitly on the kind of wave or disordered potential. This is due to the fact that, in the critical region, the localization length diverges with a power law behavior $\xi\propto |E-E_c|^{-\nu}$, where the critical exponent $\nu$ characterizes the universality class of the phase transition \cite{Ramakrishnan85,revue_kirk,revue_russe}.

The localization of waves appears as a very rich phenomenon, but it is also very delicate and complex, often difficult to observe experimentally due to parasitic effects (absorption of the wave, finite size effects, ...). A large scientific literature exists on wave localization for several physical systems. Here we deal with the localization of ultra-cold atomic matter waves. Indeed, ultra-cold atomic gases offer a unique system in terms of isolation from environment, cleanness of the system, realization of the system in several spatial dimensions (1D, 2D, 3D), development of several direct detection techniques, and control of the interaction strength. One can even imagine localizing different kinds of waves (matter or light) in different kinds of disordered potentials (created by light or matter). Recently, several experimental investigations on the localization of atomic matter waves have been performed: in a 1D genuine disordered potential made by a laser speckle \cite{Aspect08}, in a 1D bi-chromatic optical lattice  \cite{Inguscio08}, and in an atomic quasi-periodic kicked rotor where a localization in momentum space has been reported  \cite{Chabe08} (true disorder is absent, but the system, which is 1D in  real space, can be mapped into an Anderson model with several effective dimensions in momentum space, giving the possibility to extract the critical exponents of a 3D Anderson transition).

A different way to realize a disordered potential for atomic matter waves (atoms of species $A$), using atoms (another species $B$) pinned at random positions  at the nodes of an optical lattice, was proposed in \cite{Castin05}. This proposal has potentially several advantages over the laser speckle route. On the experimental side, it may be realized in principle as easily in 1D, in 2D, and in 3D, and also 
since an optimization of the matter wave interaction with each individual scatterer may be performed, one may hope to
reach very short mean free paths and localization lengths. On the theoretical side, it allows an exact numerical study for a large number of scatterers - as many as in typical experiments as we shall see. A quantitative study of this model was done for a 1D system \cite{Castin05}, predicting minimal localization lengths $\lesssim 10 \mu$m. A first investigation of the model for 3D systems was performed in \cite{Castin06}, revealing the existence of a large density of localized states with a very short localization length (a few microns) for positive energies. 
Recently, similar models have been considered in the case where the matter wave experiences both a periodic potential and interactions among the $A$ atoms \cite{DisorderPinned}.

\begin{figure}[htb]
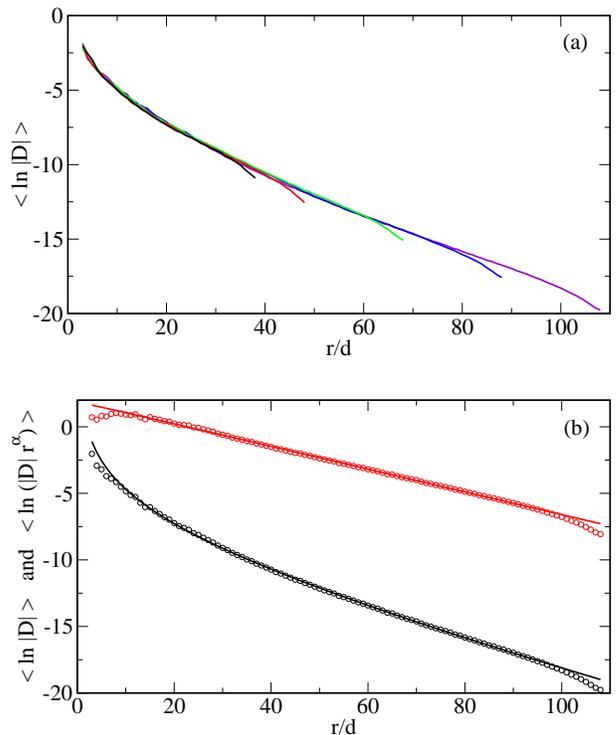

\begin{center}
\includegraphics[width=8cm,clip=]{fig1a.eps}

\vspace{0.5cm}

\includegraphics[width=8cm,clip=]{fig1b.eps}
\end{center}
\caption{(Color online) For a source placed at the origin $\rr_0=\mathbf{0}$, coarse grained histogram
giving $\langle \ln |D|\rangle$ as a function of the distance from the source, averaged over 100
realizations of disorder. The lattice for the $B$ atoms  is cubic with a lattice constant $d$,
and a filling factor $p_{\rm occ}=1/10$.
The energy of the emitting source is $E=\hbar^2 k^2/2m$ with $kd=0.94$.
The effective scattering length is $a_{\rm eff}=0.7d$.
(a) Coarse grained histogram for 5 different
radii $R$ of the sphere containing the scatterers: $R/d=40$ (black), $R/d=50$ (red),
$R/d=70$ (green), $R/d=90$ (blue), $R/d=110$ (violet). (b) For $R/d=110$, coarse grained
histogram (black circles) and its fit by the functional form Eq.~(\ref{eq:ff}) (black lower solid line),
resulting in $\kappa d=0.086$ and $\alpha=2.5$;
the same histogram (red circles) and the same fit (red upper solid line) with $D_i$ multiplied by $r_i^\alpha$.
The fit was performed over the interval $10 < r/d < 90$.
Note that the largest considered value of $R/d=110$ corresponds to a number of scatterers
$N\simeq 5.6\times 10^5$.}
\label{fig:fig1}
\end{figure}

In this paper, we study more extensively  the behavior of the matter wave localization for this model, for both 3D and 2D systems, for both positive and negative energies, and for a broad range of values of the matter wave-pinned
atom interaction strength. We explicitly study the features of the localization length and of the density of localized states, the occurrence of mobility edges, and the effect of the presence of an underlying periodic lattice. We find that, both in 2D and 3D, a strongly resonant interaction between the matter wave
and each pinned atom is required to obtain localized states with small localization lengths.
On the experimental side, this requires a tuning of this interaction by use of a Feshbach resonance.
First steps in this direction were recently taken:
confinement induced resonances between almost free
$A={}^{87}$Rb atoms and $B={}^{41}$K  atoms tightly trapped in a 1D laser standing wave were observed \cite{Inguscio10}.

The article is organized as follows. The first part deals with the 3D case: We present the model
in section \ref{sec:model}, we calculate the localization length in section \ref{sec:loclength} by studying the response
of the system to a source of atomic matter waves $A$ in the medium consisting of pinned $B$ atoms, we calculate and discuss the density of localized states
in section \ref{sec:DOS}, in the form of resonances at positive energies and bound states at negative energies.
The second part of the article contains the same analysis for the 2D case, see section \ref{sec:loc2D}.
We conclude in section \ref{sec:conclusions}.

\section{Physical system and model}
 \label{sec:model}

 We consider two atomic species, $A$ and $B$. The $B$ atoms are tightly trapped at
 the nodes of a cubic optical lattice of lattice spacing $d$,
in a regime where their tunneling among the neighboring
 lattice sites is negligible over the duration of the experiment.
 These $B$ atoms are prepared in the vibrational ground state of the local microtrap, and are randomly
 distributed among the lattice sites, with a uniform occupation probability $p_{\rm occ}$ inside a sphere of radius $R$;
 their locations are independent random variables, except
 for the constraint that there is nowhere more than one $B$ atom per lattice site \cite{note_multiple}.
 
 The  $A$ atoms form the matter wave to be
 strongly localized. They are assumed to move freely in space (in particular they are insensitive
 to the optical lattice), except that they scatter on the $B$ atoms.
 This scattering is assumed to be elastic, under the condition that the kinetic energy of the $A$ atoms
 is much smaller than the quantum of oscillation  of the trapped
 $B$ atoms,
 $\frac{\hbar^2 k^2}{2m} \ll \hbar \omega_{\rm osc}$,
 where $m$ and $\hbar k$ are the mass and momentum of an $A$ atom.
 This scattering is also assumed to be in the zero-range regime $k |r_{\rm{eff}}|\ll1$ where $r_{\rm{eff}}$ is the effective range for the scattering of an $A$ atom on a trapped $B$ atom. Since $r_{\rm{eff}} \simeq a_{\rm ho}$ when the modulus of the 3D free space $A-B$ scattering length $a$ is much larger than
  the harmonic oscillator length $a_{\rm ho}$ of the $B$ atoms (see \cite{Castin06}), we assume that $k a_{\rm ho} \ll1$.  
  In this zero-range regime,
 the $B$ atoms may be considered as point-like scatterers, and the $A-B$
 scattering is characterized by the effective scattering length $a_{\rm eff}$ \cite{noteaeff}.
 In practice, at positive energy $E=\hbar^2k^2/2m>0$ of the matter wave atoms $A$, we shall consider at most an energy of the order of $E_0\equiv\hbar^2/md^2$, so that $k\lesssim 1/d$, and the zero-range condition becomes $d\gg a_{\rm ho}$, a condition well satisfied in a deep optical lattice. At negative energies $E<0$, an $AB$ dimer of energy $E=E_{\rm{dim}}=-\hbar^2/(2ma_{\rm{eff}}^2)$ exists for $a_{\rm eff}>0,$ such
that the zero-range condition becomes $a_{\rm{eff}}\gg a_{\rm{ho}}$. Remarkably, using a Feshbach resonance technique to
 adjust the free-space $A-B$ scattering length $a$, one can realize confinement-induced
 resonances leading to arbitrarily large values of $a_{\rm eff}$
  \cite{Castin06}, so that the condition $a_{\rm{eff}}\gg a_{\rm{ho}}$ may be realized.
 The $B$ atoms may thus effectively constitute a static and strong disordered potential
 for the matter wave.

The problem is thus modeled as follows: The matter wave Hamiltonian is that
of the free $A$ atom, $\mathcal{H}=-\frac{\hbar^2}{2m} \Delta_{\rr}$, where $\Delta_{\rr}$ is the Laplace operator, and the $A-B$ interaction is replaced by
the following contact conditions for the $A$ atom wave-function $\psi(\rr)$: there exist complex numbers $D_i$ such that
\be
\label{eq:contact3d}
\psi(\rr) = -\frac{m}{2\pi \hbar^2}\, D_i \left[|\rr-\rr_i|^{-1} - a_{\rm eff}^{-1}\right] + O(|\rr-\rr_i|)
\ee
in the vicinity of each $B$ scatterer location $\rr_i$. The factor $m/(2\pi\hbar^2)$ is introduced
for convenience. The Bethe-Peierls contact condition (\ref{eq:contact3d}) is equivalent to the pseudo-potential as used in \cite{Castin06}.

\begin{widetext}

\begin{figure}[htb]
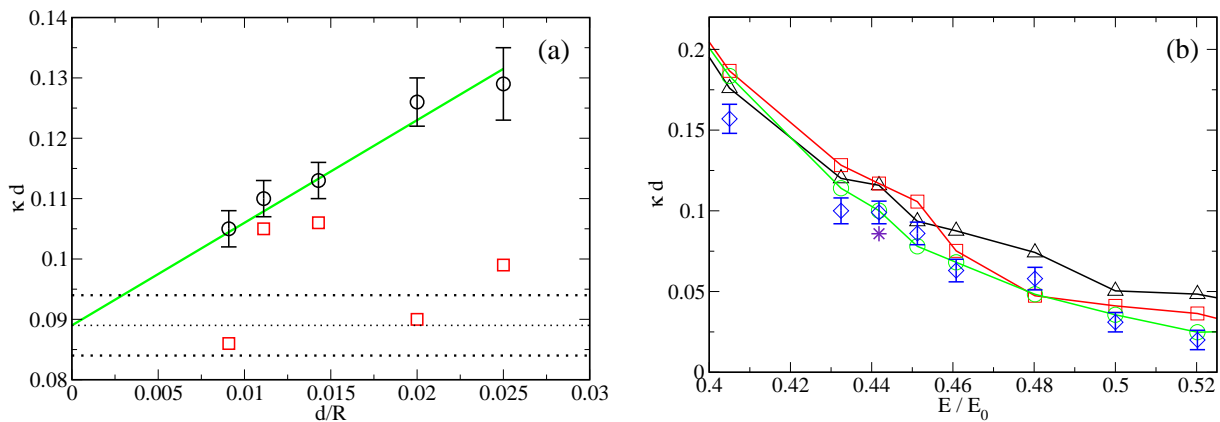

\begin{center}
\includegraphics[width=8cm,clip=]{fig2a.eps}\hspace{.5cm}\includegraphics[width=7.6cm,clip=]{fig2b.eps}
\end{center}
\caption{(Color online) For the 3D system: Comparison of the two methods of calculating the Lyapunov exponent $\kappa=1/\xi$.
As in Fig.\ref{fig:fig1}, the filling factor is $p_{\rm occ}=1/10$, and the effective scattering length is $a_{\rm eff}=0.7d$.
(a) Lyapunov exponent $\kappa$ for a fixed energy of the emitting source, $E=\hbar^2 k^2/2m$ with $kd=0.94$.
Red squares: Lyapunov exponent from the first method, see Eq.~(\ref{eq:ff}), for several values of $d/R$;
each point is the result of a fitting on the interval $10 d < r < R- 20 d$. The second method,
see Eqs.~(\ref{eq:lyapn3d}), (\ref{eq:lyapn3dneg}), and (\ref{eq:lyap3d}), leads to a Lyapunov exponent which is the extrapolation at $d/R=0$ of the black
circles; the green straight
solid line is the linear extrapolation, giving $\kappa d=0.089\pm 0.005$ at $d/R=0$;
the three horizontal dotted lines represent the value of $\kappa$ and its confidence interval.
The two methods are essentially compatible if one considers the scatter in the data for the first method
and the error bars for the second one.
(b) For the two methods, Lyapunov exponent $\kappa$ as a function of the energy of the emitting source $E=\hbar^2 k^2/2m$.
The lattice spacing $d$ of the cubic lattice, and the energy $E_0=\hbar^2/(md^2)$, are used as units. Note that $E_0=2 E_{\rm rec}/\pi^2$, where
$E_{\rm rec}$ is the atomic recoil energy  after absorption of a lattice photon.
Here the filling factor is $p_{\rm occ}=1/10$, and the effective scattering length is $a_{\rm eff}=0.7d$. The solid lines and the violet star
are obtained with the first method [see Eq.~(\ref{eq:ff})].
The three solid lines correspond to sphere radius $R=50d$ (black line, triangles), $R=60d$ (red line, squares),
$R=70d$ (green line, circles), and  an average over 500 realizations of disorder.
The violet star corresponds to a bigger sphere with radius $R=110d$,  and 100 realizations of disorder.
The blue diamond symbols with error bars, are obtained with the second method [see Eqs.~(\ref{eq:lyapn3d}), (\ref{eq:lyapn3dneg}),  and (\ref{eq:lyap3d})] where the extrapolation has been performed using  $d/R=1/50$, $d/R=1/60$, $d/R=1/70$, and  500 realizations of disorder. The second method provides results essentially in agreement with those of the first method.
Note the quite small values of the Lyapunov exponents (large localization lengths).
\label{fig:fig2}
}
\end{figure}
\end{widetext}

Here,  the disordered potential has a finite extension, so that all the positive energy eigenstates
are extended states belonging to a continuum. However, the localized states
that would exist for a strictly infinite extension disorder have \emph{precursors} in the form
of sharp resonances with a width tending exponentially to zero
with the disorder extension \cite{Bart,Castin06}.
At negative energies the matter wave is bound inside the gas of scatterers, and the corresponding bound states can be either extended or localized.
The appropriate tool to find these resonances and bound states is the matter wave Green's function for
an energy $E$, for a {\sl given} realization of the disorder,
\be
\label{eq:def_green}
G(\rr,\rr') = \langle \rr | \left(\frac{1}{E+i0^+-\mathcal{H}}\right)_{\rm c.c.}|\rr'\rangle
\ee
and its analytical continuation to complex energies in the lower half-plane. Here we have used the standard notation  $G(E+i0^+)=\lim_{\eta\to0^+}G(E+i\eta)$. 
Note that $G(\rr,\rr')$ is subject to the same contact conditions as $\psi$ in (\ref{eq:contact3d}), hence the subscript ${\rm c.c.}$ in Eq.(\ref{eq:def_green}).
This Green's function $G$ actually plays a major role in the theory of transport phenomena \cite{Ramakrishnan85}:
The large $|\rr-\rr'|$ behaviors of $\langle G(\rr,\rr')\rangle $ and $\langle \ln |G(\rr,\rr')|\rangle$
provide the scattering mean free path and the localization length, respectively,
where $\langle\ldots\rangle$ represents the average over all realizations of disorder, that is over the
$B$ locations.

A remarkable feature of the point-like scatterers is that the Green's function may be
obtained by the solution of a $N\times N$ complex linear system, where $N$ is the
number of  $B$ scatterers.
This allows an exact numerical calculation, up to $N\approx 5 \times 10^5$ in this work,
a value of the order of typical experiments with atoms in optical lattices.

This remarkable feature may be derived as follows. One starts with the fact that the Green's function
$G(\rr,\rr_0)$
obeys a Schr\"odinger equation with a point-like source term $\delta(\rr-\rr_{0})$ of matter waves at position $\rr_0$.
The diverging terms of the contact conditions (\ref{eq:contact3d}) for the Green's function
give rise to secondary point-like sources
of amplitudes $D_i$ at the scatterers' positions $\rr_i$, by virtue of the usual
relation $\Delta_{\rr} |\rr-\rr_i|^{-1} = -4\pi  \delta(\rr-\rr_i)$. The resulting wave equation
is thus
\begin{multline}
\label{eq:eqg}
\left(E+i0^++\frac{\hbar^2}{2m}\Delta_{\rr}\right ) G(\rr,\rr_0) =
\delta(\rr-\rr_0) \\ + \sum_{i=1}^{N} D_i \delta(\rr-\rr_i).
\end{multline}
This may be integrated using the Green's function $g_0(\rr)$  for $\mathcal{H}$ in the absence of scatterers.
We set
\be
\label{eq:def_k}
E=\frac{\hbar^2 k^2}{2m}.
\ee
For $E>0$, we impose $k>0$, and
\be
\label{eq:g03d}
g_0(\rr) = -\frac{m}{2\pi\hbar^2} \frac{e^{ikr}}{r}.
\ee
For $E<0$, we take $k=iq$, with $q>0$ in Eq.~(\ref{eq:g03d}).
The resulting solution of Eq.~(\ref{eq:eqg}) in thus
\be
\label{eq:gfd3d}
G(\rr,\rr_0) = g_0(\rr-\rr_0) + \sum_{i=1}^{N} D_i g_0(\rr-\rr_i).
\ee
The secondary sources amplitudes $D_i$ are then determined by imposing on (\ref{eq:gfd3d}) the contact
conditions (\ref{eq:contact3d}) at the order $O(1)$. That is, for the non-diverging term $1/a_{\rm eff}$
\be
\label{eq:ls3d}
\sum_{j=1}^{N} M_{ij} D_j =
\frac{2\pi\hbar^2}{m} g_0(\rr_i-\rr_0) , \ \ \ \forall i\in\{1,\ldots,N\}
\ee
where we have introduced the $N\times N$ matrix
\be
\label{eq:def_M}
M_{ij} = \left\{ \begin{array}{ccc} \displaystyle -  \frac{2\pi\hbar^2}{m} g_0(\rr_i-\rr_j) & \, {\rm if}\,  & i\neq j, \\
&& \\
            ik+a_{\rm eff}^{-1} &  \, {\rm if}\,  & i=j.\end{array} \right.
\ee
Eq.~(\ref{eq:ls3d}) constitutes the aforementioned $N\times N$ linear system, and its formal solution gives \cite{Castin06}
\be
\label{eq:gfd3dexp}
G(\rr,\rr_0) = g_0(\rr-\rr_0) + \frac{2\pi\hbar^2}{m} \sum_{i,j=1}^{N} g_0(\rr-\rr_i) [M^{-1}]_{ij} g_0(\rr_j-\rr_0).
\ee

\begin{figure}[htb]
\begin{center}
\includegraphics[width=8cm,clip=]{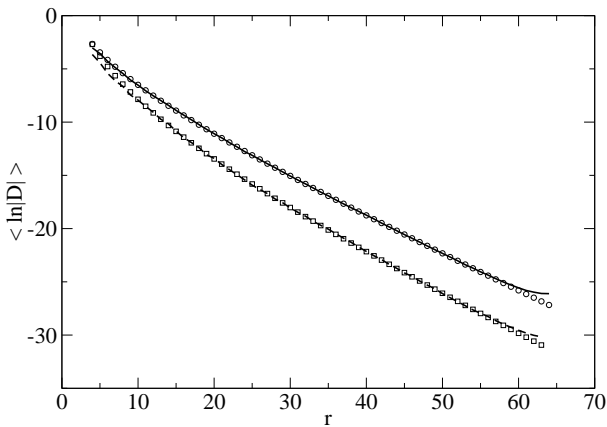}
\caption{For the 3D system: Coarse grained histogram $\langle \ln |D|\rangle$ as a function of the distance from the source, for two different kinds of disorder. The first kind of disorder (solid line) is obtained by a random filling of the cubic lattice inside the sphere of radius
$R=70 d$.
A second kind of disorder (dashed line) is obtained first by a random filling of the cubic lattice inside only one sector (1/8)
of the sphere, $x>0, y>0, z>0$, and then by filling the remaining seven sectors by reflecting the positions in the first sector
with respect to planes $x=0$, $y=0$, $z=0$. The average is taken over 50 realizations of disorder.
The filling factor is $p_{\rm occ}=1/10$, the effective scattering length is $a_{\rm eff}=d$ and the energy of the emitting source is $E=\hbar^2 k^2/2m$ with $kd=0.1$. The circles represent,  for the two kinds of disorder, the fit of the histogram
on the interval $10 d < r < 55 d$  to extract the Lyapunov exponent $\kappa$ with the first method [see Eq.~(\ref{eq:ff})].
We obtain  ($\kappa d=0.31$, $\alpha=2.2$), and ($\kappa d=0.31$, $\alpha=3.6$), for the first and second kind of disorder, respectively. The  Lyapunov exponents appear to be identical, so that one can use the second kind of disorder to calculate $\kappa$
with a substantial gain in the computational effort (a factor 64 on the memory size and a factor 512 in the CPU time).}
\label{fig:fig3}
\end{center}
\end{figure}

The matrix $M$ will play a crucial role in what follows. Indeed, we will relate the localized eigenmodes of the matter wave to the eigenvectors of the matrix $M$. This matrix shows a purely off-diagonal disordered coupling between different scatterers, which is long-range at positive energies, and short-range at negative energies. This kind of disorder does not coincide with the one originally introduced by Anderson, characterized by diagonal disorder and short range couplings \cite{Anderson58}. Finally, it is worth noticing that this model can be exactly mapped to the case of a scalar light wave scattered by a disordered ensemble of two-level atoms, where the same matrix $M$ appears \cite{ScalarLight}.

\begin{widetext}

\begin{figure}[htb]
\begin{center}
\includegraphics[width=16cm,clip=]{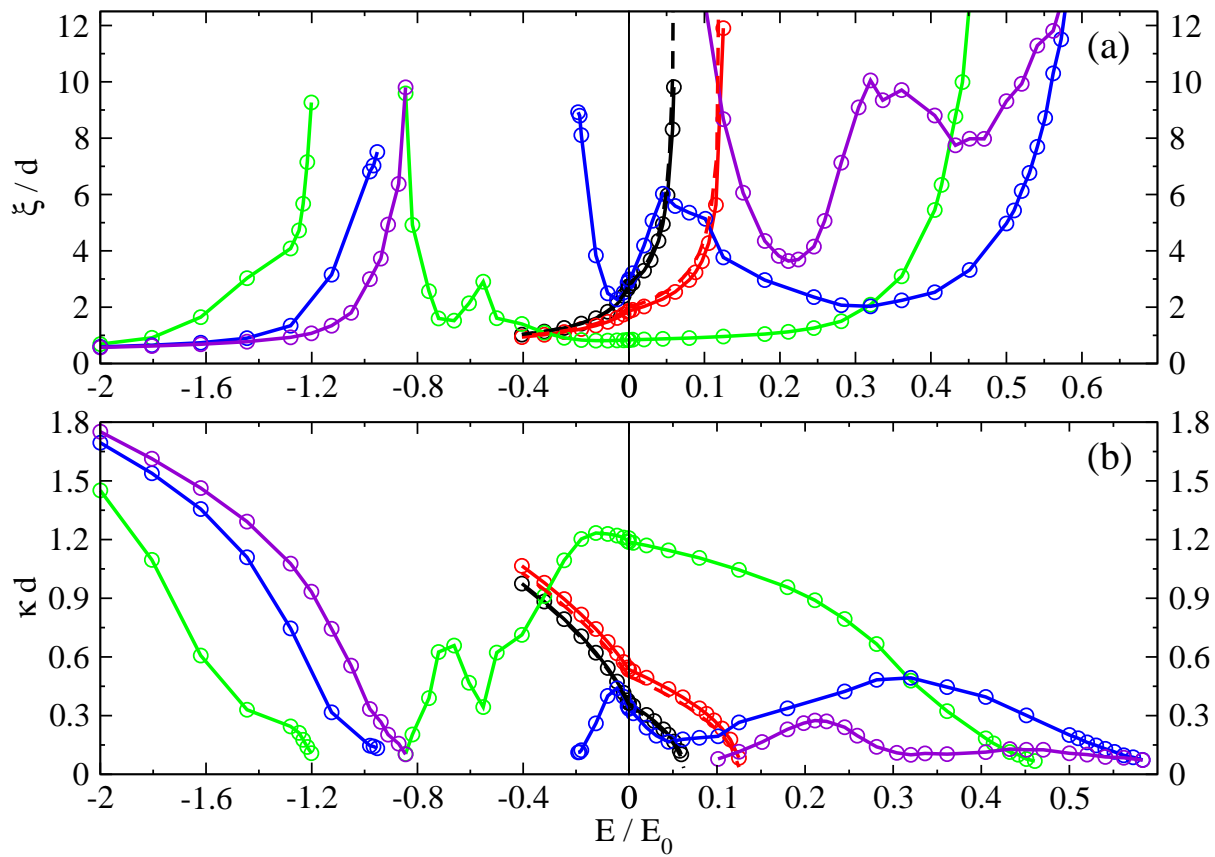}
\end{center}
\caption{(Color online) For the 3D system: Localization length $\xi$ [(a)], and Lyapunov exponents
$\kappa=1/\xi$ [(b)], as a function of the energy
$E$ of the emitting source
for different values of the effective scattering length $a_{\rm eff}$:
$a_{\rm eff}=0.1d$ (black solid line),
$a_{\rm eff}=0.2d$ (red solid line) ,
$a_{\rm eff}=0.7d$ (green solid line),
$a_{\rm eff}=d$ (blue solid line),
$a_{\rm eff}=1.3d$ (violet solid line).
The mean field calculation [see Eq.~(\ref{eq:kappa_mf})]
is also shown for $a_{\rm eff}=0.1d$ (black dashed line),
and for $a_{\rm eff}=0.2d$ (red dashed line).
In (a), for $E/E_0=-1.2$, one has $a_{\rm eff}/d=1.3, 1, 0.7$ from bottom
to top, and for $E/E_0=0.4$, one has $a_{\rm eff}/d=1,0.7,1.3$ from bottom to top.
The lattice spacing $d$ of the cubic lattice,
and the energy $E_0=\hbar^2/(md^2)$, are used as units.
Results are obtained using the first method [see Eq.~(\ref{eq:ff})], with
$500$ realizations of the disorder for $E>0$ and $100$ realizations
for $E<0$.
The filling factor is $p_{\rm occ}=1/1$,  and
the sphere radius is $R=70d$, which leads to a mean number of
scatterers $\langle N\rangle \approx 1.4 \times 10^5$.
Comparing the results for $E>0$ with 100 and 500 realizations,
we estimate that the error on $\kappa$ is $\approx 10\%$, and even smaller
for the lowest values of $\xi$.
\label{fig:fig4}
}
\end{figure}
\end{widetext}

\section{Localization length}
\label{sec:loclength}

\subsection{Defining and calculating the localization length}
\label{subsec:def_xi}

In an infinitely extended disordered three-dimensional system, a matter wave of positive energy emitted by a source is expected to be exponentially suppressed (localized) at large distances, if its energy is smaller than a critical energy $E_c$, the so-called mobility edge. This corresponds to absence of matter wave
transport. The  localization length $\xi$ is the length scale associated with this exponential decay.
It is an average quantity, which has to be calculated by taking the mean over all possible
 realizations of the disorder.
Waves emitted by a source of energy $E$   larger than  $E_c$ are instead expected to  propagate diffusively in the system, with an intensity decaying as the inverse of the distance.

As we shall see, mobility edges can also be present for negative energies. Indeed, a single matter wave $A$ atom   exhibits bound states with a few $B$ scatterers (dimers $AB$, trimers $AB_2$, tetramers $AB_3$, etc.), that have a non-zero  hopping amplitude among different scatterers.
Then, the evanescent matter wave emitted by a source at negative energy may populate such bound states. This may lead to waves of dimers, trimers, etc., bounded in the volume occupied by the scatterers, but which can be either extended or localized within that volume.

To obtain these properties we thus calculate
the Green's function $G({\rm {\bf r}},{\rm {\bf r}}_0)$ which gives the $\rr$-dependent
matter wave amplitude resulting from the source in $\rr_0$.
In practice we solve numerically the linear system Eq.~(\ref{eq:ls3d}) and we extract
the localization length $\xi$ for the amplitudes $D_i$ in two different ways.

\noindent {\it First method:} The spatial localization of the Green's function is reflected in a localization of the secondary source amplitudes $D_i$.
We thus calculate the average over disorder $\langle \ln |D_i|\rangle$,
and construct a coarse grained  histogram of the data set
$(|\rr_i -\rr_0|, \langle \ln |D_i|\rangle)$, that we fit
with the functional form
\be
\label{eq:ff}
|\rr-\rr_0| \longmapsto \ln\left[C \frac{e^{-\kappa |\rr-\rr_0|}}{|\rr-\rr_0|^\alpha}\right],
\ee
where $C$, $\alpha$ and the Lyapunov exponent $\kappa$ are the three free parameters.
The localization length is then  $\xi=1/\kappa$.
The fit is performed over the secondary source amplitudes $D_i$ with positions $\rr_i$ such that
$r_{\rm min} < |\rr_i-\rr_0| < r_{\rm max}$, in order to exclude the near-field  contribution
of the source, and to minimize the effects of the boundaries of the disorder.
Both effects appear in Fig.\ref{fig:fig1}a, which reveals
a boundary layer of $\approx 15$ lattice spacings. Apart from
these boundary effects, the results
for increasing radii $R$ of the sphere containing the scatterers are in good agreement,
and reveal that the decay of $D_i$ is not simply exponential.
On the contrary, the inclusion of the power-law factor in Eq.~(\ref{eq:ff})
provides an excellent fit to the numerical data, as shown in Fig.\ref{fig:fig1}b.
For the system sizes that we are able to treat numerically, the failure to include
 $\alpha$  as a fitting parameter would lead to unreliable values of $\kappa$
 dependent upon the system size.
Finally, we note that the functional form
Eq.~(\ref{eq:ff}) includes both the localized and the diffusive sides of the phase diagram,
with $\xi =+\infty$ in the diffusive regions.

\noindent{\it Second method:} Inspired by the usual definition of the Lyapunov
exponent in one-dimensional disordered systems, we define a direction dependent
transmission coefficient $t(\nn)$ for the field emitted by the source, for a given realization of disorder:
\be
\label{eq:asympt}
G(\rr,\rr_0) \underset{r\to+\infty}{\sim} t(\nn) g_0(\rr-\rr_0),
\ee
with the unit vector $\nn=\rr/r$. Here $f\sim g$ means $f/g\to1$. This expression, compared with Eq.~(\ref{eq:gfd3d}) for $r\to+\infty$, provides the exact relation
\be
\label{eq:t3D}
t(\nn) = 1 + \sum_{i=1}^{N} D_i e^{-i k \nn\cdot (\rr_i-\rr_0)}.
\ee
In the one-dimensional case, the quantity to consider is the logarithm of the transmission
coefficient, rather than the coefficient; the former is indeed a self-averaging
quantity contrary to the latter \cite{self_aver}.
We thus define the direction dependent three-dimensional Lyapunov exponent as
\bea
\kappa(\nn) &=& -\lim_{R\to+\infty} \frac{\langle \ln |t(\nn)|\rangle}{R} \;\;\; {\rm for}\; E>0,
\label{eq:lyapn3d}\\
\kappa(\nn) &=& q-\lim_{R\to+\infty} \frac{\langle \ln |t(\nn)|\rangle}{R} \;\;\; {\rm for}\; E<0,
\label{eq:lyapn3dneg}
\eea
where $R$ is the radius of the sphere containing the point-like scatterers
and the average is taken over the disorder \cite{variante}. The term $q$ in (\ref{eq:lyapn3dneg})
has been introduced since, for $E<0$, the field emitted by the source $g_0(\rr-\rr_0)$ taken as a reference field in Eq.(\ref{eq:asympt}) decays as $e^{-qr}$ for $r\to\infty$.

We have found numerically that $\kappa(\nn)$ only weakly depends on the direction
so that we may define the localization length in terms of its average over solid angle:
\be
\label{eq:lyap3d}
\kappa=\frac{1}{\xi} \equiv \int \frac{d^2n}{4\pi}\, \kappa(\nn).
\ee
In practice, we calculate the limit in Eq.~(\ref{eq:lyapn3d}) and (\ref{eq:lyapn3dneg}) by a linear extrapolation
to $R^{-1}=0$ of the values obtained for at least three different values of $R^{-1}$.

The first and second methods are compared, for a fixed value of the energy $E$ in Fig.\ref{fig:fig2}a and for several
values of $E$ in Fig.\ref{fig:fig2}b, and are found to give compatible values of the Lyapunov exponent.
In practice, the first method has the advantage of not requiring a calculation for different
values of the radius $R$, the border effects being eliminated by a suitable choice of the fitting range.
Also, the value of $|t(\nn)|$ is bounded from below by the numerical accuracy
$\epsilon_{\rm num}$ so that the second method may be used only if $\exp(-\kappa R) \gg \epsilon_{\rm num}$.

\noindent{\sl A symmetry trick:} In practice, to reduce the computational effort,
we have imposed some symmetry properties on the disorder used to obtain the results shown in
Figs.~\ref{fig:fig1}, \ref{fig:fig2}: We have imposed reflection symmetries with respect
to the planes $x=0$, $y=0$, $z=0$ and we have set $\rr_0=\mathbf{0}$, so that the
independent unknowns in the linear system Eq.~(\ref{eq:ls3d}) are the amplitudes
$D_i$ in the sector $x>0,y>0,z>0$. This effectively reduces the number of unknowns
by a factor of $8$, resulting in a gain of a factor 64 on the memory size and of a factor 512 in the CPU time,
without affecting the value of the Lyapunov exponent $\kappa$, as shown in Fig.\ref{fig:fig3}. We have also used this symmetry trick in Figs. \ref{fig:fig4} and \ref{fig:fig5}.

\subsection{Numerical results for the localization length}
\label{subsec:nrftll}

\begin{widetext}

\begin{figure}[htb]
\begin{center}
\includegraphics[width=16cm,clip=]{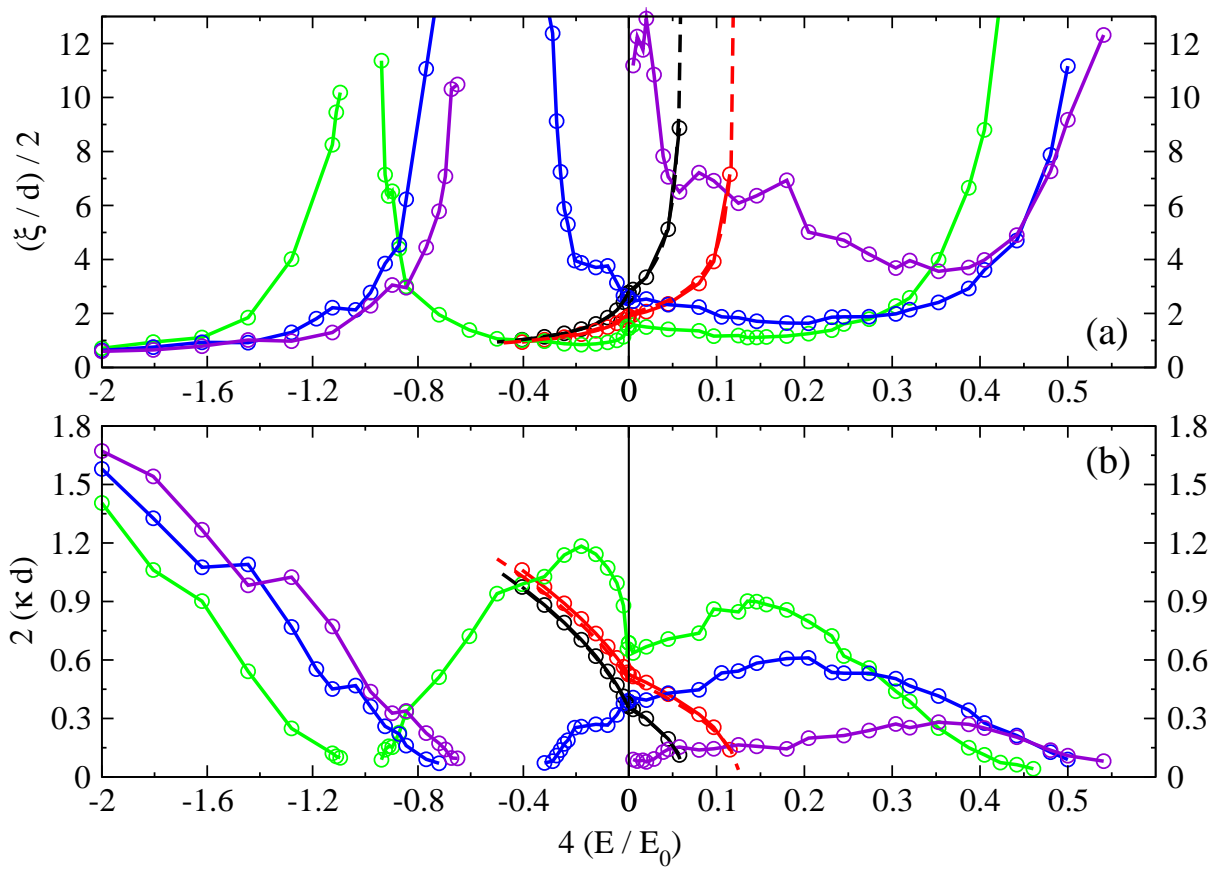}
\end{center}
\caption{
(Color online) The same as Fig.~\ref{fig:fig4}, except that: The value of
$p_{\rm occ}$ is $1/80$, the radius is $R=140 d$ (corresponding to a mean number
of scatterers $\langle N\rangle \approx 1.4\times 10^5$),
and $a_{\rm eff}$ is rescaled to have the same
values of $\rho^{1/3} a_{\rm eff}$ as in Fig.~\ref{fig:fig4},
where $\rho=p_{\rm occ}/d^3$ is the mean density of scatterers.
Since $p_{\rm occ}^{1/3}$ is two times smaller as compared to
Fig.~\ref{fig:fig4}, we have
$a_{\rm eff}=2\times 0.1d$ (black solid line),
$a_{\rm eff}=2\times 0.2d$ (red solid line) ,
$a_{\rm eff}=2\times 0.7d$ (green solid line),
$a_{\rm eff}=2\times d$ (blue solid line),
$a_{\rm eff}=2\times 1.3d$ (violet solid line).
In (a), for $4(E/E_0)=-1.2$, one has $a_{\rm eff}/d=2\times 1.3, 2\times 1, 2\times 0.7$ from bottom
to top, and for $4(E/E_0)=0.2$, one has $a_{\rm eff}/d=2\times 0.7,2\times 1,2\times 1.3$ from bottom to top.
The axes are rescaled accordingly, to allow a direct comparison
with the results of Fig.~\ref{fig:fig4}.
The number of realizations of disorder is equal to $100$ for all points.
}
\label{fig:fig5}
\end{figure}
\end{widetext}

\begin{widetext}

\begin{figure}[htb]
\begin{center}
\includegraphics[width=16cm,clip=]{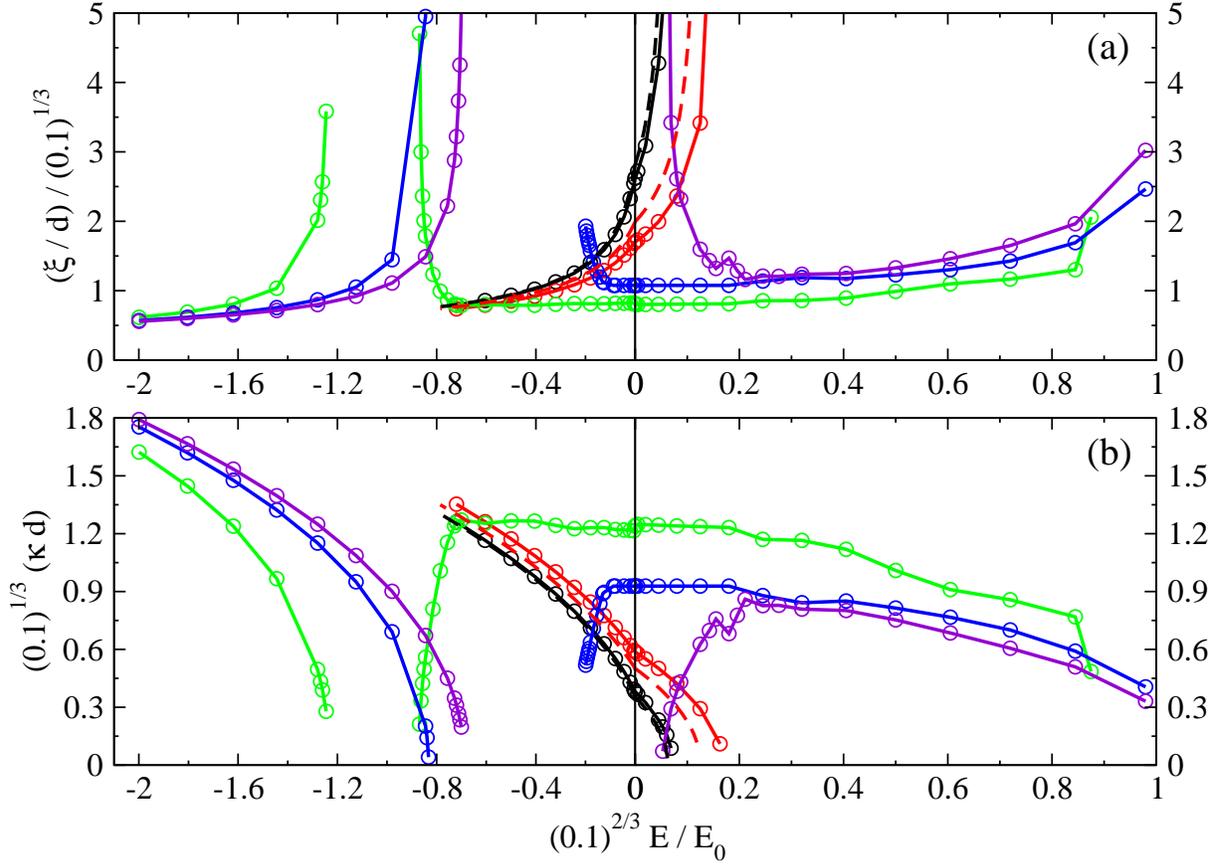}
\end{center}
\caption{
(Color online) The same as Fig.~\ref{fig:fig4}, except that: The value of
$p_{\rm occ}$ is $1$ (there is no disorder,
since all the lattice sites within the sphere of radius
$R=30 d$ are occupied by the scatterers, which leads
to $N\approx 1.1\times 10^5$ \cite{oscillations}),
and $a_{\rm eff}$ is rescaled to have the same
values of $\rho^{1/3} a_{\rm eff}$ as in Fig.~\ref{fig:fig4},
where $\rho=p_{\rm occ}/d^3$ is the mean density of scatterers.
Since $p_{\rm occ}^{1/3}$ is $(0.1)^{1/3}$ times smaller as compared to
Fig.~\ref{fig:fig4}, we have
$a_{\rm eff}=(0.1)^{1/3}\times 0.1d$ (black solid line),
$a_{\rm eff}=(0.1)^{1/3}\times 0.2d$ (red solid line) ,
$a_{\rm eff}=(0.1)^{1/3}\times 0.7d$ (green solid line),
$a_{\rm eff}=(0.1)^{1/3}\times d$ (blue solid line),
$a_{\rm eff}=(0.1)^{1/3}\times 1.3d$ (violet solid line).
In (a), for $(0.1)^{2/3}E/E_0 < -1.2$, one has $a_{\rm eff}/d=(0.1)^{1/3}\times 1.3, (0.1)^{1/3}\times 1,
(0.1)^{1/3}\times 0.7$ from bottom to top, and for $(0.1)^{2/3}E/E_0=0.5$, one has
$a_{\rm eff}/d=(0.1)^{1/3}\times 0.7,(0.1)^{1/3}\times 1, (0.1)^{1/3}\times 1.3$ from
bottom to top. The axes are rescaled accordingly, to allow a direct comparison
with the results of Fig.~\ref{fig:fig4}.
\label{fig:fig6}
}

\end{figure}
\end{widetext}

Using the first method described in the previous subsection,
we have calculated numerically the Lyapunov exponent $\kappa$
and the localization length $\xi=1/\kappa$ for different
values of the effective scattering length $a_{\rm eff}$, and of the
filling factor $p_{\rm occ}$, see Figs.\ref{fig:fig4},\ref{fig:fig5},\ref{fig:fig6}.

A central point is the choice of the value of $p_{\rm occ}$.
Our scope was to study the regime where the presence of the lattice
has a small effect on the localization properties, since we consider
here the lattice mainly as an experimental tool to realize strong disorder.
To this end we will not investigate the regime where $p_{\rm occ}$
only weakly differs from unit: This regime indeed corresponds
to a  matter wave propagating in a periodic structure with dilute
random vacancies.
On the other side, the opposite regime $p_{\rm occ}\to 0$ is not
favorable experimentally since it leads to large localization lengths
scaling at least as the mean distance between scatterers
$d/p_{\rm occ}^{1/3}$ \cite{pourquoi}.

For these reasons, we choose in Fig.\ref{fig:fig4} the reasonable value $p_{\rm occ}=0.1$.
This figure shows the localization length, see Fig.\ref{fig:fig4}a,
and the Lyapunov exponent, see Fig.\ref{fig:fig4}b, as a function of
the energy $E$ of the emitting source, both for negative and positive
values of $E$.
Several values of $a_{\rm eff}$ are considered in this figure.

The values $a_{\rm eff}/d \ll 1$ are easy to analyze in a perturbative picture:
The matter wave inside the cloud of scatterers experiences a mean field
shift equal to $\rho g_{\rm eff}$ where $\rho=p_{\rm occ}/d^3$ is the mean
density of scatterers and $g_{\rm eff}=4\pi\hbar^2 a_{\rm eff}/(2m)$
is the effective coupling constant between the matter wave and a scatterer.
In this mean field model, the Green's function is simply that of a free
matter wave in the presence of the uniform mean field shift $\rho g_{\rm eff}$
inside the sphere. We then have, for $r< R$ and $r_0 < R$
\be
\langle \rr | \frac{1}{E+i0^+-\mathcal{H}_{\rm mf}} |\rr_0\rangle
\simeq -\frac{m}{2\pi\hbar^2} \frac{e^{ik_{\rm mf}r}}{r},
\ee
with $\mathcal{H}_{\rm mf}=\mathcal{H}+\rho g_{\rm eff}$. The wave vector is then given by
\be
\label{eq:emf}
E =\frac{\hbar^2k^2}{2m}= \frac{\hbar^2 k_{\rm mf}^2}{2m} + \rho g_{\rm eff}.
\ee
The mean field prediction is thus that, for $E< \rho g_{\rm eff}$,
$k_{\rm mf}$ is purely imaginary, the matter wave cannot propagate in the gas of scatterers and
is damped with a Lyapunov exponent
\be
\label{eq:kappa_mf}
\kappa_{\rm mf}^2 = 4\pi \rho a_{\rm eff} - k^2.
\ee
The two conditions of validity for this mean field prediction
are that (i) the matter wave scattering on a single scatterer
should be in the regime of the Born approximation, $|ka_{\rm eff}| \ll 1$,
and (ii) there should be a large mean number of scatterers
within the volume $\xi_{\rm mf}^3$, that is $\rho \xi_{\rm mf}^3 \gg 1$,
so one may neglect fluctuations in the density of scatterers over the scale $\xi_{\rm mf}$.
The mean field prediction (\ref{eq:kappa_mf})
is plotted as a dashed line in Fig.~\ref{fig:fig4}
over the range $E< \rho g_{\rm eff}$, for the two lowest values
of $a_{\rm eff}/d$, where it is in good agreement with the numerical
results. For $E > \rho g_{\rm eff}$, the mean field picture cannot of course
 predict how the Green's function decays with $|\rr-\rr_0|$.
In the numerical approach, the localization length turns out
to be too large to be determined in a reliable way for the accessible system sizes.

In the regime $E > \rho g_{\rm eff}$, we have found a second manifestation
of the mean field potential $\rho g_{\rm eff}$. In an optical analogy,
this potential creates a discontinuity in the refractive index for the
matter wave at the border of the sphere containing the scatterers,
the index passing from unity outside the sphere to $n\approx
(1-\rho g_{\rm eff}/E)^{1/2}$ within.
This induces a reflection of the matter wave, with a Fresnel reflection
coefficient $r=(n-1)/(n+1)$.
Since the source position $\rr_0$ is in the center of the sphere,
this gives rise to a radial stationary matter wave with an intensity  contrast
$(I_{\rm max}-I_{\rm min})/(I_{\rm max}+I_{\rm min})=2|r|/(1+|r|^2)$,
provided that one can neglect the attenuation of the coherent part
of the matter field, that is if
the mean free path $\approx 1/(\rho \sigma_{\rm scatt})$,
with $\sigma_{\rm scatt}=4\pi a_{\rm eff}^2$, is larger than the sphere radius.
We have indeed observed this phenomenon numerically.

For larger values of $a_{\rm eff}$, of the order of the lattice
spacing $d$, Fig.~\ref{fig:fig4} shows an interesting structure of
maxima and minima of the Lyapunov exponent at both positive and negative energies.

For $E>0$, the value of $a_{\rm eff}=0.7d$ shows a wide region
of remarkably small values of the localization length,
with $\xi\approx d$ even smaller than the mean distance between scatterers.
For larger values of $E$, a sharp rise of $\xi$ is observed.
A natural question is whether  this rise may be attributed to the presence
of a mobility edge. This requires a proof of the presence
of localized states to the left of this ``edge".
This will be addressed in section \ref{sec:DOS},
where the presence of localized states will be confirmed
for $a_{\rm eff}/d=1$ and $1.3$, in contrast to the cases
$a_{\rm eff}/d=0.1, 0.2$ and $0.7$.
A careful study of the corresponding phase transition and its critical exponents
requires an examination of finite size scaling that we leave for future work.

Another interesting point is that,
further increasing the value of $a_{\rm eff}/d$,  although
intuitively it may be thought to increase the strength of disorder, actually leads
to larger values of $\xi/d$. In particular, at unitarity
($a_{\rm eff}/d\to +\infty$) our system sizes were too small
to allow a reliable determination of $\xi$. Similarly, for negative values
of $a_{\rm eff}/d$, we have not obtained evidence of a finite $\xi$.
This is consistent with the prediction, based on a perturbative calculation
of the transport mean free path, that no matter wave localization can take
place for $a_{\rm eff}<0$ \cite{Bart}.

For $E<0$, Fig.~\ref{fig:fig4} shows an energy interval
where the localization length takes very large
values. At first sight, this may be surprising, as one may naively expect
e.g. from Eq.~(\ref{eq:kappa_mf}), 
that $\kappa$ is an increasing function of $|E|$
at least equal to $(2m|E|)^{1/2}/\hbar$.
In the case where $a_{\rm eff}$ is much smaller than the mean distance between scatterers,
we relate the existence of this energy interval
to the fact that the matter wave can form a bound state (a dimer)
with wavefunction $\phi_0(\rr-\rr_i)$
for each scatterer $i$, of spatial extension $a_{\rm eff}$
and of energy \cite{Enerfewbody}
\be
\label{eq:edim}
E_{\rm dim}=-\frac{\hbar^2}{2 m a_{\rm eff}^2}.
\ee
The various probability amplitudes to have the dimer on the sites
$\rr_i$, $\rr_j$, $\ldots$, are then coupled by transition amplitudes
$t_{\rm trans}$
of the order of $E_{\rm dim}$ times overlap integrals between
$\phi_0(\rr-\rr_i)$  and $\phi_0(\rr-\rr_j)$, integrals that drop exponentially
with the distance as $\exp(-|\rr_i-\rr_j|/a_{\rm eff})$.
A precise calculation gives a transition amplitude between two scatterers
separated by a distance $r_{ij}$ \cite{noteexact}
\be
\label{eq:trans}
t_{\rm trans}(r_{ij}) = -\frac{\hbar^2}{m a_{\rm eff}} \frac{e^{-r_{ij}/a_{\rm eff}}}{r_{ij}}.
\ee
Over the energy interval roughly extending around $E_{\rm dim}$
with a width of the order of typical values of $|t_{\rm trans}|$,
we then face the problem of a bound state that can ``tunnel"
from one scatterer to another, which may result in a
localization length $\xi$. It turns out that this length may have sharp rises as a function of the energy, suggesting the occurrence of two mobility edges.
This picture cannot of course be quantitative
for values of $a_{\rm eff}$ as large as $0.7 d$, but it correctly
predicts that the negative energy interval of large values of $\xi$
shifts to higher energies and broadens, when one increases $a_{\rm eff}/d$.

In Fig.~\ref{fig:fig5}  we consider the same problem as in Fig.~\ref{fig:fig4}
with a much smaller filling factor $p_{\rm occ}=1/80$.
We considered values of $a_{\rm eff}$ such that $\rho^{1/3} a_{\rm eff}$
is the same for the two figures. We also rescaled the axes
of the figure so that two identical values of the abscissa for Fig.~\ref{fig:fig4}
and Fig.~\ref{fig:fig5} correspond to an identical value of $mE/(\rho^{2/3}\hbar^2)$,
and two identical values of the ordinate correspond to an identical
value of $\rho^{1/3}\xi$.
This allows us to see the effect of the underlying lattice \cite{pourquoi}.
There are some quantitative (not qualitative) differences between the two figures.
In particular, at negative energies, the separation between the energies where $\xi$ has a sharp rise is smaller in the case of $p_{\rm occ}=1/80$ than in the case of $p_{\rm occ}=1/10$ and the periodic case $p_{\rm occ}=1$ (as we will see in Fig.~\ref{fig:fig6}), on the rescaled energy axis. An interesting question is whether such a sharp rise is still present in the absence of the lattice.

Finally, we consider in Fig.~\ref{fig:fig6} the ``periodic" case corresponding
to $p_{\rm occ}=1$. This study can be instructive to see to what extent the disorder affects 
the energy dependence of $\kappa$. In the periodic case, of course, the Bloch theorem forbids the existence of localized states in the bulk.  The non-zero values of $\kappa$ are then  
not a signature of disordered-induced localization, but rather result from the existence
of forbidden energy bands: at forbidden energies the matter wave cannot propagate and forms an evanescent wave exponentially decreasing away from the source. 
For $E<0$, as in the disordered case, an allowed energy band is present, which may be interpreted as a dimer energy band for $a_{\rm{eff}}$ much smaller than the lattice spacing.
For $E>0$, the dependence of $\xi$ with the energy is more flat
than in the disordered case, except close to the band edges where it
diverges. We note that the forbidden band is quite broad for
values of $a_{\rm eff}$ of the order of the lattice spacing or larger.

\section{Density of localized states}
\label{sec:DOS}

Calculation of the localization length, as defined in the previous section,
is not sufficient to prove the existence of localized states at the considered energy.
E.g. it is expected that, in a spectral gap of the system, the localization
length is finite whereas there is no state available.
In the present section, we directly investigate the presence of localized
states as a function of the energy $E$.

\subsection{Method}
\label{subsec:methodDOS}

We have to distinguish two cases, according to the sign of the energy $E$.

\noindent{\it Negative energies:}
For $E<0$, the eigenstates of the matter wave are discrete bound states. Such
a bound state corresponds to the matter wave being trapped by the gas of scatterers,
but it is not necessarily a localized state.
In particular, in the limit
$0<a_{\rm eff}$ smaller than the mean distance between scatterers, we recall
the picture of an $AB$ dimer that may tunnel from one scattering site to another, see
subsection \ref{subsec:nrftll}, and this dimer, with a negative total energy, may be
delocalized over the whole gas of scatterers.

The bound state eigenenergies correspond to poles of the Green's function Eq.(\ref{eq:def_green}).
Since $G(\rr,\rr_0)$ diverges for such an eigenenergy,
the secondary source amplitudes $D_i$ diverge in Eq.(\ref{eq:gfd3d}),
or equivalently, according to Eq.(\ref{eq:gfd3dexp}), this implies that the matrix $M$ defined
in Eq.(\ref{eq:def_M}) has a zero eigenvalue.
For $E<0$, the matrix $M$ is real and symmetric,
so has $N$ real eigenvalues $m_i$.
To find the bound state energies, there are two strategies.

In the first strategy, keeping the effective scattering
length $a_{\rm eff}$ fixed, one numerically calculates the $N$ eigenvalues $m_i(E,a_{\rm eff})$,
$1\leq i \leq N$,
and then solves for $E$ each implicit equation
\be
\label{eq:ie3d}
m_i(E,a_{\rm eff})=0.
\ee
This may be done efficiently by dichotomy,  using the property that $m_i$ is a monotonically increasing function of $E$ (see appendix \ref{app:mon}).
In the second strategy, one inverts the problem, working for a fixed value of the energy $E$
and solving $m_i(E,a_{\rm eff})=0$ for $a_{\rm eff}$. This turns out to be straightforward,
because of the following structure of the matrix $M$ \cite{Castin06}:
\be
M = \frac{1}{a_{\rm eff}} \mbox{Id} + M_\infty(E)
\ee
where $M_\infty$ depends only on the energy, not on the effective scattering length.
For each value of $E$, one then simply has to diagonalize $M_\infty(E)$, with
resulting eigenvalues $m_i^{\infty}(E)$. From the relation
$m_i(E,a_{\rm eff})=a_{\rm eff}^{-1} + m_i^{\infty}(E)=0$, one obtains the
following parametrization of the negative eigenenergy branches:
\be
\label{eq:param}
\frac{1}{a_{\rm eff}} = - m_i^{\infty}(E).
\ee

\noindent{\it Positive energies:}
Let us consider first the ideal case of scatterers extending over the entire position space.
Then, for $E>0$ two kinds of states are expected, square integrable
localized states corresponding to a discrete spectrum, and extended states
corresponding to a continuous spectrum. It is expected that the continuous spectrum exists
at $E>E_{\rm c}$, where $E_c$ is called a mobility edge, and that the
localized states are essentially at $E<E_c$ \cite{dispersion}.

In reality, the scatterers occupy a finite volume of radius $R$, so that,
for $E>0$, the energy spectrum of the matter wave
forms a continuum extending from $0$ to $+\infty$,
and none of the positive energy eigenstates is square integrable.
Nevertheless, for a large enough radius $R$, one can still find eigenstates
which, inside the medium, exponentially decrease over several orders of magnitude
away from a central region, toward the borders of the gas of scatterers \cite{Castin06}.
For all practical purposes, these are localized states.
From a spectral point of view,
these localized states are expected to correspond to very narrow resonances of the matter wave
inside the scattering medium, that is to {\sl complex} poles
\be
\label{eq:zrescompl}
z_{\rm res}=E_{\rm res} -i\hbar \Gamma/2
\ee
of the Green's function
Eq.(\ref{eq:def_green}) analytically continued to the lower half of the complex plane,
$\Gamma >0$. Such a resonance state, localized close to the center of the sphere containing the scatterers, has a width $\Gamma$, i.e.  an inverse lifetime,
that vanishes exponentially with the radius $R$ as $\exp(-2R/\xi)$, where $\xi$
is the localization length of the localized states \cite{Bart,Castin06},
reproducing for $R=+\infty$ the aforementioned ideal case of poles on the real axis.

To find such resonances, one has to analytically extend the Green's function
Eq.(\ref{eq:def_green}) to complex energies $z$. Then, setting $k=(2mz/\hbar^2)^{1/2}$
in Eq.(\ref{eq:g03d}), taking the real and negative axis as the line cut for
$z^{1/2}$, one finds that Eq.(\ref{eq:gfd3dexp}) still holds.
Then the complex poles $z_{\rm res}$ are such that the matrix $M$
has a zero eigenvalue.
Adopting the second strategy used for bound states, one actually
diagonalizes $M^\infty(z)$ so that one has to solve the equations for $z_{\rm res}$:
\be
\label{eq:tbs}
\frac{1}{a_{\rm eff}} = - m_i^{\infty}(z_{\rm res}).
\ee
In practice, to solve Eq.(\ref{eq:tbs}), we implement Newton's method as follows.
We choose some real and positive value for $E$. We diagonalize
$M^\infty(E)$, calculating the eigenvalues and the eigenvectors;
we select an eigenvalue $m_i^{\infty}(E)$ and we set the effective
scattering length to the value such that $a_{\rm eff}^{-1} + m_i^{\infty}(E)$
is purely imaginary,
\be
a_{\rm eff}= -\frac{1}{\mathrm{Re}\, m_i^{\infty}(E)}.
\ee
This constitutes the initial guess in Newton's method.
The successive steps of the method involve the calculation
of the derivative of $m_i^{\infty}(z)$ with respect to $z$,
which can be done thanks to the extension of the Hellmann-Feynman
theorem to non-hermitian matrices \cite{newton}.

\begin{widetext}

\begin{figure}[htb]
\begin{center}
\begin{tabular}{cc}
\hspace{5cm} (a) & \hspace{5cm} (b) \\
\vspace{-0.4cm} \\
\includegraphics[width=8.5cm,clip=]{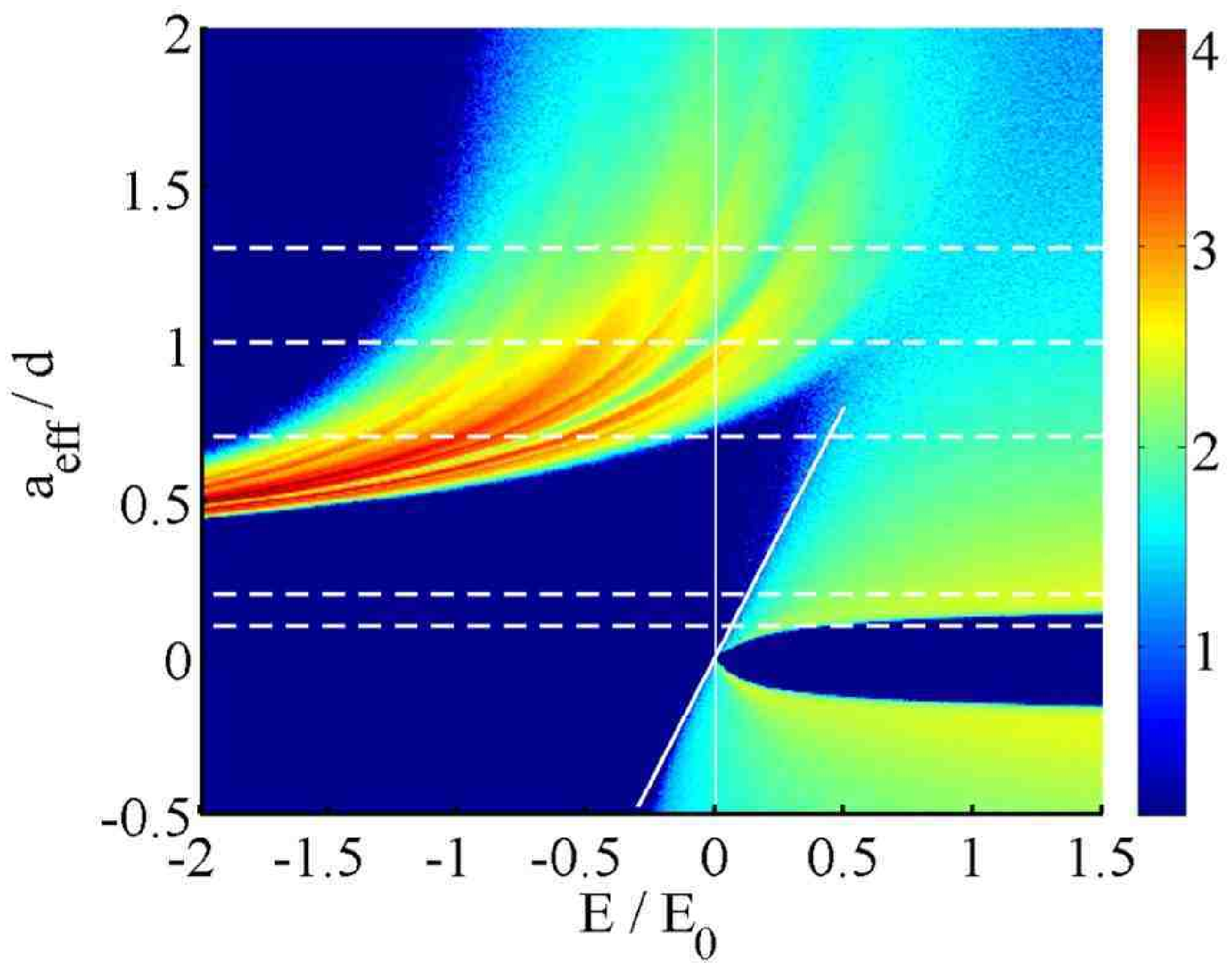} &
\includegraphics[width=9cm,clip=]{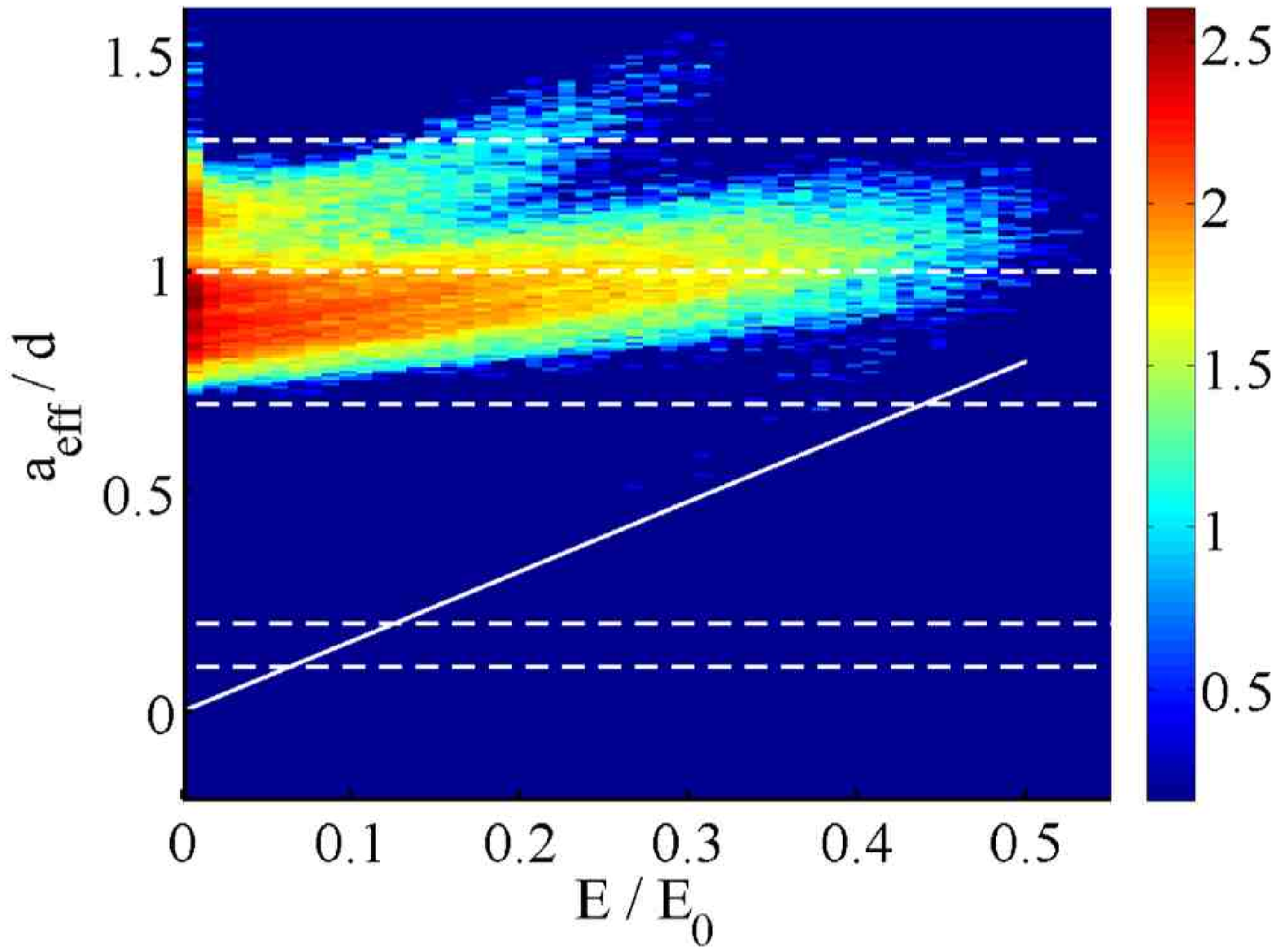} \\
& \\
\hspace{5cm} (c) & \hspace{5cm} (d) \\
\vspace{-0.4cm} \\
\includegraphics[width=8.5cm,clip=]{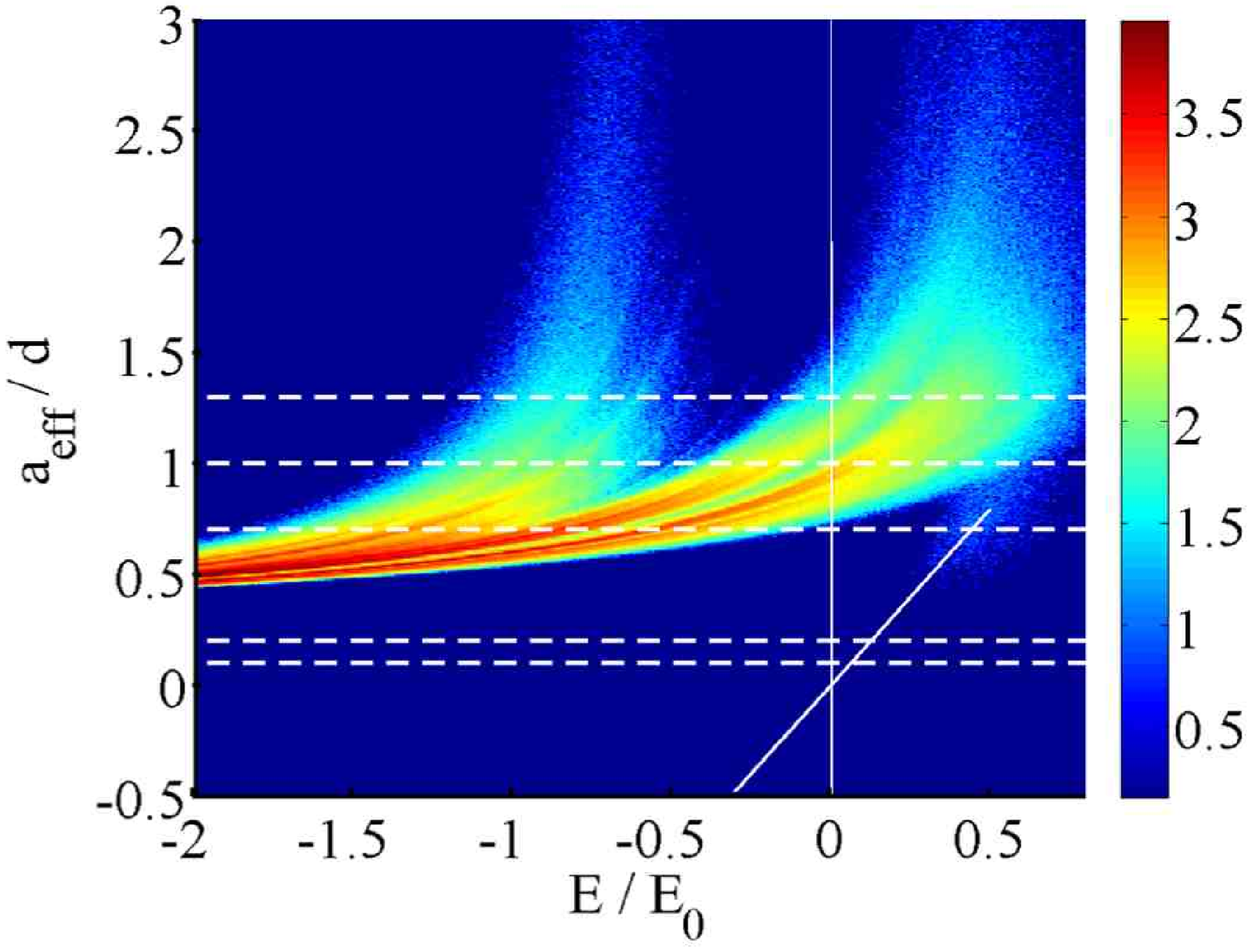}  &
\includegraphics[width=8.5cm,clip=]{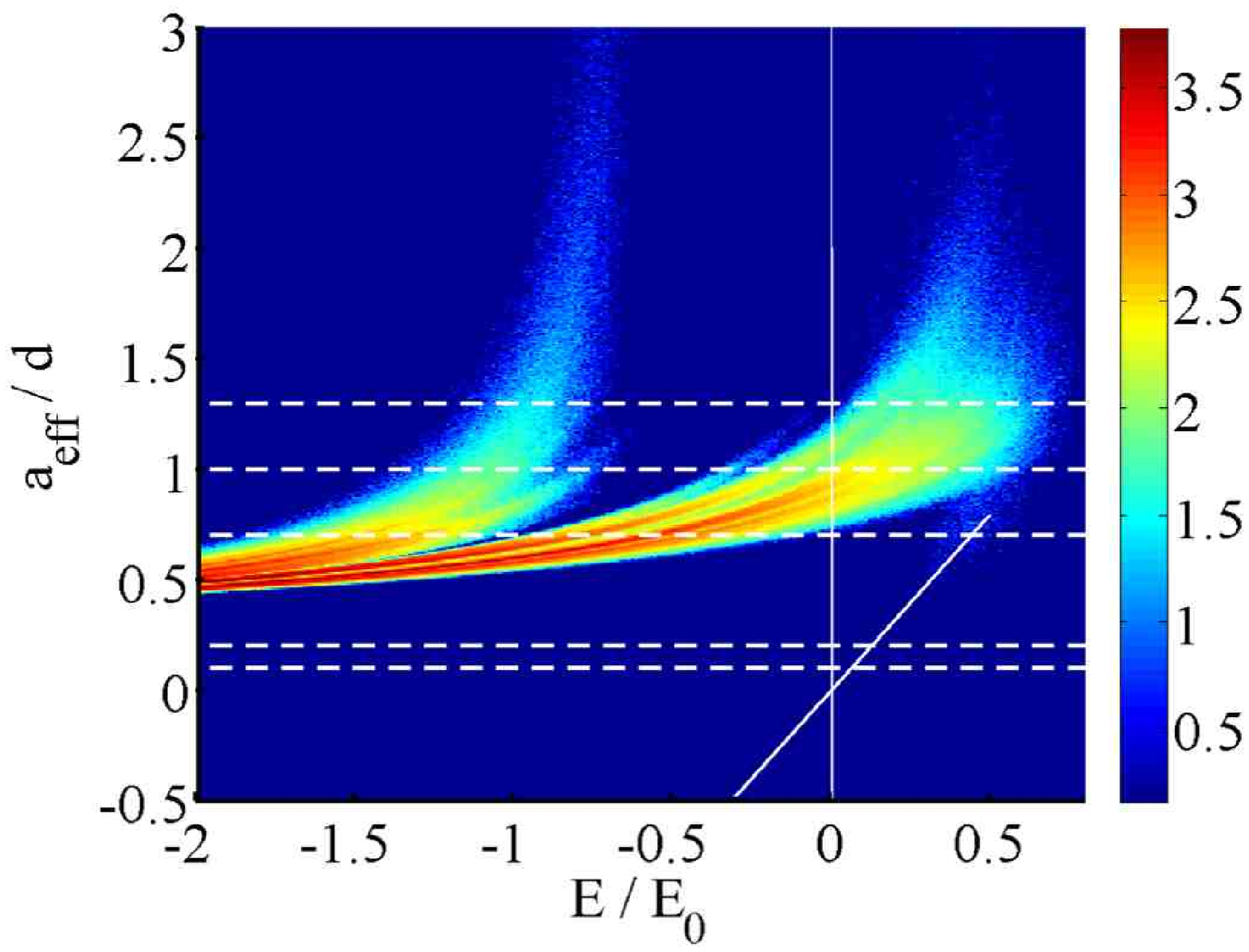}
\end{tabular}
\end{center}
\caption{(Color online) For the 3D system: Density of resonances and bound states
per scatterer in the plane
(energy $E$, effective scattering length $a_{\rm eff}$),
obtained as explained in subsection \ref{subsec:methodDOS}.
$E$ on the horizontal axis is either the real part of $z_{\rm res}$ on the positive energy side,
or the bound state eigenenergy on the negative energy side.
The filling factor is $p_{\rm occ}=1/10$ within a sphere
of radius $R=29 d$, so that the mean number of scatterers
is $\langle N\rangle\approx 10^4$.
The value of $E$ is discretized with a step $0.0025 E_0$ for both positive and negative values of $E$, and we use  $E_0=\hbar^2/(m d^2)$ as unit of energy.
For each value of $E$
one different realization of disorder is used, without imposing
any reflection symmetry (see end of subsection \ref{subsec:def_xi}).
A logarithmic scale is used:
The color map (see bar on the right)
is applied to the quantities  $\log_{10} \frac{N_{\rm res}}{N} \frac{E_0 d}{\delta S}$, for $E>0$, and  $\log_{10} \frac{N_{\rm bound}}{N} \frac{E_0 d}{\delta S}$, for $E<0$,
where $N_{\rm res}$ and $N_{\rm bound}$ are the number of resonances and bound states respectively, within
each rectangular bin of area $\delta S=\delta E \delta a_{\rm eff}$
($\delta E= 0.01 E_0$ and $\delta a_{\rm eff}=0.007 d$, for both positive and negative values of $E$). The plotted quantity is then $\simeq\log_{10} (1.4 N_{\rm res})$, and $\simeq\log_{10} (1.4 N_{\rm bound})$, respectively.
The horizontal dashed lines correspond to the values of $a_{\rm eff}/d
\in\{0.1,0.2,0.7,1,1.3\}$ used in Fig.~\ref{fig:fig4}.
The oblique solid line is the border of the energy gap
$E=\rho g_{\rm eff}$ predicted by the mean field theory
(see Eq.(\ref{eq:kappa_mf}) with $\kappa_{\rm mf}=0$).
(a) No selection is applied to the resonances ($E>0$). As explained in the text,
most of the displayed resonances (the ones with too large a width)
are not expected to be meaningful.
(b) Restricting to $E>0$, only the resonances with a width $\Gamma < \Gamma_{\rm max}=10^{-6} E_0/\hbar$ are kept in
the density of resonances. The value of $\Gamma_{\rm max}$ is essentially infinite
with respect to the duration of typical experiments: For a matter wave
of ${}^{87}$Rb atoms and an optical lattice with $d=0.4\mu$m,
as in \cite{Inguscio10}, one has indeed $1/\Gamma_{\rm max} \simeq 200$ seconds.
(c) Only the resonances and bound states corresponding to a small enough participation volume
$V_p$ are kept in the density [$V_p^{1/3}/d<6.5$, see Fig.\ref{fig:fig9}a and c].
(d) Only the resonances and bound states corresponding to a small enough r.m.s.\ size $\sigma$
in real space are kept in the density ($\sigma/d< 4.2$, see Fig.\ref{fig:fig9}b and d).
}
\label{fig:fig7}
\end{figure}
\end{widetext}

Of particular interest is the approximate value of $z_{\rm res}$
obtained after the first step:
\be
\label{eq:fs}
z_{\rm res} = E - i \frac{\mathrm{Im}\, m_i^\infty(E)}{dm_i^\infty(E)/dE},
\ee
which, if compared with (\ref{eq:zrescompl}), provides aproximate values of $E_{\rm res}$ and $\Gamma$. We have checked that, if the imaginary part of $m_i^\infty(E)$
is small enough, typically $\lesssim 10^{-3}/d$ for $p_{\rm occ}=0.1$,
the first step Eq.(\ref{eq:fs}) has in practice already converged to the exact pole location.
To obtain Fig.\ref{fig:fig7}, we used this first step approximation.

\subsection{Results for the density of bound states and resonances}

Using the techniques presented in the previous subsection, we have calculated the density of resonances and of bound states
in the plane $(E,a_{\rm eff})$, $E$ being the real part of the Green's function complex poles for resonances and the eigenenergy
for bound states. The raw result is presented in Fig.\ref{fig:fig7}a.

For large and negative $E$, and {\sl positive} $a_{\rm eff}$,
a first class of bound states is observed,
with a density of states
concentrated in a narrow interval of
values of $a_{\rm eff}$. This was expected: For $a_{\rm eff}>0$ the matter wave can form a bound state
(a dimer) with a single scatterer, of energy $E_{\rm dim}$ given by Eq.(\ref{eq:edim}). In the presence of the disordered ensemble of scatterers,
this dimer subsists in the limit $a_{\rm eff} \ll \rho^{-1/3}$, where $\rho$ is the mean density of scatterers. Simply, the dimer state
may tunnel from one scatterer in location $\rr_i$ to another one in $\rr_j$, with a tunneling amplitude close to Eq.(\ref{eq:trans}).
This tunneling broadens the eigenenergy interval around $E_{\rm dim}$
by an amount of the order of the tunneling amplitude, a small amount in relative value if $a_{\rm eff} \ll \rho^{-1/3}$.

\begin{figure}[htb]
\includegraphics[width=9cm,clip=]{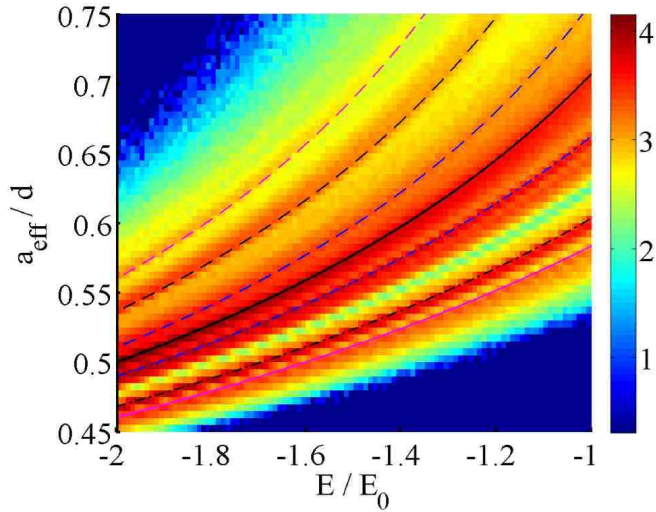}
\caption{
(Color online) Zoom of figure \ref{fig:fig7}a, in the region of $E<0$ and small and positive values of
$a_{\textrm{eff}}$ (here $\delta E= 0.01 E_0$, and $\delta a_{\rm eff}=0.005 d$). On the color map representing the density of states (see
caption of Fig.\ref{fig:fig7}a), are plotted the energies of bound states of an $A$
atom with one, two or three $B$ atoms, denoted respectively dimer $AB$, trimer $AB_2$, and
tetramer $AB_3$ states. More precisely are plotted the energies of a dimer Eq.(\ref{eq:edim})
(black solid line), of a trimer Eq.(\ref{eq:trim})
with two $B$ atoms separated by a distance $r_{12}=d$ (dashed black
and dash-dotted black lines), of a trimer with two $B$ atoms separated by a
distance $r_{12}=\sqrt{2}d$ (dashed and dash-dotted blue lines), and of a tetramer Eq.(\ref{eq:tetram})
with three $B$ atoms
separated by distances $r_{13}=r_{23}=d$ and $r_{12}=\sqrt{2}d$ (solid and dashed
magenta lines). It is worth noting that the prediction for
the energy of such isolated bound states correspond to the higher density lines
of the color map plot.
}
\label{fig:fig8}
\end{figure}

As one moves to larger values of the energy, the aforementioned
interval of eigenenergies moves toward larger values of $a_{\rm eff}$ and broadens, as expected from the previous reasoning.
In addition, an internal structure appears in the figure, see the zoom in Fig.\ref{fig:fig8},
the density of states being peaked over several subintervals, corresponding to various branches.
These branches may be identified physically by calculating the eigenenergies of trimers, tetramers,
... of the matter wave with two, three, ... scatterers on the lattice, see the lines
in the corresponding Fig.\ref{fig:fig8}. One simply has to apply the method described in the previous subsection,
for $N=2, 3, \ldots$, see Eq.(\ref{eq:param}),
and the corresponding matrices $M(E)$ may be diagonalized analytically.
Setting $E=-\hbar^2 q^2/(2m), q>0$, one finds that $a_{\rm eff}$ is given as a function
of the energy of an $AB_2$ trimer by the two branches 
\be
\label{eq:trim}
\frac{1}{a_{\rm eff}} = q \pm \frac{e^{-q r_{12}}}{r_{12}}
\ee
where $r_{12}$ is the distance between the two $B$ scatterers.
The calculation may be done also for the $AB_3$ tetramers, giving rise to three branches
for each particular set of values of the inter-scatterer distances $r_{12}, r_{13}, r_{23}$.
The analytical expressions are simple
in the relevant case $r_{13}=r_{23}$, where one sets
$\beta\equiv e^{-q r_{12}}/r_{12}$ and $\delta\equiv e^{-q r_{13}}/r_{13}$.
One then finds that
one of the branches reduces to one of the trimer branches, $1/a_{\rm eff} = q+ \beta$.
The other two branches are then given by \cite{Enerfewbody}:
\be
\label{eq:tetram}
\frac{1}{a_{\rm eff}} = q - \frac{1}{2} \left[\beta\pm (\beta^2 + 8 \delta^2)^{1/2}\right].
\ee
The values of $a_{\rm{eff}}$ for trimer and tetramer thresholds are readily obtained from (\ref{eq:trim}) and (\ref{eq:tetram}) by taking $q\to0$. Remarkably some of these threshold values are negative.  
 
For negative values of $E$ close to the origin, another class of bound states  is observed
for {\sl negative} $a_{\rm eff}$. For values of $a_{\rm eff}$ close to zero,
this class may be interpreted by a simple mean field effect. The matter wave
experiences an effective attraction inside the volume containing the scatterers, represented
by the negative mean field potential $\rho g_{\rm eff}$, see Eq.(\ref{eq:emf}), so that the gas of
scatterers produces an effective square well potential of radius $R$ that can support
matter wave bound states. This mean field interpretation is in good agreement with the numerical
results for small $|a_{\rm eff}|$. The oblique mean field white line in Fig.\ref{fig:fig7} giving the bottom
$\rho g_{\rm eff}$ of the effective
square well potential accurately reproduces the lower boundary of the bound state energies.
We note that expressions (\ref{eq:trim}) and (\ref{eq:tetram}) predict the presence of bound states also at negative values of $a_{\rm{eff}}$, for low enough values of $q$. In such an energy region the values of $a_{\rm{eff}}$ are large and negative ($a_{\rm{eff}}<-d/2$). Then the effective scattering length is typically larger that the mean distance between scatterers $\rho^{-1/3}\lesssim|a_{\rm{eff}}|$. This may explain why no well delineated internal structure is observed in the density of states for negative $a_{\rm{eff}}$.

Each of the two components that we discussed for $E<0$ continuously develops in two corresponding
components on the $E>0$ side of Fig.\ref{fig:fig7}a.

The $E>0$ component in the lower part of Fig.\ref{fig:fig7}a has a first boundary (delimiting a blue
triangular zone containing no resonances) which is well described by mean field theory at low values of $E$
and $a_{\rm eff}$, see the oblique white line. It has a second boundary on the right,
corresponding to a blue oval region again with no resonances.
There is no physical interpretation of this blue oval region at the present moment.
A word of caution may be useful at this stage. For $E>0$, we calculate a density of
resonances, with an approximate method valid for long-lived resonances
only (see subsection \ref{subsec:methodDOS}),
so the absence of resonances does not imply the absence of eigenstates
(in particular {\sl extended} states) of the system. The blue oval region thus does
not necessarily correspond to a spectral gap in the density of states.
In particular, the calculation of the localization length $\xi$ performed in the previous
section for $a_{\rm eff}/d=0.1$ shows a divergence of $\xi$ at the mean field border,
with small values of $\xi$ at energies below this border,
which is compatible with a spectral gap. However it does not show any sharp decrease of $\xi$
at the border with the oval region, $\xi$ retains very high values even
for energies in the oval region, raising doubts about the existence of a spectral gap in that region.
As we shall see in Fig.\ref{fig:fig7}b, the component in the lower part of Fig.\ref{fig:fig7}a contains no long
lived resonances, hence no localized states.

The $E>0$ component in the upper part of Fig.\ref{fig:fig7}a is more promising for observing localized
states. In Fig.\ref{fig:fig7}b we have kept only very long-lived resonances, with a decay rate
$\Gamma < 10^{-6} \hbar/(m d^2)$ \cite{SmallE}. This corresponds to a lifetime larger than
$\simeq 200$s for ${}^{87}$Rb atoms in an optical lattice of spacing $d=0.4\mu$m
\cite{Inguscio10}, and can then be considered as infinite with respect to the typical duration of ultracold gases  experiments. We see that, after this selection, the upper component still gives
a significant contribution, while the lower component essentially disappears.

In Fig.\ref{fig:fig7}c and Fig.\ref{fig:fig7}d, we have kept only the bound states and resonances
that are localized in real space, according to the two following criteria respectively:
In Fig.\ref{fig:fig7}c we keep states with a participation volume $V_{\rm p}$ smaller than $(6.5d)^3$,
where the participation volume is defined as:
\be
\label{eq:Vp}
V_{\rm p} \equiv \frac{1}
{\rho\sum_{i=1}^{N} |D_i|^4}
\ee
where the amplitudes $D_i$ are the components of the eigenstates of the $M$ matrix with a zero eigenvalue,
and are normalized such that $\sum_{i=1}^{N} |D_i|^2=1$. The participation volume (also called participation ratio) provides an estimate of the volume over which the amplitudes $D_i$ take significant values \cite{revue_russe}.
In Fig.\ref{fig:fig7}d we keep states with a root-mean square size $\sigma$ in real space
smaller than $4.2d$, with the definition
\be
\label{eq:sigma}
\sigma^2 \equiv \left(\sum_{i=1}^{N} r_i^2 |D_i|^2\right)
 - \left(\sum_{i=1}^{N} \rr_i |D_i|^2\right)^2,
\ee
where the $D_i$ are defined as for $V_{\rm p}$.

For computational convenience, in defining these two quantities we used  the amplitudes $D_i$, rather than the wavefunctions $\psi$ of the resonances and the bound states. We verified for several examples that localization of $D_i$ implies a localization of $\psi$, and vice versa. This can be easily understood since the matter wave problem is analogous to that of a scalar wave of light scattered by pinned atomic dipoles. In this case, the values of $D_i$ correspond to the values of the electric dipole moment on each atom, and the matter wave wavefunction $\psi$ corresponds to the electric field $\mathcal{E}$. In this scalar light analogy it is intuitive that the spatial extension of the dipoles reflect that of the electric field, and vice versa \cite{ScalarLight}. For completeness, we provide the expression of $\psi$ in terms of $D_i$ in the appendix \ref{AppPsiD}.

\begin{figure}[htb]
\begin{center}
\includegraphics[width=4cm,clip=]{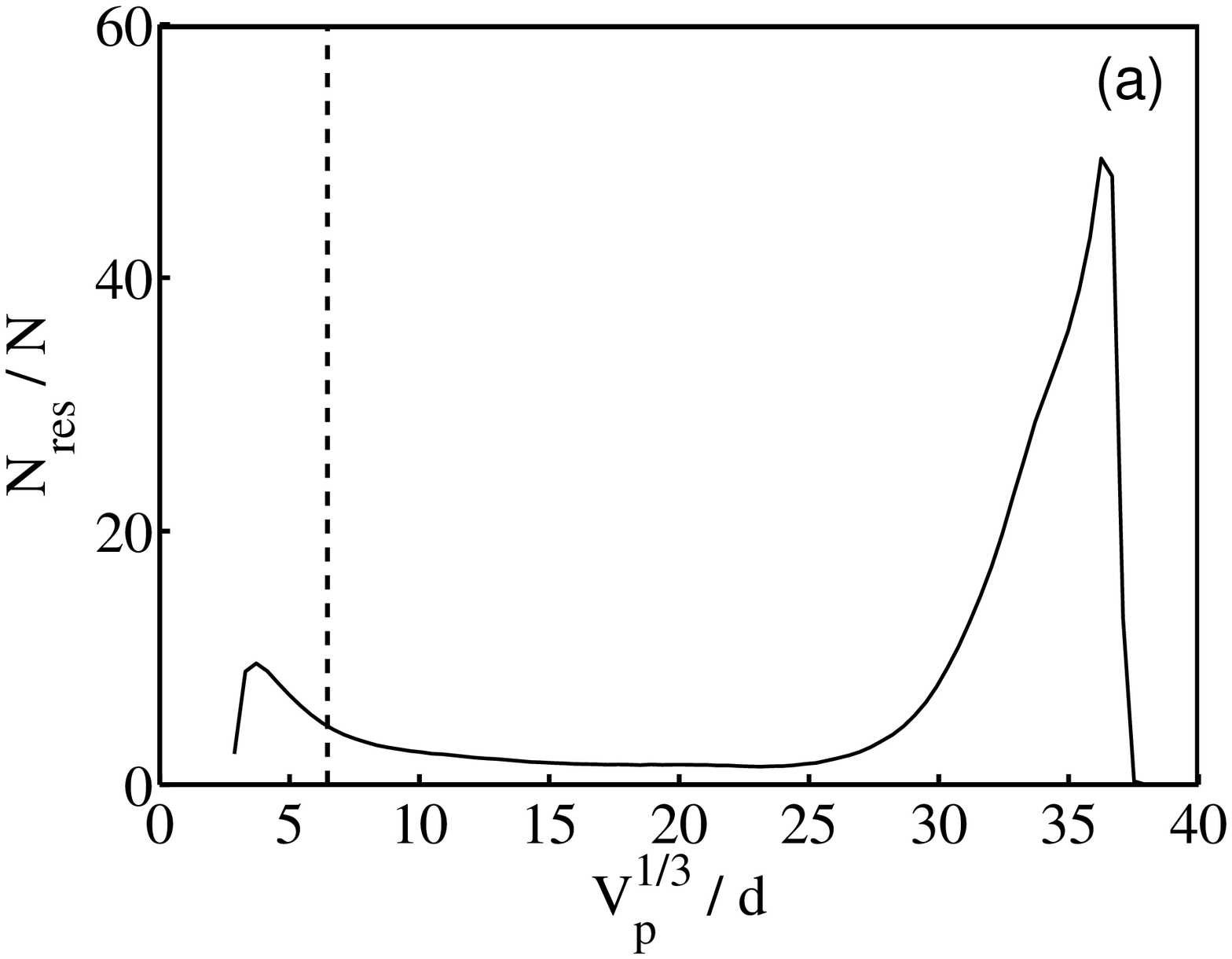}
\includegraphics[width=4cm,clip=]{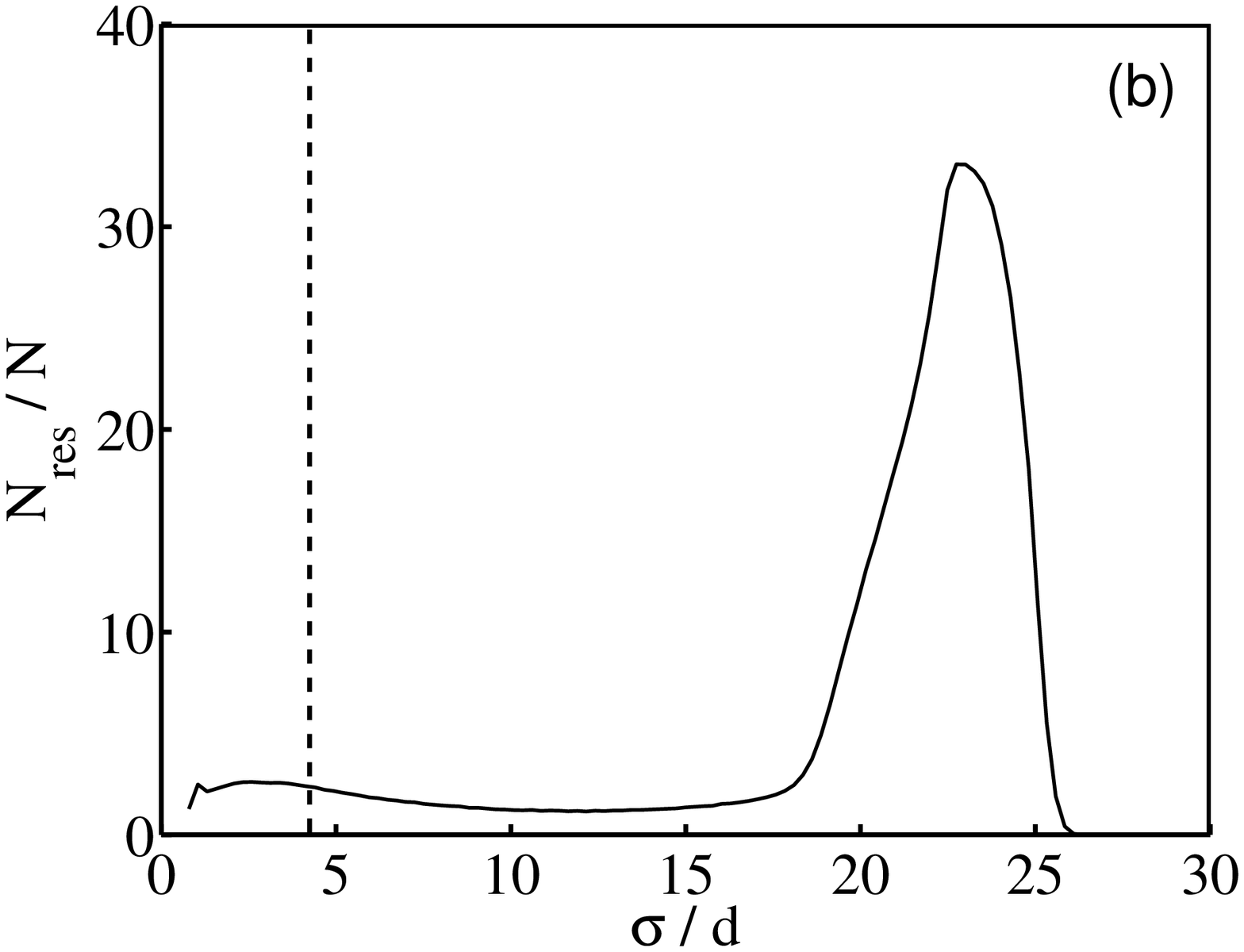} \\
\includegraphics[width=4cm,clip=]{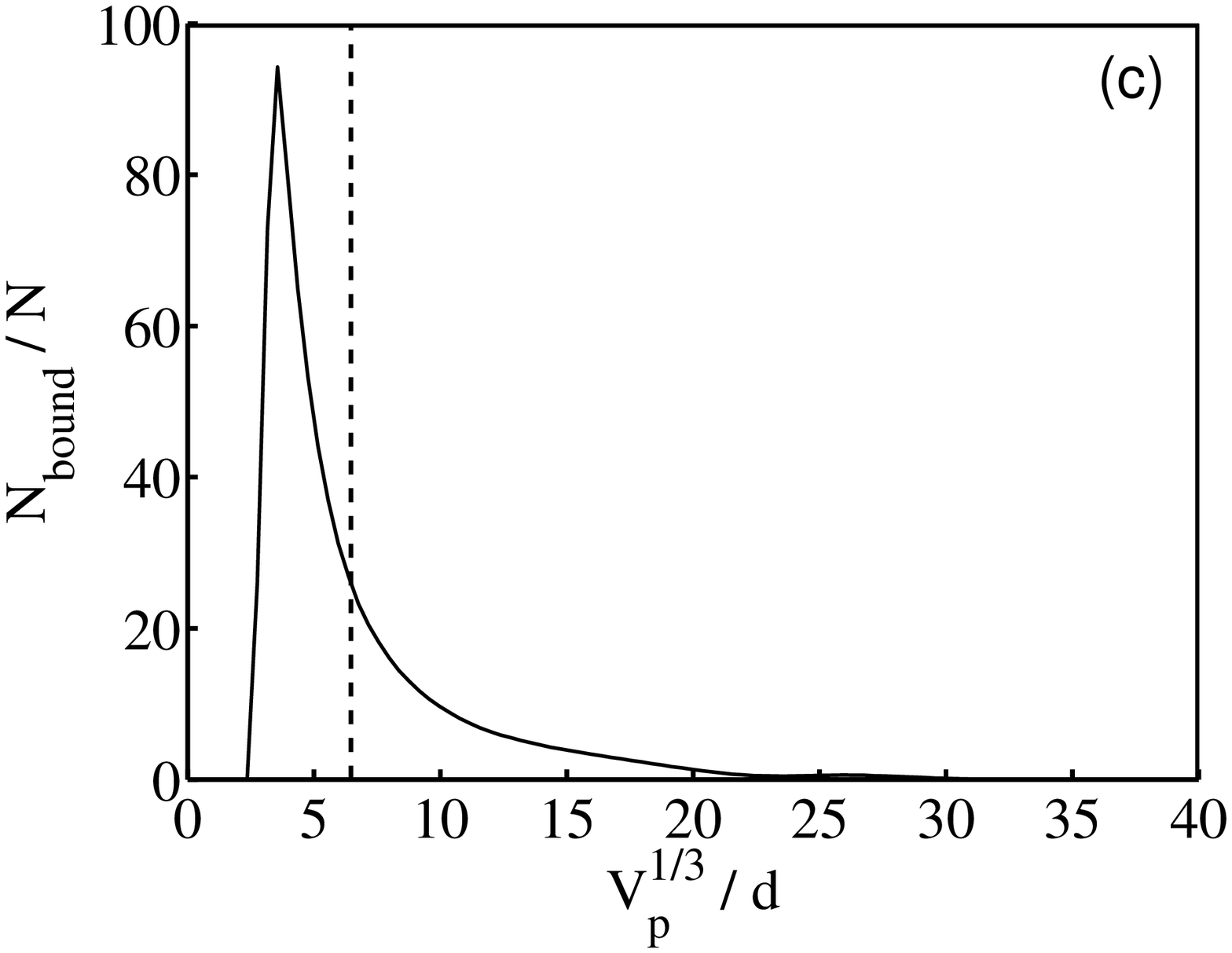}
\includegraphics[width=4cm,clip=]{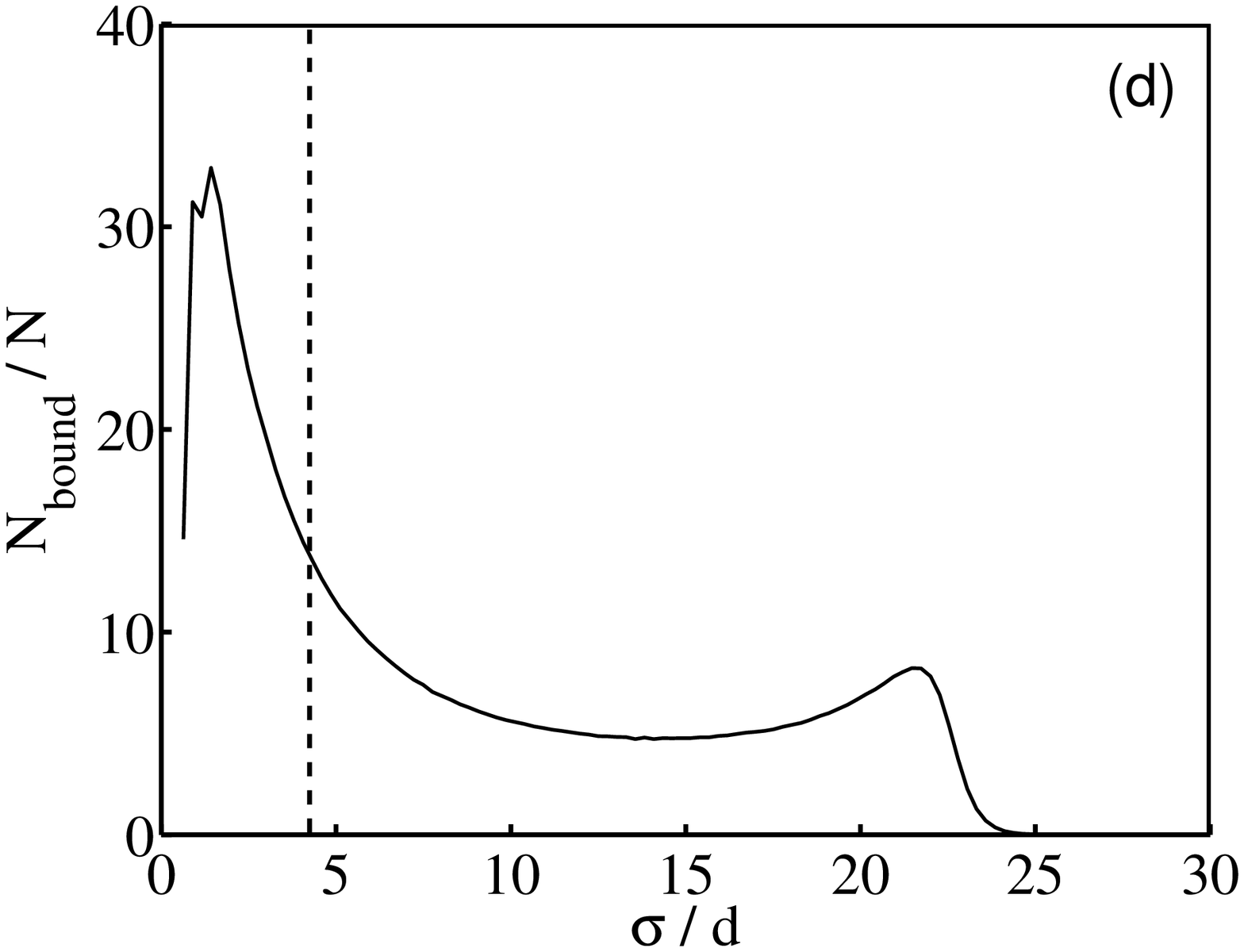}
\caption{For the 3D system: For all resonances with $E_{\rm res}/E_0\in (0,2)$ and $a_{\rm eff}/d \in (-3,3)$, and all bound states
with $E/E_0\in(-2,0)$ and $a_{\rm eff}/d \in (-3,3)$, the figures show an
histogram giving the number of resonances (in (a) and (b)) and bound states
(in (c) and (d)) per scatterer (the number of scatterers is $N=10^4$) as a function
of:  in (a) and (c), the cubic root of the participation volume defined in Eq.(\ref{eq:Vp});
in (b) and (d), the r.m.s. size in real space defined in Eq.(\ref{eq:sigma}).
The bin size is $\sim0.2d$ for (a) and (c),  and $\sim0.26d$ for (b) and (d).
The parameters are the same as in Fig.~\ref{fig:fig7}a. The dashed vertical lines
are the values $V_{\rm p}^{1/3}/d\simeq 6.5$ and $\sigma/d\simeq 4.2$  \cite{unif}
used in Fig.~\ref{fig:fig7}c and Fig.~\ref{fig:fig7}d to select the bound states and resonances
that are spatially localized and to filter out the bound states and resonances that are
spatially extended.}
\label{fig:fig9}
\end{center}
\end{figure}

The choice of above mentioned limiting values for $V_{\rm p}$ and $\sigma$ are motivated by
the histograms of Fig.\ref{fig:fig9}.  For $E>0$, the histograms reveal a bimodal
structure (i.e. two maxima) in the probability distribution of $V_{\rm p}$ and $\sigma$ 
(the bimodal structure is quite clear in Fig. \ref{fig:fig9}a, 
and less clear in Fig. \ref{fig:fig9}b), which suggests
the coexistence of localized resonances (small values of $V_{\rm p}^{1/3}$ and $\sigma$)
and extended resonances (large values of $V_{\rm p}^{1/3}$ and $\sigma$, of the order
of the radius $R=29 d$ of the sphere containing the scatterers).
For $E<0$, the bimodal structure still appears for $\sigma$, showing that some bound states
are extended. Interestingly, the bimodal structure is not present for $E<0$ on the histogram
of $V_{\rm p}^{1/3}$. Actually, it may happen
that a state which is a coherent superposition of a few-body bound state
(e.g.\ a dimer, a trimer) at different locations is considered as a localized state with
the criterion based on $V_{\rm p}$, whereas it is considered as an extended state with the one based on
$\sigma$.
An extreme example of such a case corresponds to a dimer delocalized over two sites
of positions $\rr$ and $-\rr$. Then $D_i=(\delta_{\rr_i,\rr}+\delta_{\rr_i,-\rr})/\sqrt{2}$,
leading to $V_{\rm p}^{1/3}=(2/\rho)^{1/3}$ (insensitive to the distance $2r$ between
the two possible dimer locations), and to $\sigma=r$ \cite{notemat}.

Coming back to Fig.\ref{fig:fig7}d, we thus see two main streams of localized states, one restricted
to negative energies, and the other extending to positive energies.
The existence of these two streams explains the structure of the localization length in
Fig.\ref{fig:fig4}a.
It is possible to establish a correspondence between the occurrence of divergences of the localization length, and boundaries between an energy interval with localized states and an energy interval with extended states.
A particularly rich example corresponds to the effective scattering length value
$a_{\rm eff}=d$, for which by combining the information coming from the localization length $\xi$ in Fig.\ref{fig:fig4}a, and the density of resonances/bound states in Figs.\ref{fig:fig7}a and \ref{fig:fig7}d, we can distinguish $5$ energy intervals:
\begin{itemize}
\item $E/E_0<-1.3$: the density of bound states is zero in Fig.\ref{fig:fig7}a, and $\xi$ takes small values in Fig.\ref{fig:fig4}a. This reflects the occurrence of an energy gap.
\item $-1.3<E/E_0<-1$: the density of localized bound states is non-zero in Fig.\ref{fig:fig7}d, and $\xi$ increases with a divergent behavior at $E/E_0\simeq-1$ in Fig.\ref{fig:fig4}a. This suggests the presence of an energy band of localized states with an upper mobility edge at $E/E_0\simeq-1$.
\item $-1<E/E_0<-0.2$: there are bound states in Fig.\ref{fig:fig7}a, but no localized bound states in Fig.\ref{fig:fig7}d, and $\xi$ assumes very large values (presumably larger than the system size). This suggests the presence of an energy band of extended bound states delimited by two mobility edges at  $E/E_0\simeq-1$ and $E/E_0\simeq-0.2$.
\item $-0.2<E/E_0<0.6$: the density of localized bound states and localized resonances is non-zero in Fig.\ref{fig:fig7}d, and $\xi$ assume small values away from divergent behaviors at $E/E_0\simeq-0.2$ and $E/E_0\simeq 0.6$ in Fig.\ref{fig:fig4}a.  This suggests the presence of an energy band of localized states delimited by two mobility edges at  $E/E_0\simeq-0.2$ and $E/E_0\simeq0.6$.
\item $E/E_0>0.6$: the density of localized resonances is zero in Fig.\ref{fig:fig7}d, and $\xi$ assumes very large values (presumably larger than the system size). This suggests the presence of an energy band of extended states delimited by a lower mobility edges at  $E/E_0\simeq0.6$.
\end{itemize}

\begin{figure}[htb]
\begin{center}
\begin{tabular}{c}
\hspace{5cm} (a) \\
\vspace{-0.4cm} \\
\includegraphics[width=9cm,clip=]{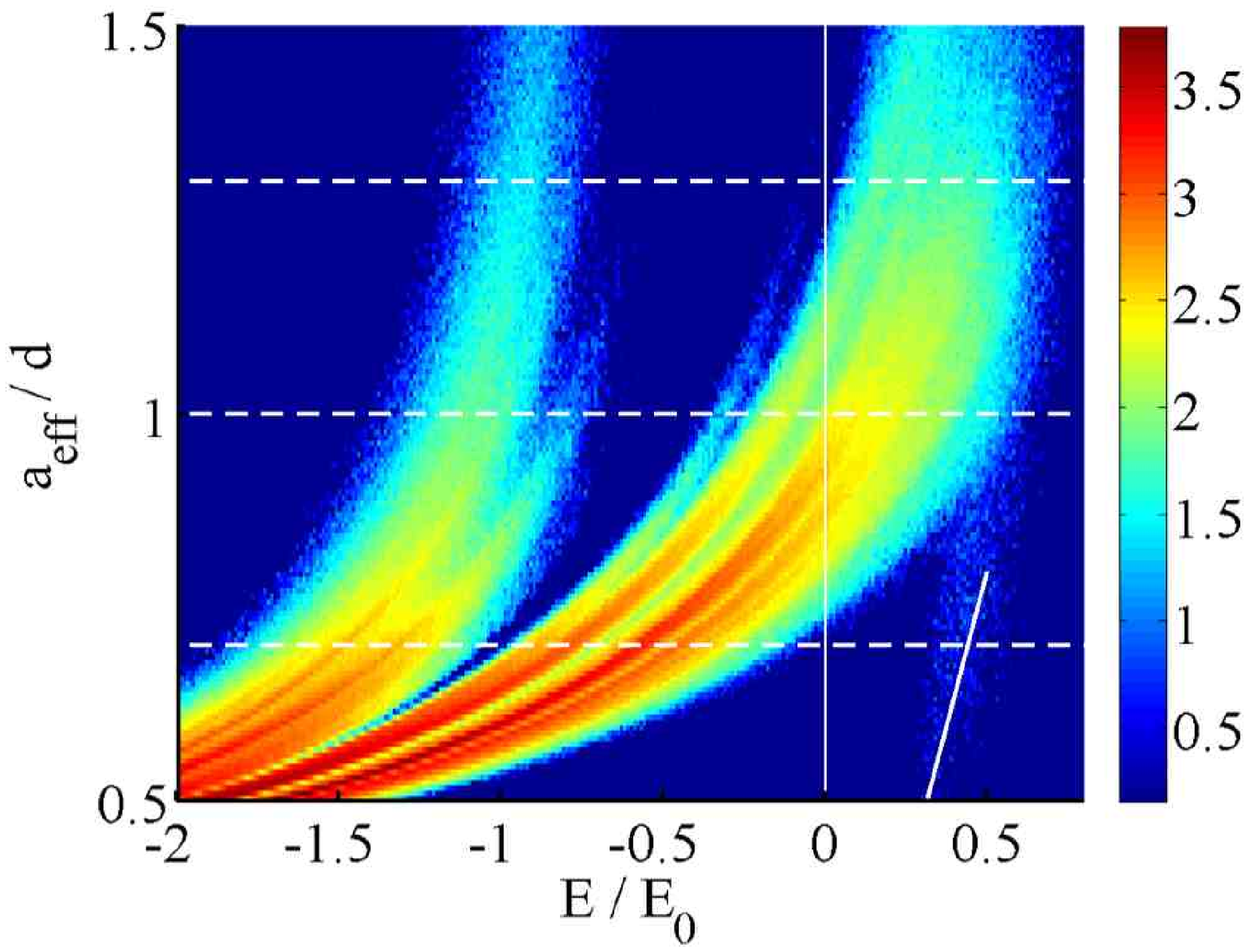} \\
\hspace{5cm} (b) \\
\vspace{-0.4cm} \\
\includegraphics[width=9cm,clip=]{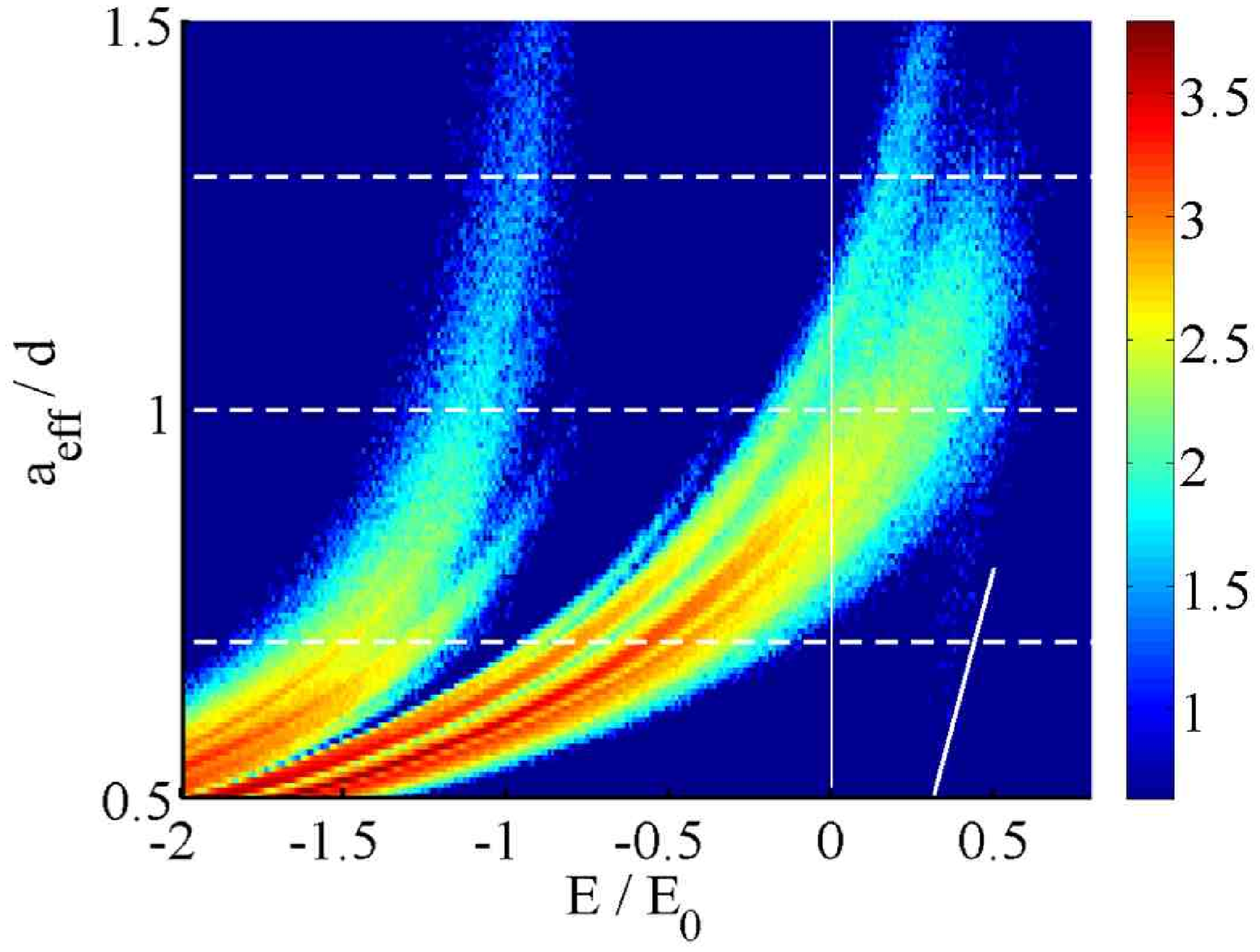}
\end{tabular}
\caption{(Color online) For the 3D system: Finite size effect on the density of localized states. (a) Zoom of the figure \ref{fig:fig7}d, which shows the density of resonances and bound states
per scatterer in the plane ($E$,$a_{\rm eff}$) having a r.m.s.\ size
in real space $\sigma< 4.2 d$. We recall that there are $N\simeq10^4$ scatterers in a sphere of radius $R=29d$. (b) Same as (a), with an additional filtering: we keep only bound states and resonances having their maximum $|D_i|$ at a distance less than $R'=20d$ from the origin (in the normalization, we considered the number of atoms $N'\simeq3.3\times 10^3$ in the sphere of radius $R'$).}
\label{fig:fig10}
\end{center}
\end{figure}

One can apply the same analysis to other values of $a_{\rm eff}$ considered in Fig.\ref{fig:fig4}a.
In particular, it is interesting to note that for $a_{\rm eff}=0.7d$ the correspondence between the positions
of mobility edges deduced from the localization length $\xi$,  and deduced from the density of resonances/bound states, is only qualitative: for negative energies, $\xi$ diverges at $E/E_0\simeq-1.2$ and $E/E_0\simeq-0.8$ in Fig.\ref{fig:fig4}a, while one finds the transition region between localized and extended states at $E/E_0\simeq-1.0$ and $E/E_0\simeq-0.9$ in Fig.\ref{fig:fig7}d. This discrepancy can be explained in terms of finite size effects, as appears from Fig.\ref{fig:fig10}. Indeed, Fig.\ref{fig:fig10}a shows a zoom of the figure \ref{fig:fig7}d (i.e. density of resonances and bound states
per scatterer in the plane ($E$,$a_{\rm eff}$) having a r.m.s.\ size
in real space $\sigma< 4.2 d$). In Fig.\ref{fig:fig10}b we impose an additional filtering with respect to that of Fig.\ref{fig:fig10}a: we keep only bound states and resonances having their maximum $|D_i|$ at a distance less than $R'=20d$ from the origin, whereas the disorder fills a sphere of radius $R=30d$. In this way only localized states located far from the border of the system are selected. In figure \ref{fig:fig10}b the transition region between localized and extended states is now at $E/E_0\simeq-1.2$ and $E/E_0\simeq-0.8$, in quantitative agreement with the mobility edges deduced from $\xi$ in Fig.\ref{fig:fig4}a.

\begin{figure}[htb]
\begin{center}
\includegraphics[width=9cm,clip=]{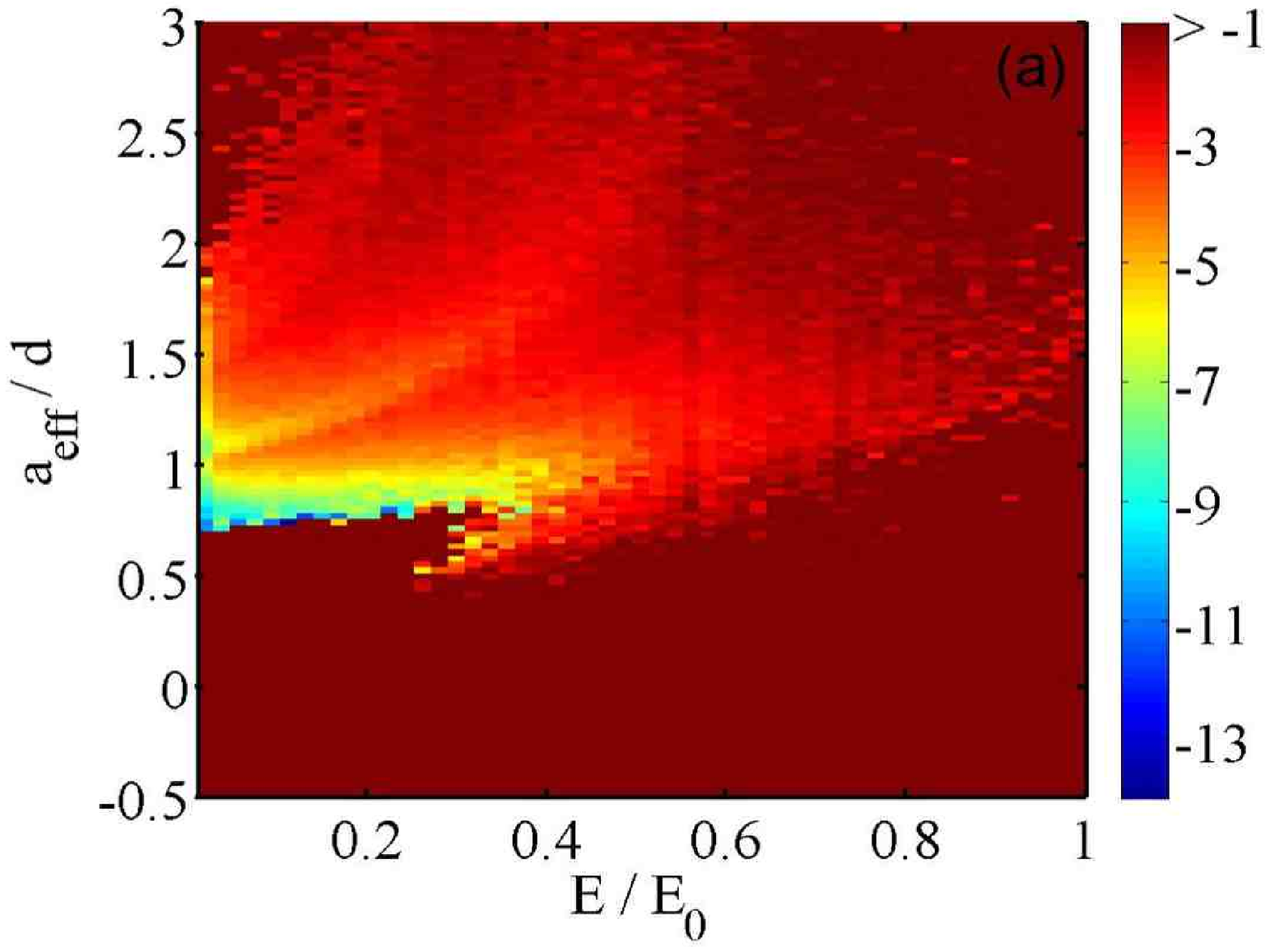}
\includegraphics[width=9cm,clip=]{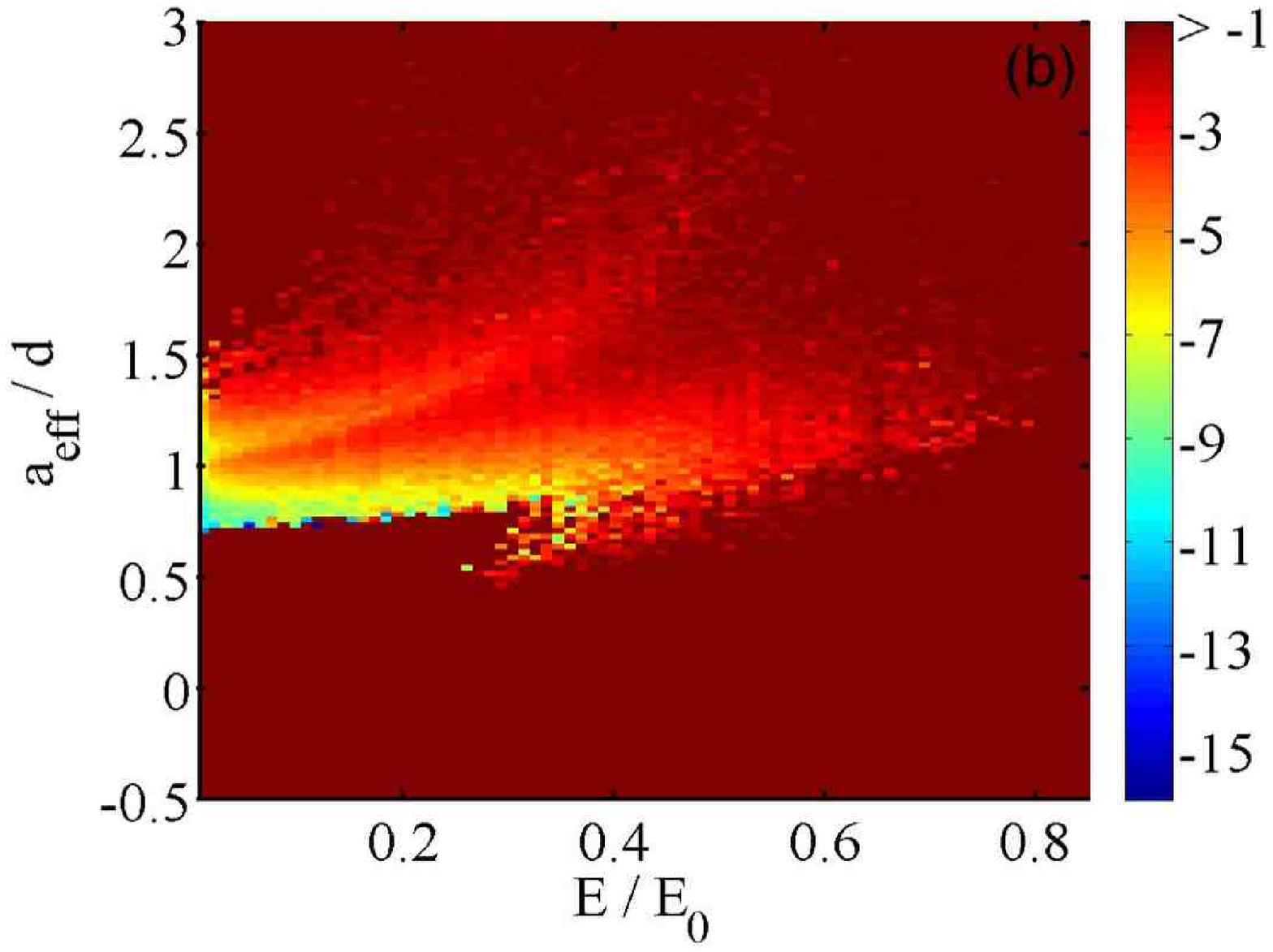}
\caption{(Color online) For the 3D system: Width $\Gamma$ of the resonances as a function of the energy $E$
and the effective scattering length $a_{\rm eff}$ [see Eqs. (\ref{eq:zrescompl}) and (\ref{eq:fs})]. The physical parameters
are the same as in Fig.~\ref{fig:fig7}. The plane $(E,a_{\rm eff})$ is decomposed
in rectangular bins of widths $\delta E$ and $\delta a_{\rm eff}$.
The color map (see bar on the right)
is applied to the quantity  $\log_{10} \hbar \langle \Gamma\rangle/E_0$
where $\langle \Gamma\rangle$ is the mean value of $\Gamma$ for the resonances
within a given bin.
The resonances are filtered in (a) over the participation volume
as in Fig.~\ref{fig:fig7}c and in (b) over the r.m.s.\ size $\sigma$
as in Fig.~\ref{fig:fig7}d.
In (a) one has $\delta E= 0.009 E_0$ and $\delta a_{\rm eff}= 0.02 d$,
and in (b) one has
$\delta E= 0.008 E_0$ and $\delta a_{\rm eff}= 0.02 d$.}
\label{fig:fig11}
\end{center}
\end{figure}

To conclude the analysis of the 3D system, we give in Fig.\ref{fig:fig11} the values of the width $\Gamma$ for the resonances in the plane $(E,a_{\rm eff})$. A filtering was applied to these resonances: In Fig.\ref{fig:fig11}a we considered only resonances with a participation volume $V_{\rm p}^{1/3}/d\leq6.5$, and in Fig.\ref{fig:fig11}b only resonances with a r.m.s size in real space $\sigma/d\leq4.2$. This analysis shows the presence of numerous  extremely long-lived resonances with $\Gamma\leq10^{-6} E_0/\hbar$. This figure may be useful for experimental purposes to identify optimal values of $(E,a_{\rm eff})$ for the observation of maximally localized resonances. These occur for values of $a_{\rm eff}\approx0.8d$ and $0<E/E_0<0.4$, for the considered density of scatterers $p_{\rm occ}=1/10$.

\section{Localization in a 2D geometry}
\label{sec:loc2D}

It is now experimentally possible to realize systems where the atomic matter wave species $A$ is confined in a 2D
geometry. One can apply a strong confinement along the direction $z$, freezing the motion in its ground
state along $z$, while leaving a free motion in the $xy$ plane. In analogy with the 3D disordered system
studied in previous sections, we study here the 2D case, where both $B$ scatterers and matter wave atoms $A$ are
confined along the $z$ direction (not necessarily with the same potential), and $B$ scatterers are randomly
distributed at the nodes of a $2D$ optical lattice in the $xy$ plane (the $A$ atoms being insensitive to that
lattice). As for the 3D case, the low energy scattering of an $A$ atom with a single trapped $B$ atom can be
characterized by a $2D$ effective scattering length, accounting for the effect of the atomic motion of the $B$
atom during the scattering process. This allows us to replace the $B$ atoms by point-like scatterers at fixed
positions, described by contact conditions on the $A$ atom wavefunction. For simplicity we use for the effective scattering length in  2D the same notation $a_{\rm eff}$ as in 3D. To our knowledge, the dependence of the effective 2D scattering length on the free space 3D scattering length $a$,  on the atomic masses and on the oscillation frequencies, has not yet been investigated.

On the theoretical side there has been significant investigation of disordered systems in 2D since this is the lower critical dimension for the occurrence of the metal-insulator transition. In 2D it is expected that all states are localized in an infinite system, but the localization length increases rapidly with the energy \cite{Ramakrishnan79,Ramakrishnan85}. It is thus important to identify the range of parameters (matter wave energy $E$ and $a_{\rm eff}$) for which the localization length $\xi$  takes small values.
In subsection \ref{sub:2Dmodel} we explain how to adapt the 3D formalism to the 2D case, and in subsection \ref{sub:res2d} we present the results for the localization length and the density of localized states.

\subsection{Formalism for 2D systems}
\label{sub:2Dmodel}

In 2D the contact condition on the matter wave wavefunction
in the vicinity of each scatterer of position $\rr_i$ is
\be
\label{eq:contact2d}
\psi(\rr) = \frac{m}{\pi \hbar^2}\, D_i
\ln (|\rr-\rr_i|/a_{\rm eff}) + O(|\rr-\rr_i|)
\ee
where $a_{\rm eff}$ is now the 2D effective scattering length
(always positive) \cite{Olshanii07}.
Comparing to the 3D contact conditions Eq.(\ref{eq:contact3d}),
we see that $-1/\ln (a_{\rm eff}/d)$ in 2D plays
the role of $a_{\rm eff}/d$ in 3D.
We recall that, in 2D, the limit $a_{\rm eff}\to 0$  does in fact correspond to a
weakly repulsive limit, while $a_{\rm eff}\to +\infty$ corresponds  to a weakly
attractive limit.

The 2D Green's function still obeys Eq.(\ref{eq:eqg}), where $\Delta_{\rr}$ is now the 2D Laplace operator,
so that formally Eq.(\ref{eq:gfd3d}) still holds provided that
one takes for $g_0$ the 2D free matter wave Green's function.
For $E>0$, we set $E=\hbar^2 k^2/(2m)$, $k>0$, and we obtain
\be
\label{eq:g02d}
g_0(\rr) = -\frac{im}{2\hbar^2} H_0^{(1)}(kr)
\ee
with the Hankel function expressed in terms of Bessel functions
as $H_0^{(1)}(z) = J_0(z) + i N_0(z)$. From \cite{Gradshteyn} one obtains the limiting
behaviors
\bea
\label{eq:han_far}
H_0^{(1)}(kr) & \underset{r\to +\infty}{\sim} & \left(\frac{2}{\pi k r}\right)^{1/2} e^{i(kr-\pi/4)},\\
\label{eq:han_close}
H_0^{(1)}(kr) & \underset{r\to 0}{=} & 1 + \frac{2i}{\pi} \ln
\left(\frac{kr e^\gamma}{2}\right) + o(1),
\eea
where $\gamma=0.57721566\ldots$ is Euler's constant. The numerical factor in Eq.(\ref{eq:g02d}) thus results from the
fact that $\Delta_{\rr} \ln r = 2 \pi \delta(\rr)$ in 2D.
For $E<0$, we set $E=-\hbar^2 q^2/(2m)$, $q>0$. That is, $k=iq$
is now purely imaginary, and we obtain

\be
g_0(\rr) = -\frac{m}{\pi \hbar^2} K_0(qr)
\ee
where $K_0$ is a modified Bessel function of the second kind.

The secondary source amplitudes $D_i$ are then determined by
imposing the contact conditions (\ref{eq:contact2d}) at the order $O(1)$.
That is, for the non-diverging term $\ln a_{\rm eff}$,
\be
\label{eq:ls2d}
\sum_{j=1}^{N} M_{ij} D_j =
-\frac{\pi\hbar^2}{m} g_0(\rr_i-\rr_0) , \ \ \ \forall i\in\{1,\ldots,N\}
\ee
where we have introduced the $N\times N$ matrix
\be
\label{eq:def_M2d}
M_{ij} = \left\{ \begin{array}{ccc} \displaystyle  \frac{\pi\hbar^2}{m} g_0(\rr_i-\rr_j) & \, {\rm if}\,  & i\neq j, \\
&& \\
            -i\frac{\pi}{2}+\ln\left(\frac{k a_{\rm eff} e^\gamma}{2}\right)
&  \, {\rm if}\,  & i=j.\end{array} \right.
\ee
For $E<0$, where $k=iq$, $q>0$, this holds with the determination
$\ln i = i \pi/2$, leading to the diagonal
element $M_{ii}=\ln\left(\frac{q a_{\rm eff} e^\gamma}{2}\right)$.

For the calculation of the localization length $\xi$, we use the second method already discussed for the 3D case. By following the same derivation as in 3D, and using Eq.(\ref{eq:han_far}), we obtain the same expression (\ref{eq:t3D}) for the values of the transmission amplitude $t(\mathbf{n})$ of a source located at the position $\rr_{0}$. Then, we define the Lyapunov exponent $\kappa$ as
\bea
\kappa &=& -\lim_{R\to+\infty}\int_0^{2\pi}\frac{d\theta}{2\pi} \frac{\langle \ln |t(\nn)|\rangle}{R} \; {\rm for}\;E>0, \label{eq:lyapn2d}\\
\kappa &=& q-\lim_{R\to+\infty}\int_0^{2\pi}\frac{d\theta}{2\pi} \frac{\langle \ln |t(\nn)|\rangle}{R} \; {\rm for}\;E<0, \label{eq:lyapn2dneg}
\eea
where $\nn$ has coordinates $(\cos\theta,\sin\theta)$ in the $xy$ plane, and $R$ is the radius of the disk containing the disorder.

To obtain the density of bound states and resonances in 2D, we use the same procedure as in 3D, and split
 the matrix $M$ as
\be
\label{eq:M2ds}
M=\ln(a_{\rm eff}/d)\mathrm{Id}+M_{\infty},
\ee
where $M_{\infty}$ is the value of $M$ for $1/\ln(a_{\rm eff}/d)=\infty$, i.e. comes from setting $a_{\rm eff}=d$ in Eq.(\ref{eq:def_M2d}).
To evaluate the density of bound states for a fixed value of $a_{\rm eff}$ one can solve by dichotomy the 2D equivalent of Eq.(\ref{eq:ie3d}), using the fact that the eigenvalues of $M(E)$, for $E<0$, are a monotonically decreasing function of $E$ (see appendix \ref{app:mon}).

\begin{widetext}

\begin{figure}[htb]
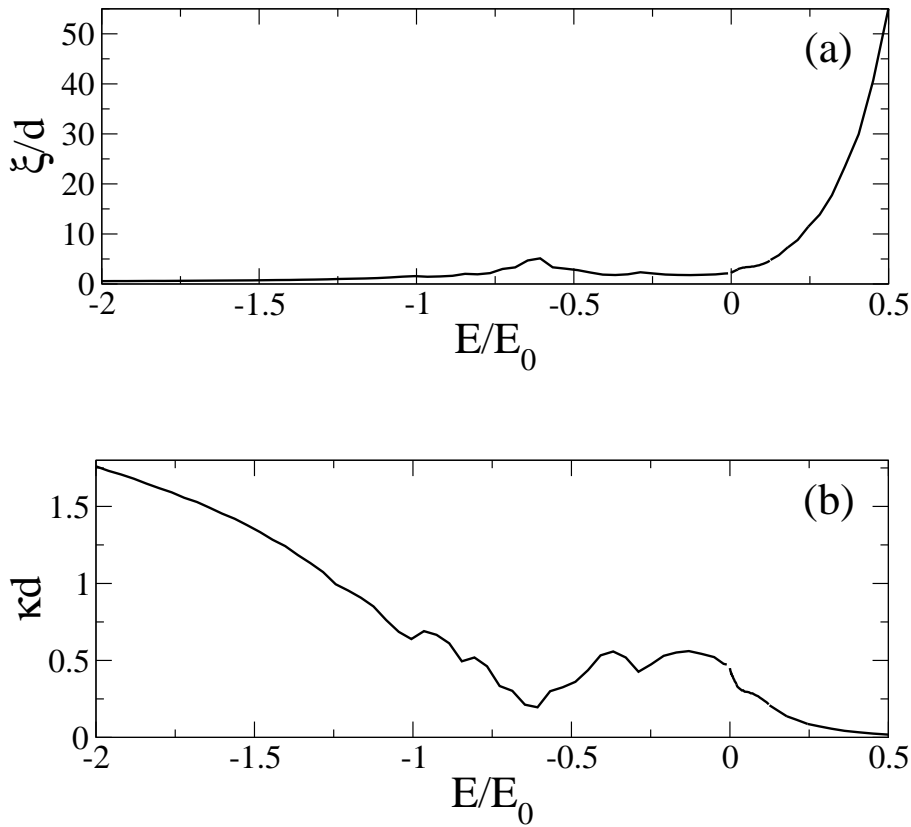

\begin{center}
\includegraphics[width=12cm,clip=]{fig12a.eps}\\
\includegraphics[width=12cm,clip=]{fig12b.eps}
\caption{For the 2D system: localization length $\xi$ (a), and Lyapunov exponent $\kappa$ (b), as a function of the energy $E$. Here
the 2D effective scattering length is $a_{\rm eff}=d$, the lattice filling factor is $p_{\rm occ}=1/10$, and an average over $100$ realizations for the positions of the scatterers has been performed. The value of $\kappa$
is obtained using the second method [see Eqs.(\ref{eq:lyapn2d}) and (\ref{eq:lyapn2dneg})]. For $0<E/E_0<0.24$, we considered larger and larger
system sizes $2R/d\in\{100,150,250,350\}$ until a convergent value for
$\kappa$ is reached, within the statistical error bars, which are smaller than $5\%$. For $0.24<E/E_0<0.5$ we considered system sizes $2R/d\in\{350,600\}$, and found a maximal deviation of $10\%$ in $\kappa$ between the two sizes, finally  plotting the values for $2R/d=600$. For negative energies $-2<E/E_0<0$ we considered
system sizes $2R/d\in\{100,250\}$ (except for $E/E_0=-0.02$ and $-0.01$, where we considered $2R/d\in\{50,75,100\}$), from which we extrapolated linearly to the case $1/R\to0$. In this region we obtain error bars smaller than $1\%$. The absence of the $E/E_0=0$ point is due to the fact that the matrix $M$ is logarithmically divergent for such a value.}
\label{fig:fig12}
\end{center}
\end{figure}
\end{widetext}

\begin{figure}[htb]
\begin{center}
\includegraphics[width=8cm,clip=]{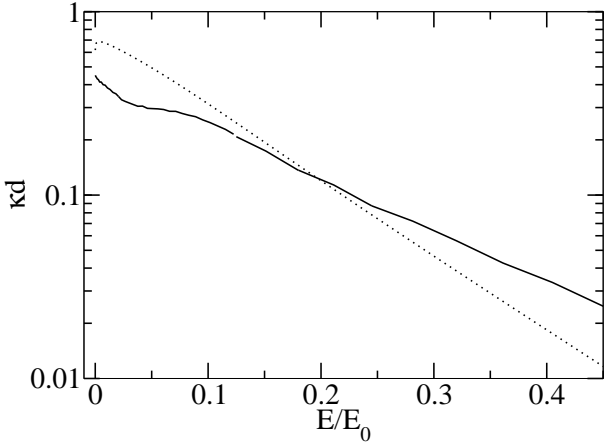}
\caption{For the 2D system: Lyapunov exponent $\kappa$ of Fig.\ref{fig:fig12}b, plotted in logarithmic scale (solid black line). The prediction of the independent scattering approximation (ISA) theory is also shown (dotted black line).}
\label{fig:fig13}
\end{center}
\end{figure}

\subsection{Results for the localization length and the density of states for the 2D system}
\label{sub:res2d}
In figures \ref{fig:fig12}a and b we show the value of the localization length $\xi$ and of the Lyapunov exponent $\kappa\equiv1/\xi$ of Eqs.(\ref{eq:lyapn2d}) and (\ref{eq:lyapn2dneg}), as a function of the energy $E$ of the matter wave, for the 2D effective scattering length $a_{\rm eff}=d$. For $E>0$ we note a monotonically increasing behavior of $\xi$, and a monotonically decreasing behavior of $\kappa$ with the energy $E$, with the occurrence of a flatter region (a ``shoulder") around $E/E_0=0.05$. The logarithmic plot for $\kappa$ in Fig.\ref{fig:fig13} suggests an exponential decay of $\kappa$ at large energies. However, being over less than one decade, it is not fully conclusive. We also performed calculations of $\kappa$  for $a_{\rm eff}/d=0.3$, obtaining similarly large values of $\kappa$ at low energy. In contrast, for $a_{\rm eff}/d=3$ the values of $\kappa$ drop by an order of magnitude for the same energy interval.
In Fig.\ref{fig:fig13} we also plot the prediction coming from the frequently used independent scattering approximation (ISA) (see, for instance \cite{isa}), according to which the localization length $\xi=l\exp{(\pi  l \mathrm{Re} k_{\rm eff} / 2)}$, is expressed in terms of the mean free path $l\equiv 1/(2\mathrm{Im}k_{\rm eff})$, and an effective wavevector $k_{\rm eff}$ defined as
\be
E\equiv\frac{\hbar^2k^2}{2m}=\frac{\hbar^2k_{\rm eff}^2}{2m}+\rho \langle{\kk}|T(E+i0^+)|\kk\rangle
\ee
where $T$ is the T-matrix in 2D for a single scatterer \cite{NoteYvan}:
\be
\langle{\kk}|T(E+i0^+)|\kk\rangle=\frac{-\pi\hbar^2/m}{\ln(ka_{\rm eff}e^{\gamma}/2)-i\pi/2}.
\ee
The figure shows that the ISA is not accurate for the considered parameters, i.e. for large values of $a_{\rm eff}$, of the order of the mean distance between scatterers.

\begin{widetext}

\begin{figure}[htb]
\begin{center}
\begin{tabular}{cc}
\hspace{5cm} (a) & \hspace{5cm} (b) \\
\vspace{-0.4cm} \\
\includegraphics[width=8.5cm,clip=]{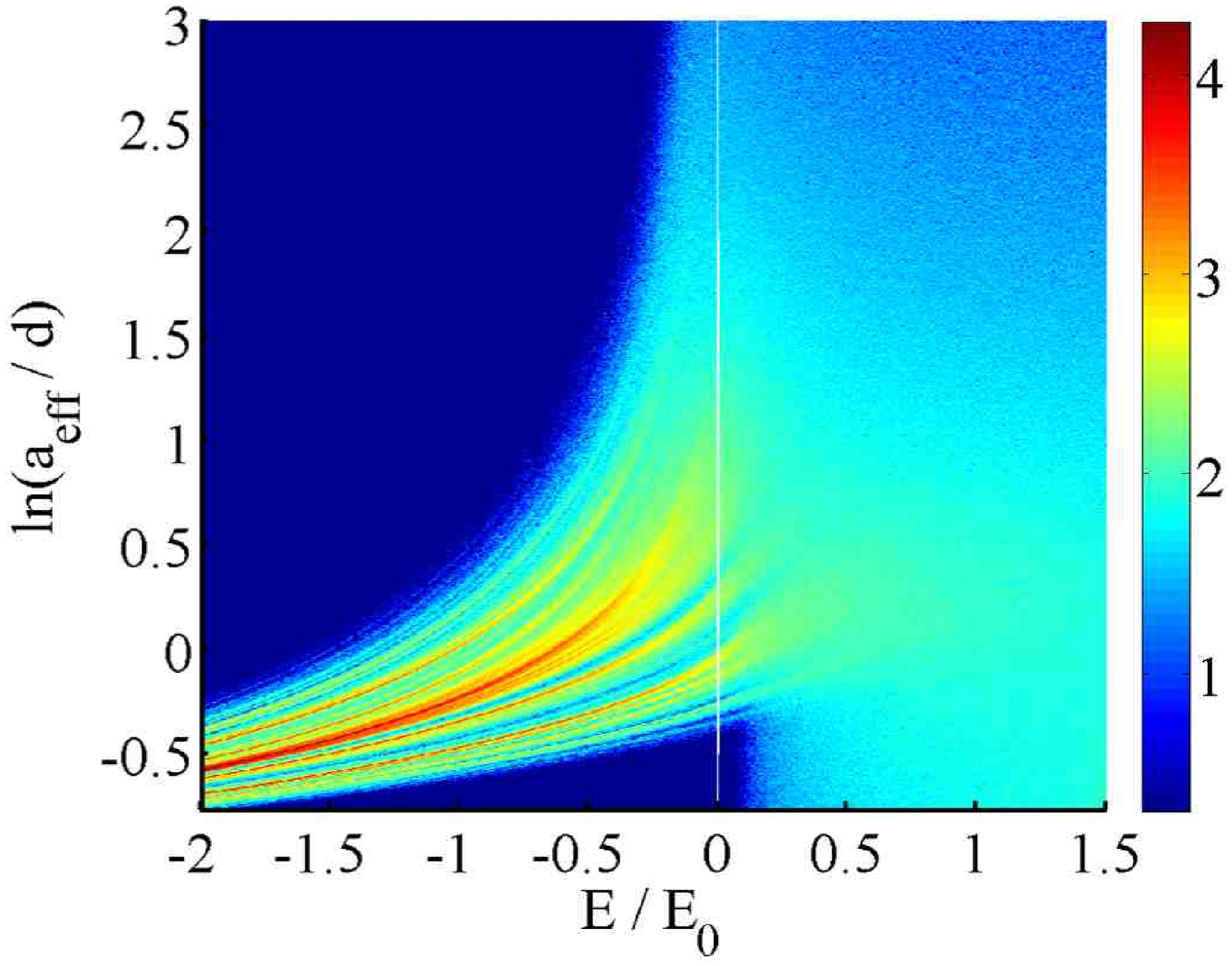} &
\includegraphics[width=9cm,clip=]{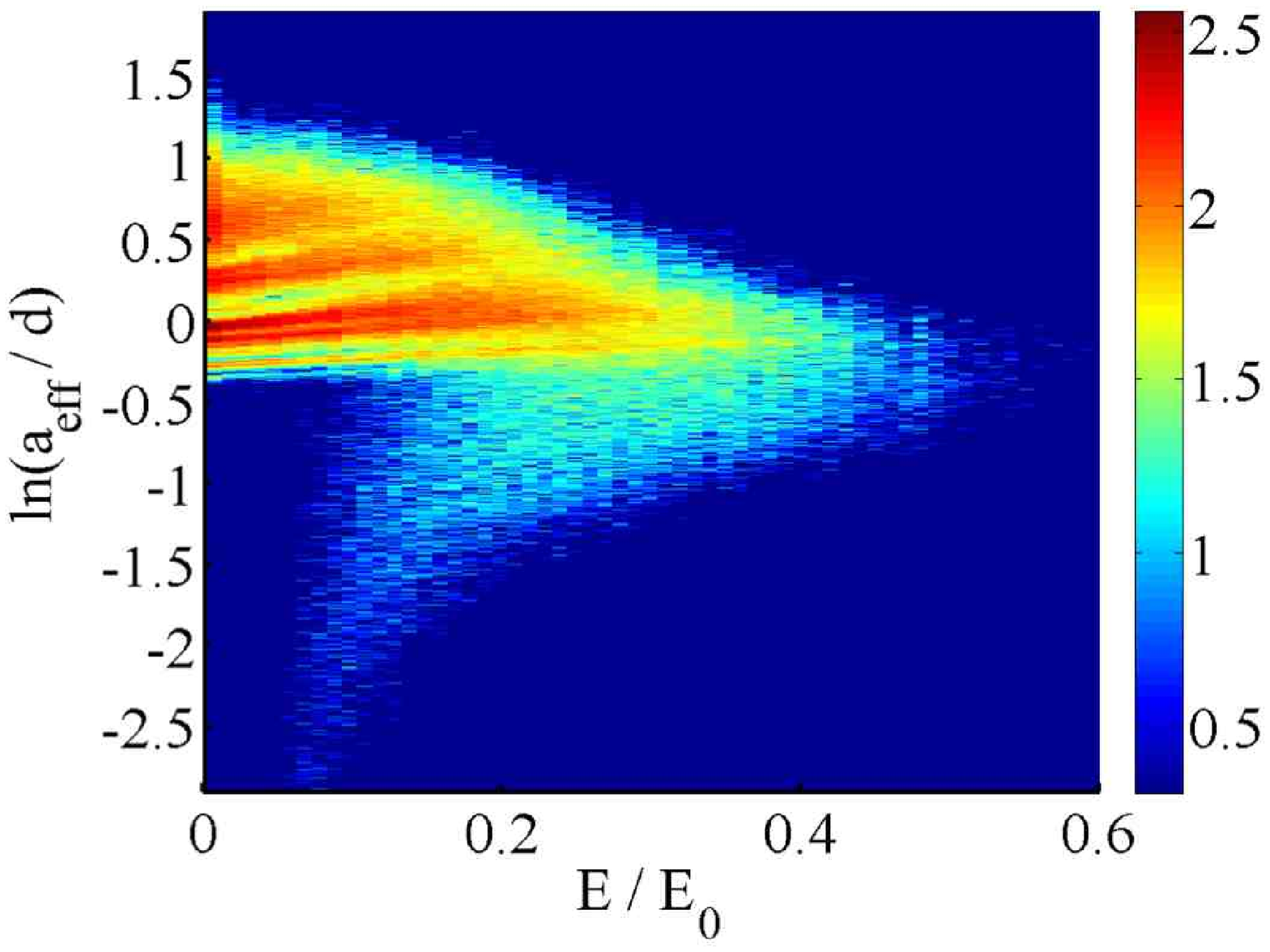} \\
& \\
\hspace{5cm} (c) & \hspace{5cm} (d) \\
\vspace{-0.4cm} \\
\includegraphics[width=8.5cm,clip=]{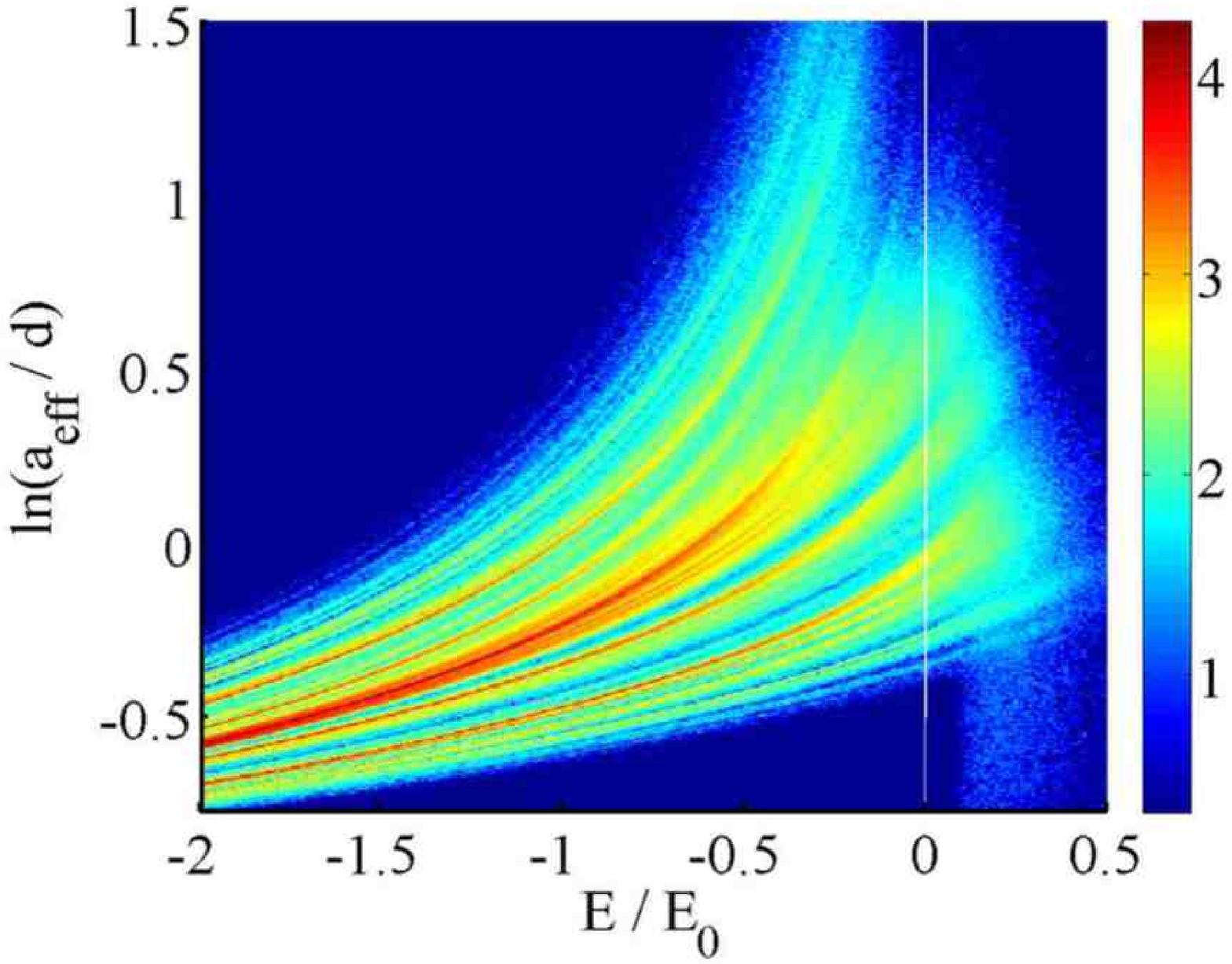}  &
\includegraphics[width=8.5cm,clip=]{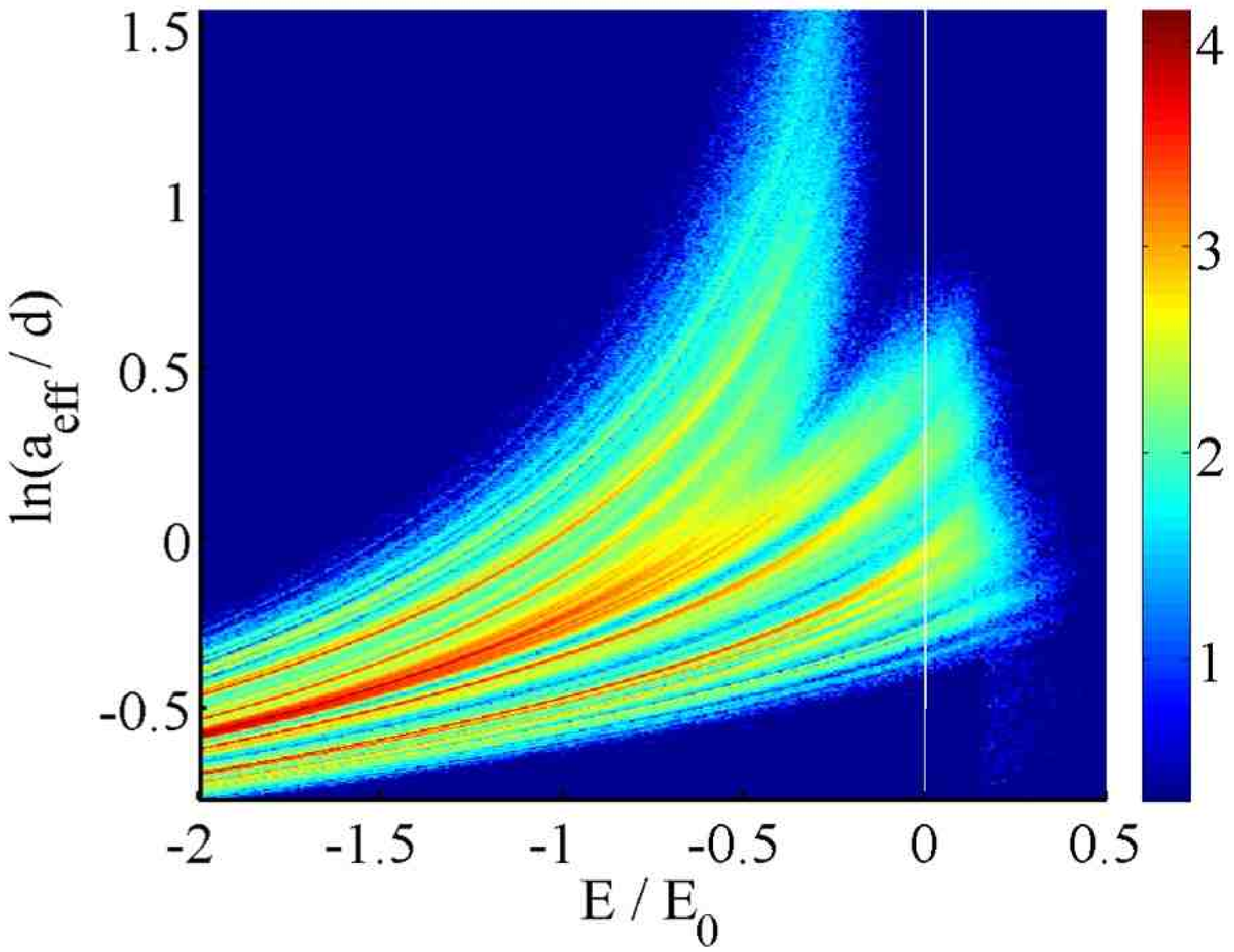}
\end{tabular}
\end{center}
\caption{(Color online) For the 2D system: Density of resonances and bound states
per scatterer in the plane
(energy $E/E_0$, logarithm of the effective scattering length $\ln(a_{\rm eff}/d)$),
obtained as explained in the subsection \ref{sub:2Dmodel}.
$E$ on the horizontal axis is either the real part of $z_{\rm res}$ on the positive energy side,
or the bound state eigenenergy on the negative energy side.
The filling factor is $p_{\rm occ}=1/10$ within a disk
of radius $R=150 d$, so that the mean number of scatterers
is $\langle N\rangle\approx 7 \times 10^3$.
The value of $E$ is discretized with a step $0.0025 E_0$ for both positive and negative values of $E$, and we used  $E_0=\hbar^2/(m d^2)$ as unit of energy.
For each value of $E$
one different realization of disorder is used, without imposing
any reflection symmetry (see end of subsection \ref{subsec:def_xi}).
A logarithmic scale is used:
The color map (see bar on the right)
is applied to the quantities  $\log_{10} \frac{N_{\rm res}}{N} \frac{E_0 d}{\delta S}$, for $E>0$, and  $\log_{10} \frac{N_{\rm bound}}{N} \frac{E_0 d}{\delta S}$, for $E<0$,
where $N_{\rm res}$ and $N_{\rm bound}$ are the number of resonances and bound states respectively, within
each rectangular bin of area $\delta S=\delta E \delta \ln(a_{\rm eff}/d)$
($\delta E= 0.01 E_0$ and $\delta \ln(a_{\rm eff}/d)=0.007$, for $E>0$, and $\delta E= 0.01 E_0$ and $\delta \ln(a_{\rm eff}/d)=0.004$ for $E<0$).
(a) No selection is applied to the resonances ($E>0$). As explained in the text,
most of the displayed resonances (the ones with a too large width)
are not expected to be meaningful.
(b) Restricting to $E>0$, only the resonances with a width $\Gamma < \Gamma_{\rm max}=10^{-6} E_0/\hbar$ are kept in
the density of resonances. The value of $\Gamma_{\rm max}$ is essentially infinite
with respect to the duration of typical experiments.
(c) Only the resonances and bound states corresponding to a small enough participation surface
$S_{\rm{p}}$ are kept in the density ($S_{\rm{p}}^{1/2}/d<9.5$, see Fig.\ref{fig:fig17}a and c).
(d) Only the resonances and bound states corresponding to a small enough r.m.s.\ size $\sigma$
in real space are kept in the density ($\sigma/d< 4.2$, see Fig.\ref{fig:fig17}b and d).
}
\label{fig:fig14}
\end{figure}

\end{widetext}

In figure \ref{fig:fig14} we show the density of bound states and resonances in the $(E/E_0,\ln(a_{\rm eff}/d))$ plane, in a 2D system (see sections \ref{subsec:methodDOS} and \ref{sub:2Dmodel}). Calculations have been performed for a disk of radius $R=150d$, and a density of scatterers $p_{\rm occ}=1/10$.
In Fig.\ref{fig:fig14}a we show raw data without any selection, in Fig.\ref{fig:fig14}b we present only resonances
with a width $\Gamma < 10^{-6} E_0/\hbar$, in Fig.\ref{fig:fig14}c we present only resonances and bound states with a participation surface $S_p^{1/2}/d<9.5$ (this value is motivated by Figs.\ref{fig:fig17}a and c, see below), and in Fig.\ref{fig:fig14}d we present only resonances and bound states with r.m.s.\ size $\sigma/d< 4.2$ (this value is motivated by Figs.\ref{fig:fig17}b and d, see below). Here the participation surface is
\be
\label{eq:Sp}
S_{\rm p} \equiv \frac{1}
{\rho\sum_{i=1}^{N} |D_i|^4}
\ee
where the $D_i$ are normalized as $\sum_{i=1}^N |D_i|^2=1$, and $\sigma$ is defined by the 2D version of Eq.(\ref{eq:sigma}).

\begin{figure}[htb]
\includegraphics[width=9cm,clip=]{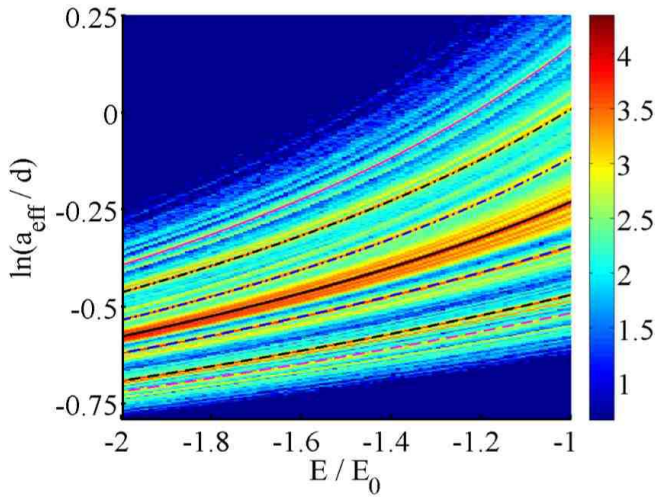}
\caption{
(Color online) Zoom of figure \ref{fig:fig14}a, in the region of $E<0$ (here $\delta E= 0.01 E_0$, and $\delta \ln(a_{\rm eff}/d)=0.003$). On the color map representing the density of states (see
caption of Fig.\ref{fig:fig14}a), are plotted the energies of bound states of an $A$
atom with one, two or three $B$ atoms, called respectively dimer $AB$, trimer $AB_2$, and
tetramer $AB_3$. In particular are plotted the energies of a dimer Eq.(\ref{eq:dim2D})
(black solid line), of a trimer Eq.(\ref{eq:trim2D})
with two $B$ atoms separated by a distance $r_{12}=d$ (dashed black
and dash-dotted black lines), of a trimer with two $B$ atoms separated by a
distance $r_{12}=\sqrt{2}d$ (dashed and dash-dotted blue lines), of a tetramer Eq.(\ref{eq:tetr2D})
with three $B$ atoms
separated by distances $r_{13}=r_{23}=d$ and $r_{12}=\sqrt{2}d$ (solid and dashed
magenta lines). It is worth noting that the predictions for
the energies of such isolated bound states correspond to the higher density lines
of the color map plot.
}
\label{fig:fig15}
\end{figure}

\begin{figure}[htb]
\begin{center}
\includegraphics[width=8cm,clip=]{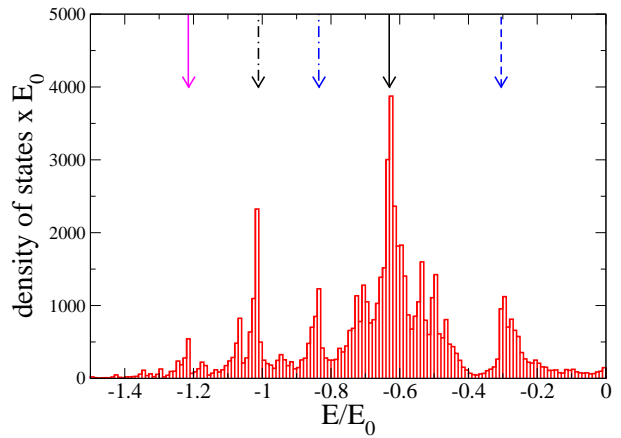}
\caption{(Color online) For the 2D system: density of bound states for a given value of the 2D effective scattering length
$a_{\rm eff}=d$. The histogram is the result of the average over 100 random realizations of the scatterers positions in a
disk of radius $R=50d$, and a filling factor $p_{\rm occ}=1/10$. On average, we find $0.85$ bound states per scatterer. As done in Fig.\ref{fig:fig15}, we also plot
 the energies of bound states of an $A$
atom with one, two or three $B$ atoms. The vertical lines with arrows indicate the energies of few body bound states $AB_n$. From left to right are shown the solid magenta line: energy of the tetramer $AB_3$ Eq.(\ref{eq:tetr2D})
with three $B$ atoms
separated by distances $r_{13}=r_{23}=d$ and $r_{12}=\sqrt{2}d$,
dash-dotted black line: energy of a $AB_2$ trimer Eq.(\ref{eq:trim2D})
with two $B$ atoms separated by a distance $r_{12}=d$,
dash-dotted blue line: energy of $AB_2$ with two $B$ atoms separated by a distance $r_{12}=\sqrt{2}d$,
black solid line: energy of a $AB$ dimer Eq.(\ref{eq:dim2D}),
dashed blue line: energy of $AB_2$ with two $B$ atoms separated by a distance $r_{12}=\sqrt{2}d$.}
\label{fig:fig16}
\end{center}
\end{figure}

In figure \ref{fig:fig15} we show a zoom of Fig.\ref{fig:fig14}a in the region of lowest energies. The streams of higher densities of bound states are shown to be in correspondence with the energy of some few body bound states ($AB$ dimers,  $AB_2$ trimers,  $AB_3$ tetramers). We recall that, in 2D, the matter wave has a single bound state $AB$ on a isolated scatterer, with an energy \cite{Olshanii07,Enerfewbody}
\be
\label{eq:dim2D}
E_{\rm dim} = -\frac{2\hbar^2}{m a_{\rm eff}^2\;e^{2\gamma}}.
\ee
The energy $E_{\rm trim}=-\hbar^2q^2/(2m)$ of the $AB_2$ trimer is given by \cite{Enerfewbody}
\be
\label{eq:trim2D}
\frac{a_{\rm eff}\;e^{\gamma}}{2}=\frac{1}{q}\exp{\left[\pm K_0(qr_{12})\right]},
\ee
where $r_{12}$ is the distance between the two $B$ scatterers.
The energy $E_{\rm tetra}=-\hbar^2q^2/(2m)$ of the $AB_3$ tetramer is given \cite{Enerfewbody}, in the particular case of $r_{13}=r_{23}$, by the expression
\be
\label{eq:tetr2D}
\frac{a_{\rm eff}\;e^{\gamma}}{2}=\frac{1}{q}\exp{\left\{\frac{1}{2}[\beta\pm(\beta^2+\delta^2)^{1/2}]\right\}},
\ee
where we only considered the branches $E_{\rm tetra}\neq E_{\rm trim}$, $\beta=K_0(qr_{12})$, $\delta=K_0(qr_{13})=K_0(qr_{23})$, and $r_{ij}$ is the distance between the two $B$ scatterers $i$ and $j$.
The signature left by few body bound states in the density of states can be further appreciated in Fig.\ref{fig:fig16}, where we plot the density of states as a function of the energy $E<0$, at the fixed value of $a_{\rm eff}=d$. The energies of some few body bound states (vertical arrows in the figure) are shown to correspond to peaks in the density of states. Other peaks can be identified with other few body bound states (not shown). The same calculation, performed for $a_{\rm eff}=3d$, shows a broad structure where essentially all the peaks are washed out.

In figure \ref{fig:fig17} we show histograms of the number of states as a function of the participation surface $S_{\rm p}$ and of the r.m.s. size $\sigma$, for positive energies (Figs.\ref{fig:fig17}a and b) and for negative  energies (Figs.\ref{fig:fig17}c and d). The structure of the histograms at positive energies shows broad structures corresponding to states extending over the whole disordered system. This fact does not contradict the general statement that all states are localized in a 2D system \cite{Ramakrishnan79}. Indeed, at large energies and large values of $a_{\rm eff}$, the localization length $\xi$ becomes very large, larger than the system size. For negative energies, the histograms show one narrow peak with no tails, indicating that all the states are localized with a localization length of few lattice spacings only.

In figures \ref{fig:fig18}a and b, we plot the value of the width $\Gamma$ of the resonances, in the $(E/E_0,\ln(a_{\rm eff}/d))$ plane. In Fig.\ref{fig:fig18}a, we considered only resonances with a participation surface $S_{\rm p}^{1/2}/d<9.5$, while in Fig.\ref{fig:fig18}b,  we considered only resonances with a r.m.s size $\sigma/d<4.2$. In both figures wide regions are present  corresponding to very small values of $\Gamma$, i.e. to extremely long-lived resonances (in practice infinitely long-lived at the scale of the experiments).

\section{Conclusion}
\label{sec:conclusions}
We performed a quantitative study of the 3D and 2D strong localization of matter waves in a random potential realized by point-like scatterers (atoms) pinned at the nodes of a cubic or square lattice. This model allows for an exact numerical analysis of both the localization length and density of states, as functions of the matter wave energy and of the effective scattering length between the matter wave and a single scatterer. We considered systems having a number of scatterers of the same order as that achieved in current experiments with ultracold atomic gases ($N\sim 10^5$), corresponding to systems with a diameter $\sim140$ lattice spacings in 3D for a lattice filling factor $=1/10$.

In 3D, we found evidence for the occurrence of several energy mobility edges, for both positive and negative matter wave energies $E$. For $E>0$, we found a mobility edge for a positive effective scattering length of the order of the mean distance between scatterers. For a too large positive, or for a negative value of $a_{\rm eff}$ we found no evidence of any mobility edge, in agreement with the predictions in \cite{Bart}. For $a_{\rm eff}$ small and positive we found the occurrence of an energy gap between $E=0$ and an upper bound coinciding with a mean field effect $E=g_{\rm eff}\rho$ where $\rho$ if the density of scatterers, and $g_{\rm eff}=2\pi\hbar^2 a_{\rm eff}/m$. Within this energy gap, there is a small localization length with no evidence of localized states
\cite{noteinp}. For $E<0$, where the matter wave is bound inside the gas of scatterers (a case not explored in previous works to our knowledge), we found localized bound states and evidence of two mobility edges separated by an energy interval where only extended bound states are present. We also found, for $a_{\rm eff}$ small and negative, extended bound states in the energy interval between the mean field shift $E=\rho g_{\rm eff}$ and $E=0$. To ascertain the presence of the mobility edges, and to identify the universality class of this disorder induced transition, a dedicated analysis should be undertaken    using finite size scaling techniques to determine the critical exponent. For experimental purposes, large values of the effective scattering length may be obtained by using Feshbach resonances to control the free space scattering length $a$ \cite{Castin06}. Recently  $a>1\mu$m has been obtained with Li gases \cite{Hulet09}.

In 2D, no evidence of mobility edges is found for either positive or negative energies. Contrary to the 3D case, at negative energies, all bound states are localized. We identified regions in the $(E/E_0,\ln(a_{\rm eff}/d))$ plane  where the localization length is small and the density of states is high. This happens for $a_{\rm eff}$ of the order of the average distance between scatterers, for a wide interval of energies. At high energies we find a rapid, exponential-like increase of the localization length with  energy. At negative energies the localization length shows no such  rapid increase of $\xi$.
 To our knowledge, large values of the 2D effective scattering length have not yet been observed. It is thus   interesting to study the dependence of the 2D $a_{\rm eff}$ with respect to the free space scattering length $a$, and the harmonic confinement in the microtrap.

\begin{figure}[htb]
\begin{center}
\includegraphics[width=4cm,clip=]{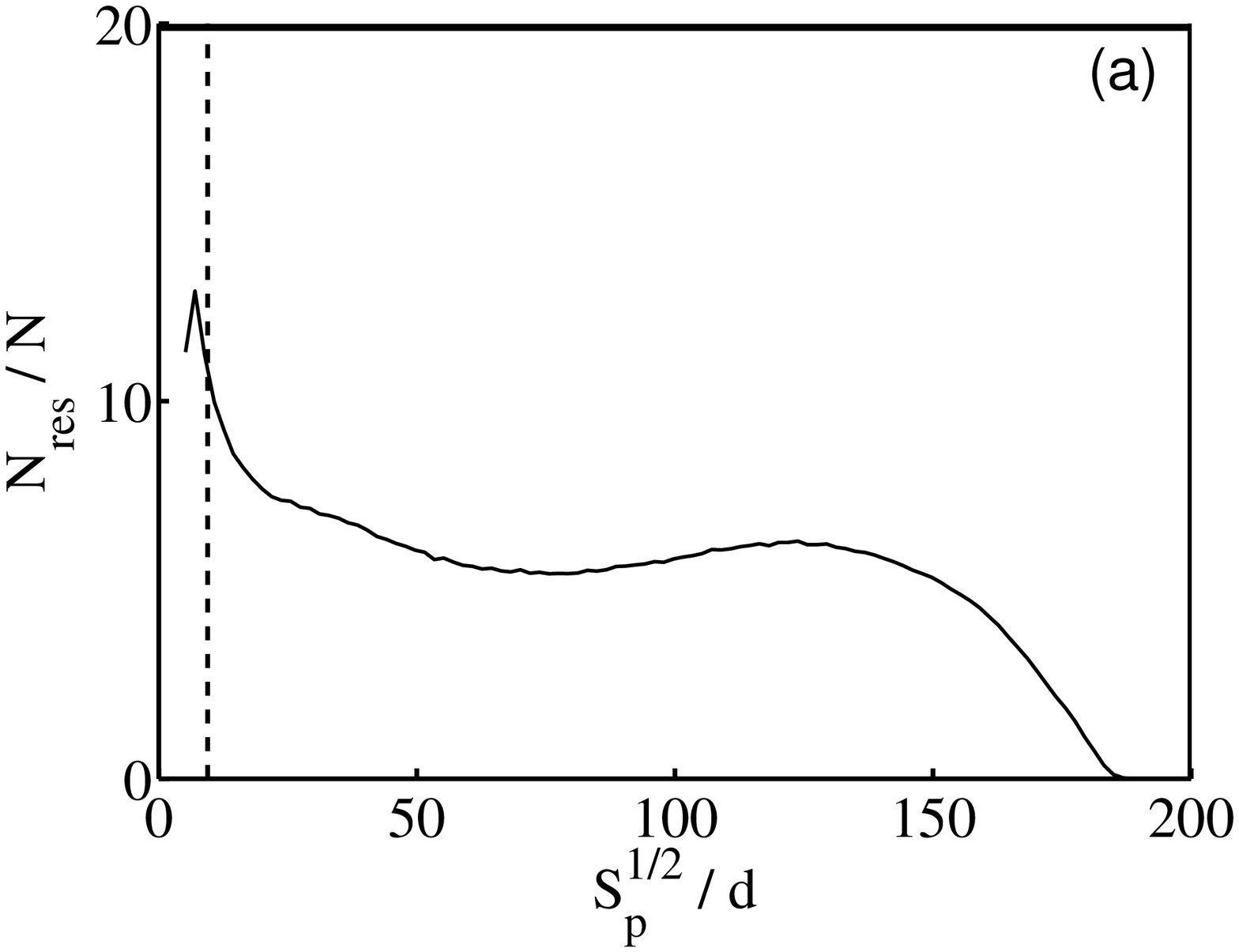}
\includegraphics[width=4cm,clip=]{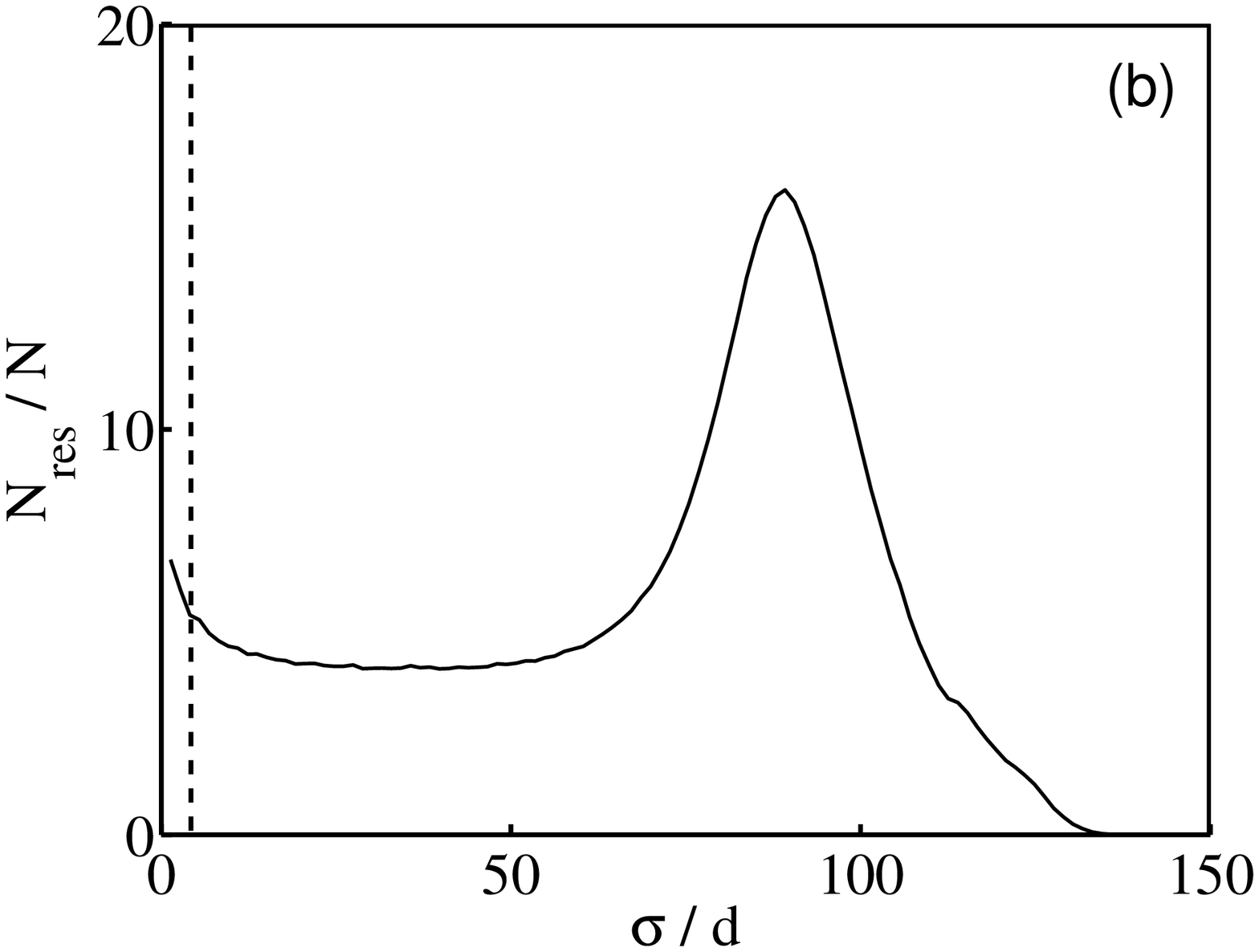} \\
\includegraphics[width=4cm,clip=]{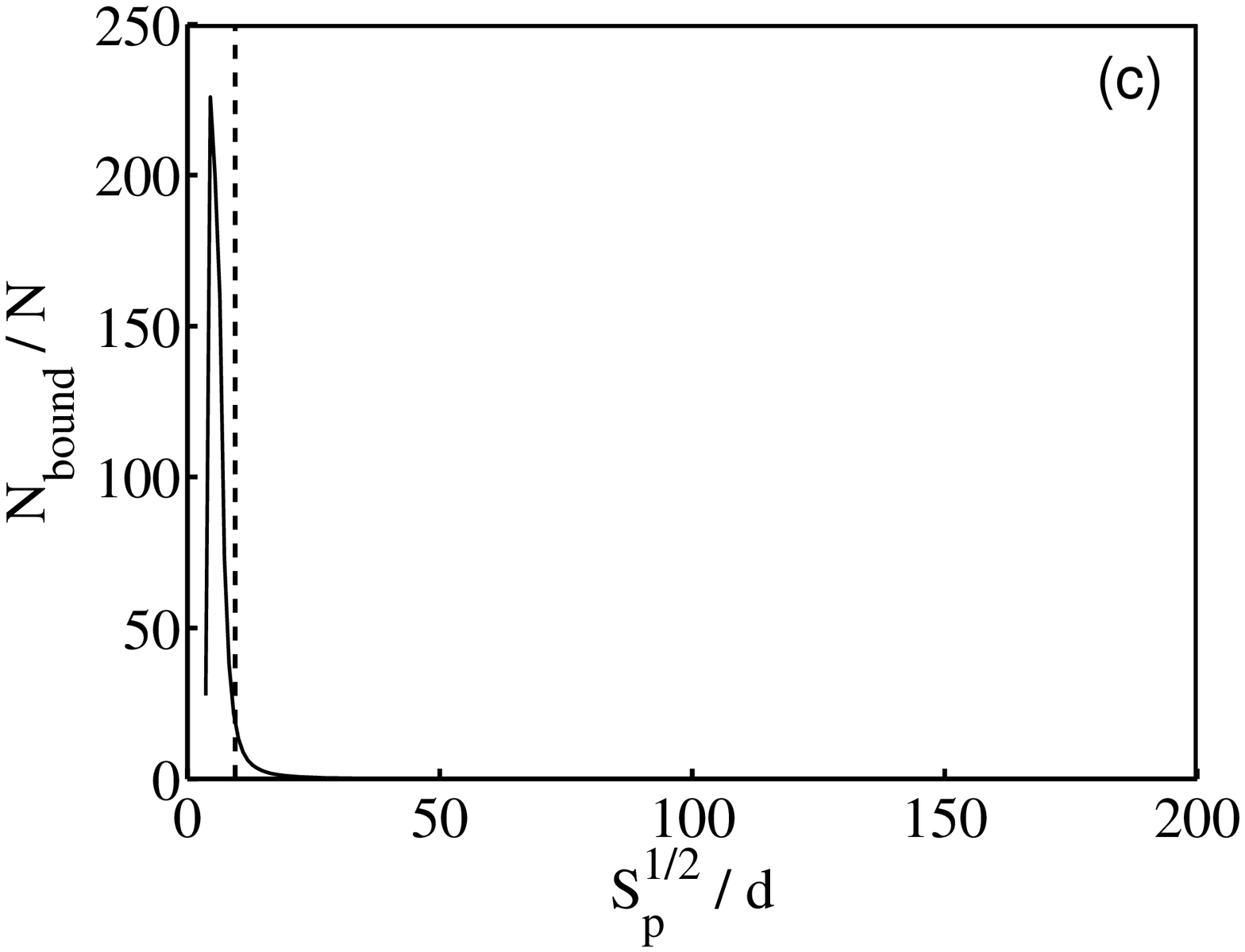}
\includegraphics[width=4cm,clip=]{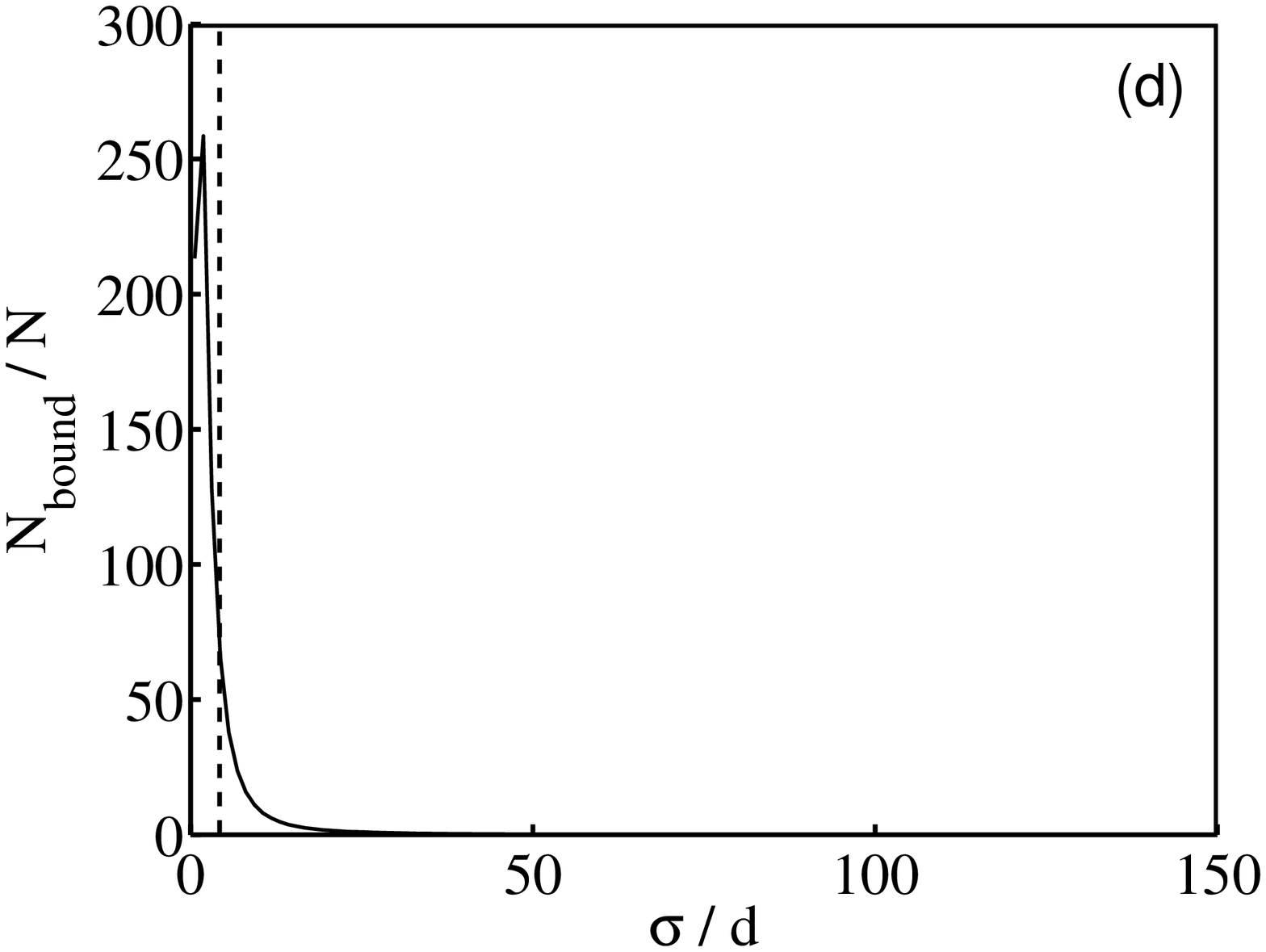}
\caption{For the 2D system: For all the resonances with $E_{\rm res}/E_0\in (0,2)$
and $\ln(a_{\rm eff}/d) \in (-3,3)$, and for all the bound states
with $E/E_0\in(-2,0)$ and $\ln(a_{\rm eff}/d) \in (-3,3)$, the figures show an
histogram giving the number of resonances (in (a) and (b)) and bound states (in (c) and (d)) per scatterer (the number of scatterers is $N=7 \times 10^3$) as a function
of:  in (a) and (c), the square root of the participation surface defined in Eq.(\ref{eq:Sp});
in (b) and (d), the r.m.s. size in real space defined by the 2D version of Eq.(\ref{eq:sigma}).
The bin size is $\sim0.6d$ for (a), $\sim1.37d$ for (b), $\sim0.3d$ for (c), and $\sim1.37d$ for (d).
The parameters are the same as in Fig.~\ref{fig:fig14}a. The dashed vertical lines
are the values $S_{\rm p}^{1/2}/d\simeq 9.5$ and $\sigma/d\simeq 4.2$  
used in Fig.~\ref{fig:fig14}c and Fig.~\ref{fig:fig14}d to select the bound states and resonances
that are spatially localized and to filter out the bound states and resonances that are
weakly localized.
}
\label{fig:fig17}
\end{center}
\end{figure}

\acknowledgments
We acknowledge useful discussion with D. Delande. Numerical calculations have been performed on the IFRAF cluster (the entire work needed several CPU months per core, on 96 Intel Xeon Quadcore processors). M.A. acknowledges financial support from the ERC Project FERLODIM N.228177. D.H. acknowledges financial support from CNRS, UPMC and IFRAF during his stay in Paris.

\appendix

\section{Monotonic behavior of the eigenvalues of $M(E)$}
\label{app:mon}

We show that, for $E=-\frac{\hbar^2 q^2}{2m}<0$,
the eigenvalues $m_i(E)$ of the real
symmetric matrix $M(E)$ are monotonic functions of the energy $E$.
This results from the Hellmann-Feynman theorem, and from the fact
that the matrix $dM(E)/dE$ is positive in 3D (respectively negative
in 2D), so that $dm_i/dE>0$ in 3D (respectively $dm_i/dE<0$ in 2D).

Let us start with the 3D case.
The idea is to show that, for all $1\leq i,j \leq N$,
\be
\label{eq:form_deriv3D}
\frac{d}{dE} M_{ij}(E)=\frac{2\pi\hbar^2}{m} \langle \rr_i|
\frac{1}{(E-h_0)^2}|\rr_j\rangle,
\ee
where $h_0=-\frac{\hbar^2}{2m} \Delta^{\rm 3D}_\rr$.
Since $(E-h_0)^2$ is a positive operator, the positivity of
$dM/dE$ readily follows.
For $i\neq j$ one has, from Eq.(\ref{eq:def_M}), that
\be
M_{ij} = -\frac{2\pi\hbar^2}{m}\langle \rr_i| \frac{1}{E-h_0}
|\rr_j\rangle,
\ee
since $(E-h_0) g_0(\rr)=\delta(\rr)$.
Taking the derivative with
respect to $E$ gives Eq.(\ref{eq:form_deriv3D}).
It remains to check that Eq.(\ref{eq:form_deriv3D}) also holds for $i=j$
by a direct calculation. On one hand, $M_{ii}=a_{\rm eff}^{-1}
-q$ so that $dM_{ii}/dE=m/(\hbar^2 q)$. On the other hand, introducing
a closure relation in the plane wave basis, one indeed finds
\begin{multline}
\frac{2\pi\hbar^2}{m}\langle \rr_i| \frac{1}{(E-h_0)^2}|\rr_i\rangle= \\
\frac{2\pi\hbar^2}{m}\int\frac{d^3k}{(2\pi)^3} \frac{1}{(E-\hbar^2k^2/2m)^2}
=\frac{m}{\hbar^2 q}.
\end{multline}

\begin{figure}[htb]
\begin{center}
\includegraphics[width=9cm,clip=]{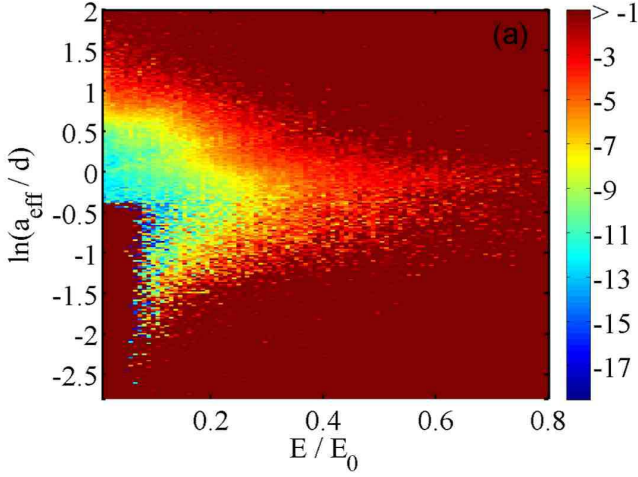}
\includegraphics[width=9cm,clip=]{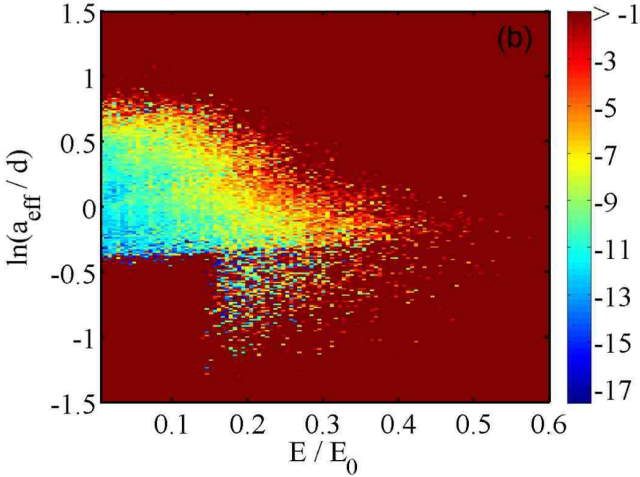}
\caption{(Color online) For the 2D system: Width $\Gamma$ of the resonances as a function of the energy $E$
and the logarithm of the effective scattering length $\ln(a_{\rm eff}/d)$ [see Eqs. (\ref{eq:zrescompl}) and (\ref{eq:fs}), applied to the eigenvalues $m_i^{\infty}(E)$ of the matrix $M_{\infty}$ of Eq.(\ref{eq:M2ds})]. The physical parameters
are the same as in Fig.~\ref{fig:fig14}. The plane $(E,\ln(a_{\rm eff}/d))$ is decomposed
in rectangular bins of widths $\delta E$ and $\delta \ln(a_{\rm eff}/d)$.
The color map (see bar on the right)
is applied to the quantity  $\log_{10} \hbar \langle \Gamma\rangle/E_0$
where $\langle \Gamma\rangle$ is the mean value of $\Gamma$ for the resonances
within a given bin.
The resonances are filtered in (a) over the participation volume
as in Fig.~\ref{fig:fig14}c and in (b) over the r.m.s.\ size $\sigma$
as in Fig.~\ref{fig:fig14}d.
In (a) one has $\delta E= 0.019 E_0$ and $\delta \ln(a_{\rm eff}/d)= 0.027$,
and in (b) one has
$\delta E= 0.011 E_0$ and $\delta \ln(a_{\rm eff}/d)= 0.024$.}
\label{fig:fig18}
\end{center}
\end{figure}

In 2D, the proof is quite similar. One simply has to show that
\be
\frac{d}{dE}M_{ij}(E) = -\frac{\pi\hbar^2}{m} \langle
\rr_i| \frac{1}{(E-h_0)^2}|\rr_j\rangle
\ee
for all $1\leq i,j\leq N$, with
$h_0=-\frac{\hbar^2}{2m} \Delta^{\rm 2D}_\rr$.
For $i\neq j$, one has from Eq.(\ref{eq:def_M2d}) that
\be
M_{ij} = \frac{\pi\hbar^2}{m}\langle \rr_i| \frac{1}{E-h_0}
|\rr_j\rangle,
\ee
since $(E-h_0) g_0(\rr)=\delta(\rr)$.
For $i=j$, one again performs a direct calculation. First, $dM_{ii}/dE=
d\ln q/dE=1/(2E)$. Second, a closure relation in the plane wave basis indeed gives
\begin{multline}
-\frac{\pi\hbar^2}{m} \langle
\rr_i| \frac{1}{(E-h_0)^2}|\rr_i\rangle
=\\
 -\frac{\pi\hbar^2}{m} \int\frac{d^2k}{(2\pi)^2} \frac{1}{(E-\hbar^2k^2/2m)^2}=\frac{1}{2E}.
\end{multline}

A simple consequence of the monotonic behavior of the eigenvalues $m_i$, is that the total number of bound states, for a given
realization of disorder,
is given by the number of positive eigenvalues of $M(E=0)$
in 3D, and by the number of negative eigenvalues of $M(E\to 0^-)$
in 2D.

\section{Matter wave wavefunction in terms of $D_i$}
\label{AppPsiD}
We restrict here to the three-dimensional case, the generalisation to the two-dimensional case
being straightforward.
Let us consider first a matter wave bound state with eigenenergy $E_0<0$.
$E_0$ is a pole of the Green's function $G(\rr,\rr_0)$ defined in Eq.(\ref{eq:def_green}),
so that, according to Eq.(\ref{eq:gfd3dexp}),  an eigenvalue $m_0(E)$ of the matrix $M(E)$ vanishes for $E=E_0$.
Since $M(E)$ is real symmetric, its inverse has the spectral decomposition
\be
\left[M^{-1}\right]_{ij}(E) = \sum_{n=0}^{N-1} \frac{1}{m_n(E)} D_i^{(n)}(E)  D_j^{(n)}(E)
\ee
where $(D_i^{(n)})_{1\leq i \leq N}$ is the orthonormal eigenvector of $M(E)$ with real components and eigenvalue
$m_n(E)$.
In particular, $\sum_{i=1}^N (D_i^{(n)})^2 =1$.
When this spectral decomposition is injected in Eq.(\ref{eq:gfd3dexp}), together with the expansion $m_0(E)=(E-E_0) m'(E_0)+\ldots$, it leads to
\be
\label{eq:divG}
G(\rr,\rr_0) \underset{E\to E_0}{\sim} \frac{\psi_0(\rr)\psi_0(\rr_0)}{E-E_0}
\ee
with
\be
\psi_0(\rr) = \frac{(2\pi\hbar^2/m)^{1/2}}{[m_0'(E_0)]^{1/2}} \sum_{i=1}^{N} D^{(0)}_i(E_0) g_0(\rr-\rr_i).
\ee
As it appears from Eq.(\ref{eq:divG}),
$\psi_0$ is the wavefunction of the bound state of energy $E_0$. Note that $m_0'(E_0)$ is positive according
to Appendix \ref{app:mon}.

Let us consider now a a resonance $z_{\rm res}$ of the system with a complex energy $z_{\rm res}$.
Then $z_{\rm res}$ is a pole of the analytic continuation of the Green's function from the upper half-plane to the lower half-plane,
so that, according to the corresponding analytic continuation of Eq.(\ref{eq:gfd3dexp}), an eigenvalue of the matrix
$M(z)$ vanishes for $z=z_{\rm res}$.
Since $M(z)$ is complex symmetric, if it is diagonalizable its inverse has the spectral decomposition
\be
[M^{-1}]_{ij}(z) = \sum_{n=0}^{N-1} \frac{1}{m_n(z)} D_i^{(n)}(z)  D_j^{(n)}(z)
\ee
where $(D_i^{(n)})_{1\leq i \leq N}$ is the right eigenvector of $M(z)$ with complex components and eigenvalue $m_n(z)$,
the corresponding left eigenvector is simply its complex conjugate, so that the normalisation condition
is $\sum_{i=1}^N (D_i^{(n)})^2 =1$.
When this spectral decomposition is injected in the analytic continuation of Eq.(\ref{eq:gfd3dexp}), together with the expansion $m_0(z)=(z-z_{\rm res}) m'(z_{\rm res})+\ldots$, it leads to
\be
\label{eq:divG2}
G(\rr,\rr_0) \underset{z\to z_{\rm res}}{\sim} \frac{\psi_0(\rr)\psi_0(\rr_0)}{z-z_{\rm res}}
\ee
with
\be
\psi_0(\rr) = \frac{(2\pi\hbar^2/m)^{1/2}}{[m_0'(z_{\rm res})]^{1/2}} \sum_{i=1}^{N} D^{(0)}_i(z_{\rm res}) g_0(\rr-\rr_i).
\ee
In the limit of an infinitely extended disordered system $\psi_0$ would correspond to the wavefunction of a localized state of energy $z_{\rm res}$.



\begin{thebibliography}{99}
\bibitem{noteaeff}
The effective scattering length generally differs from the free space scattering
 length $a$, due to the effect of the confining potential experienced by the $B$ atoms: $a_{\rm eff}/a_{\rm ho}$ is a function
 of $a/a_{\rm ho}$ and of the $A-B$ mass ratio, as calculated in
 \cite{Castin06}.
\bibitem{variante} An alternative definition of $\kappa$,
not used in this work, is to consider
the logarithm of the angular average of $|t(\nn)|^2$ rather than the angular average
of the logarithm, setting (for $E>0$) $\kappa_{\rm alt}=  -\lim_{R\to+\infty} (2R)^{-1}
\langle \ln \langle |t(\nn)|^2\rangle_{\nn}\rangle$ where $\langle\ldots\rangle_{\nn}$
is the angular average. From the convexity of the function $x\to -\ln x$
one has $\kappa_{\rm alt}\leq \kappa$. Using the conservation of the flux of probability
one can show that
\be
\langle |t(\nn)|^2\rangle_{\nn} = 1 -  k^{-1} \mathrm{Im}\, \sum_{i=1}^{N} D_i s_i
\label{eq:var}
\ee
with $s_i=(2\pi\hbar^2/m) g_0(\rr_i-\rr_0)$. To this end one considers the flux of the probability
current $\phi_{\rm far}$ through a sphere of arbitrarily large radius and the flux
$\phi_{\rm near}$ through a sphere of arbitrarily small radius around the source location
$\rr_0$. One finds from Eq.~(\ref{eq:asympt}) that $\phi_{\rm far}= [m/(2\pi \hbar^2)]^2 (4\pi \hbar k/m)
\langle |t(\nn)|^2\rangle_{\nn}$. Expanding the Green's function close to the
source up to order $|\rr-\rr_0|^0$, one finds $\phi_{\rm near}$ and one obtains Eq.~(\ref{eq:var}). A practical consequence
is that the Lyapunov exponent $\kappa_{\rm alt}$ may be deduced from measurements of the Green's function
close to the source.
\bibitem{pourquoi}
In the limit $p_{\rm occ}\to 0$
the fact that the scatterers positions are distributed
on a lattice rather than uniformly in continuous space is expected to be
no longer relevant, so that the lattice spacing  $d$ drops
out of the problem. Three length scales then remain,
the mean distance $\rho^{-1/3}=d/p_{\rm occ}^{1/3}$
between scatterers, $1/k$ and $a_{\rm eff}$.
From dimensional analysis, $\rho^{1/3} \xi$ is a function
of $\rho^{1/3} a_{\rm eff}$ and of $\rho^{-1/3} k$.
The minimal achievable $\xi$ is then reached for values of $a_{\rm eff}$
and $1/k$ proportional to $\rho^{-1/3}$, and scales as $\rho^{-1/3}$. This also motivates the choice of $\rho^{-1/3}$ and $\hbar^2\rho^{2/3}/m$ as units of length and energy, respectively.
\bibitem{Enerfewbody}
The energy of the $AB_{n}$ bound states between an $A$ atom and $n$ $B$ atomic scatterers can be calculated by using equation (\ref{eq:param}) in 3D, or its 2D equivalent $\ln(a_{\rm{eff}}/d)=-m_i^{\infty}(E)$ from \ref{eq:M2ds}, where the eigenvalues $m^{\infty}$ of $M_{\infty}$ can be calculated analytically in the case of $n\leq 4$.
\bibitem{noteexact}
One starts with the exact property that, at an eigenenergy $E$, the matrix $M$ defined
in Eq.(\ref{eq:def_M}) has a zero eigenvalue. Setting $E=-\hbar^2 q^2/(2m)$, where $q>0$,
$q$ will deviate from $1/a_{\rm eff}$ by an exponentially small amount $\delta q$ if
$a_{\rm eff}\rho^{1/3} \to 0^+$. Then to first order in $\delta q$, $E-E_{\rm dim}=
-\hbar^2\delta q/(m a_{\rm eff}).$ Replacing $q$ with its zeroth order value $1/a_{\rm eff}$
in the off-diagonal terms of $M(E)$, one obtains the eigenvalue problem
$(E-E_{\rm dim}) D_i = \sum_{j\neq i} t_{\rm trans}(r_{ij}) D_j$.
\bibitem{oscillations}
As there is no averaging over disorder, the spatial oscillations
in the Green's function are not washed out. For values
of $\xi$ larger than the ones presented in the figure, that is
for $\xi$ getting larger than the lattice constant $d$,
our fitting procedure based on Eq.(\ref{eq:ff}) is no longer
appropriate.
\bibitem{dispersion}
This picture is expected for a quadratic matter wave dispersion relation,
as considered here. For more complicated dispersion relations,
as in the Hubbard model, there may be several mobility edges \cite{revue_kirk}.
\bibitem{newton}
The generalized Hellmann-Feynman theorem is
$d m^{\infty}_i(z)/dz = \vec{v}_i\,^* \cdot dM^{\infty}(z)/dz \vec{u}_i$,
where $\vec{u}_i$ and $\vec{v}_i$ are the right and left eigenvectors,
respectively, corresponding to the eigenvalue $m^{\infty}_i(z)$,
and normalised as $\vec{v}_i\,^*\cdot\vec{u}_i=1$.
Since $M^{\infty}(z)$ is complex symmetric, one has
$\vec{v}_i=\vec{u}_i\,^*$ so that it suffices to calculate
the right eigenvector numerically.
\bibitem{SmallE} We note the presence of a narrow vertical band of resonances for small values of the energy $E$ in the first energy pixel of Fig.\ref{fig:fig7}b. Indeed, for very small values of $E$ an extra-density of long-lived resonances appear, with values of $a_{\rm eff}$ spreading over a large interval (including negative values out of the range of Fig.\ref{fig:fig7}b). Nonetheless, these resonances are not spatially localized, and are eliminated by the filtering used in Fig.\ref{fig:fig7}d. These long-lived extended states correspond to $kR\leq1$, which suggest that they are related to finite size effects.
\bibitem{unif}
For a uniformly distributed state in the sphere of radius $R$,
one has $V_{\rm p}^{1/3}/R=(4\pi/3)^{1/3}\simeq 1.6$ and
$\sigma/R=(3/5)^{1/2}\simeq 0.77$.
\bibitem{notemat}
Mathematically, it is possible to construct extended wavefunctions with a finite participation
ratio. Restricting for simplicity to the 1D case,  we define the unnormalized wavefunction
$\psi(x)=\sum_{n=1}^{N} (n/d)^{1/2} \chi[(x-nd)/(d/n)]$, where $d$ is the spatial period of a
1D lattice, and $\chi(x)=1$ if $|x|< 1/2$, zero otherwise. None of the $N$ compact support
spatial components of $\psi$ overlap, so that $\int_\mathbb{R} dx |\psi(x)|^{2k}
= \sum_{n=1}^{N} (n/d)^{k-1}$. The participation length $l_p=(\int_\mathbb{R} dx |\psi(x)|^2)^2
/(\int_\mathbb{R} dx |\psi(x)|^4)$ is thus $l_P = 2 N d/(N+1)$ and has a finite limit $2 d$
when $N\to +\infty$. On the contrary, the root mean square size $\sigma$ diverges
as $d N/\sqrt{12}$.
\bibitem{noteinp}
In this regime, one could however imagine a highly improbable realization of disorder where
a cavity of radius $l$ with no scatterers exists inside the system. If $\hbar^2/(m l^2) < \rho g_{\rm eff}$
there may be a localized state inside this cavity with $0<E<\rho g_{\rm eff}$.
\end{thebibliography}
\end{document}